\documentclass[a4paper,twoside,openany,11pt]{memoir} 
\pdfoutput=1

\title{
    Next-to-Leading Order Running in the SMEFT
}
\author{
    L. Born, J. Fuentes-Martín, and A.E. Thomsen
}

\def\fullheadfoot{0} 
\usepackage[english]{babel}
\usepackage{microtype,xspace}
\usepackage[utf8]{inputenc}	

\usepackage{amsfonts,amsmath,amssymb,bm,mathrsfs,mathtools,dsfont} 	
\allowdisplaybreaks 	
\usepackage[table]{xcolor}
\usepackage{hyperref}
\usepackage{slashed}
\usepackage{multicol}

\usepackage{tikz}
\usetikzlibrary{shapes, arrows, positioning, decorations.pathmorphing, decorations.pathreplacing, decorations.shapes, decorations.text, calc} 
\tikzset{every picture/.style={line width=.8pt}}

\usepackage{siunitx}
\usepackage{geometry}
\usepackage[sort&compress,numbers,merge]{natbib}
\usepackage{chapterbib}
\addto\captionsenglish{\renewcommand*{\bibname}{References}}
\makeatletter
\renewcommand{\@memb@bchap}{ 
\bibmark \prebibhook
}
\makeatother

\usepackage{graphicx}
\graphicspath{{./Figures/}}
\usepackage{caption,subcaption}
\usepackage{multirow}
\renewcommand{\arraystretch}{1.2}
\usepackage{tabularx}
\usepackage{hhline}
\newcolumntype{Y}{>{\centering\arraybackslash}X}
\newcolumntype{b}{>{\columncolor{blue!30}}l}

\usepackage[shortlabels]{enumitem}
\setlist{itemsep=.1em,topsep=.5em}
\SetEnumerateShortLabel{i}{\textit{\roman*}}

\definecolor{red}{rgb}{0.6,.0706,.1373}
\definecolor{blue}{rgb}{0,0.396,0.741}
\definecolor{green}{rgb}{0.25,0.6,0.2}
\definecolor{teal}{rgb}{0.11,0.6,0.6}
\definecolor{orange}{rgb}{.8, .4806, 0.173}
\definecolor{yellow}{rgb}{.8, .7, 0.05}
\colorlet{blueref}{blue!80!black}
\colorlet{bluelink}{blue!90!black}
\hypersetup{
	colorlinks, 
	bookmarksopen, 
	bookmarksnumbered,
	citecolor=blueref, 		
	linkcolor=bluelink,	
	urlcolor=bluelink,			
	pdftitle={\thetitle},    
	pdfauthor={\theauthor},    	
}

\addto\captionsenglish{}

\renewcommand{\contentsname}{Contents}
\renewcommand{\printtoctitle}[1]{}

\settocdepth{subsection}
\newcommand{\toc}{ {
	\hypersetup{linkcolor = black} 
	\vspace*{-.06\textheight}	
	\tableofcontents*
	\thispagestyle{empty} 
} }

\makeatletter
\newcommand*\ifthispageodd{%
  \checkoddpage
  \ifoddpage
    \expandafter\@firstoftwo
  \else
    \expandafter\@secondoftwo
  \fi
}
\makeatother

\usepackage[noindentafter]{titlesec} 
\counterwithout{section}{chapter} 
\counterwithout{figure}{chapter} 
\counterwithout{table}{chapter} 
\numberwithin{equation}{section} 
\setsecnumdepth{subsubsection}

\DeclareMathVersion{sans}
\SetSymbolFont{operators}{sans}{OT1}{cmbr}{m}{n}
\SetSymbolFont{letters}{sans}{OML}{cmbrm}{m}{it}
\SetSymbolFont{symbols}{sans}{OMS}{cmbrs}{m}{n}
\SetMathAlphabet{\mathit}{sans}{OT1}{cmbr}{m}{sl}
\SetMathAlphabet{\mathbf}{sans}{OT1}{cmbr}{bx}{n}
\SetMathAlphabet{\mathtt}{sans}{OT1}{cmtl}{m}{n}
\SetSymbolFont{largesymbols}{sans}{OMX}{iwona}{m}{n}

\DeclareMathVersion{boldsans}
\SetSymbolFont{operators}{boldsans}{OT1}{cmbr}{b}{n}
\SetSymbolFont{letters}{boldsans}{OML}{cmbrm}{b}{it}
\SetSymbolFont{symbols}{boldsans}{OMS}{cmbrs}{b}{n}
\SetMathAlphabet{\mathit}{boldsans}{OT1}{cmbr}{b}{sl}
\SetMathAlphabet{\mathbf}{boldsans}{OT1}{cmbr}{bx}{n}
\SetMathAlphabet{\mathtt}{boldsans}{OT1}{cmtl}{b}{n}
\SetSymbolFont{largesymbols}{boldsans}{OMX}{iwona}{bx}{n}

\titleformat{\section}{\centering \needspace{5\baselineskip}\Large \bfseries \sffamily \mathversion{boldsans} \color{blue!80!black} }{\thesection}{15pt}{}{}
\titlespacing{\section}{0pt}{15pt}{5pt}
\titleformat{\subsection}{\needspace{2\baselineskip} \large \sffamily \mathversion{sans} \color{blue!70!black} }{\thesubsection}{10pt}{}{}
\titlespacing{\subsection}{0pt}{10pt}{5pt}
\titleformat{\subsubsection}{\normalsize \sffamily \itshape \mathversion{sans} \color{blue!70!black} }{\thesubsubsection}{10pt}{}{}
\titlespacing{\subsubsection}{0pt}{10pt}{0pt}

\newcommand{\sectionlike}[1]{\phantomsection \addcontentsline{toc}{section}{#1} \setcounter{subsection}{0} \sectionmark{#1}
		\begin{center}
		\needspace{5\baselineskip}
		\Large \bfseries \sffamily \mathversion{boldsans} \color{blue!80!black} #1  
		\end{center}
	\vspace{-5pt} 
}

\renewcommand{\paragraph}[1]{\needspace{1\baselineskip} \vspace{.3em} \indent {\bfseries \boldmath #1 ---}\xspace }
\newcommand{\sectionlikeParagraph}[1]{\vspace{.3em} \noindent {\bfseries \boldmath #1}\xspace }

\ifnum\fullheadfoot=1
	\makepagestyle{standardstyle}
	\makeoddhead{standardstyle}{\sffamily \mathversion{subsectionmath} \rightmark}{}{\sffamily \bfseries{\thepage}}
	\makeevenhead{standardstyle}{\sffamily \bfseries{\thepage}}{}{\sffamily \mathversion{subsectionmath} \leftmark}
	\makeheadrule{standardstyle}{\textwidth}{\normalrulethickness}
	\makepsmarks{standardstyle}{
		\nouppercaseheads
		\createmark{section}{both}{shownumber}{\ }{\hspace{1.2em}  }
		\createmark{subsection}{right}{shownumber}{}{\hspace{1.2em} }
	}
	
	\copypagestyle{appstyle}{standardstyle}
	\makeevenhead{appstyle}{\sffamily \bfseries{\thepage}}{}{ \sffamily \mathversion{subsectionmath} \leftmark}
	\makepsmarks{appstyle}{
		\nouppercaseheads
		\createmark{section}{both}{nonumber}{\ }{\ \ }
		\createmark{subsection}{right}{shownumber}{}{\ \ }
	}
\else 
	\makepagestyle{standardstyle}
	\makeoddfoot{standardstyle}{}{\sffamily -- {\bfseries\thepage} --}{} 
	\makeevenfoot{standardstyle}{}{\sffamily -- {\bfseries\thepage} --}{}
	\copypagestyle{appstyle}{standardstyle}
\fi

\createplainmark{toc}{both}{\contentsname}
\createplainmark{bib}{both}{\bibname}

\makeatletter
\let\MyIntOrig\int
\def\MyIntSpace{\hspace{-.35em}} 
\def\int{\MyInt}
\def\MyInt{\MyIntOrig\MyIntSkipMaybe}
\def\MyIntSkipMaybe{
	\@ifnextchar_{\MyIntSkipScript}{%
		\@ifnextchar^{\MyIntSkipScript}{%
			\@ifnextchar\limits{\MyIntSkipTok}{%
				\@ifnextchar\nolimits{\MyIntSkipTok}{%
					\MyIntSpace}}}}%
}
\def\MyIntSkipScript#1#2{#1{#2}\MyIntSkipMaybe}
\def\MyIntSkipTok#1{#1\MyIntSkipMaybe}

\newcommand{\pushright}[1]{\ifmeasuring@#1\else\omit\hfill$\displaystyle#1$\fi\ignorespaces}
\makeatother

\newcommand{\Tr}{\mathop{\mathrm{Tr}}}

\newcommand{\eminus}{\vcenter{\hbox{\scalebox{0.6}[1]{$ - $}}}}	
\newcommand{\ord}[1]{\mathcal{O}( #1 )}

\newcommand{\hc}{\mathrm{H.c.}}

\newcommand{\dd}{\mathop{}\!\mathrm{d}}
\newcommand{\cc}{\textsf{c}}
\newcommand{\ud}[2]{\phantom{}^{#1}\phantom{}_{#2}}
\newcommand{\du}[2]{\phantom{}_{#1}\phantom{}^{#2}} 

\newcommand{\transpose}{^\intercal}

\newcommand{\sscript}[1]{{\scriptscriptstyle \mathrm{#1}}}

\renewcommand{\L}{\mathcal{L}}
\newcommand{\LL}{\mathrm{L}}
\newcommand{\RR}{\mathrm{R}}
\newcommand{\U}{\mathrm{U}}
\newcommand{\SU}{\mathrm{SU}}

\newcommand{\UV}{\sscript{UV}}

\newcommand{\lzm}{\big(}
\newcommand{\dzm}{\big)}
\newcommand{\lrderiv}[2][\hspace{4pt}]{\overset{\hspace{1pt} \leftrightarrow}{D}_{\hspace{-0.5pt}#2}{}^{\hspace{-4pt}#1}}

\newcommand{\vdagph}{\vphantom{\dagger}}

\newcommand{\ctop}{\boldsymbol{\Delta}}
\newcommand{\kop}{\boldsymbol{K}}
\newcommand{\rstar}{\boldsymbol{R}^{\ast}}
\newcommand{\rstarbar}{\bar{\boldsymbol{R}}^{\ast}}
\newcommand{\rir}{\boldsymbol{R}_{\sscript{IR}}}
\newcommand{\Pbar}{{\bar{\mathcal{P}}}}
\newcommand{\Pf}{\mathcal{P}_{\mathrm{f}}}

\newcommand{\dbar}{\bar{d}}

\newcommand{\bef}{$ \beta $-function\xspace}
\newcommand{\befs}{$ \beta $-functions\xspace}
\newcommand{\msbar}{$ \overline{\text{\small MS}} $\xspace}
\newcommand{\rstarop}{$ \boldsymbol{R}^{\ast} $-operation\xspace}

\colorlet{redref}{red!80!violet}

\usepackage{accents}
\usepackage{trimclip}

\DeclareRobustCommand*{\ptilde}[1]{{\accentset{(\!\trimbox{0pt 1.1ex}{\ensuremath{\string~}}\!)}{#1}}}

\DeclareRobustCommand*{\tildepm}{%
  \mathbin{%
    \ooalign{%
      \hfil\raisebox{1.1ex}{$\scriptscriptstyle(+)$}\hfil\cr
      \hfil\raisebox{-0.3ex}{$-$}\hfil\cr
    }%
  }%
}

\DeclareRobustCommand*{\tildemp}{%
  \mathbin{%
    \ooalign{%
      \hfil\raisebox{1.1ex}{$\scriptscriptstyle(-)$}\hfil\cr
      \hfil\raisebox{-0.3ex}{$\scriptstyle+$}\hfil\cr
    }%
  }%
}

\usepackage{arydshln}
\makeatletter
  \renewcommand*\env@matrix[1][*\c@MaxMatrixCols c]{%
    \hskip -\arraycolsep
    \let\@ifnextchar\new@ifnextchar
  \array{#1}}
\makeatother

\usepackage{graphicx}

\begin{document}

\thispagestyle{empty}
\renewcommand*{\thefootnote}{\fnsymbol{footnote}}

\begin{center}
    {\sffamily \bfseries \fontsize{22}{22}\selectfont \mathversion{boldsans}
    Next-to-Leading Order Running in the SMEFT\\[-.005\textheight]
    \textcolor{blue!80!black}{\rule{.25\textwidth}{.7pt}}\\[.015\textheight]}
    {\sffamily \mathversion{sans} \Large 
    Lukas Born,$^{1}$\footnote{lukas.born@unibe.ch}
    Javier Fuentes-Martín,$^{2}$\footnote{javier.fuentes@ugr.es} \\[.2em] 
    and Anders Eller Thomsen$^{1}$\footnote{anders.thomsen@unibe.ch}
    }\\[1.25em]
    { \small \sffamily \mathversion{sans} 
        $^{1}\,$Albert Einstein Center for Fundamental Physics, Institute for Theoretical Physics,\\ University of Bern,  Sidlerstrasse 5, CH-3012 Bern, Switzerland\\[5pt] 
        $^{2}\,$Departamento de Física Teórica y del Cosmos, Universidad de Granada,\\
        Campus de Fuentenueva, E–18071 Granada, Spain
    }
    \\[.01\textheight]{\itshape \sffamily \today}
    \\[.01\textheight]
    \textcolor{blue!80!black}{\rule{.25\textwidth}{.7pt}}\\[.01\textheight]
\end{center}
\setcounter{footnote}{0}
\renewcommand*{\thefootnote}{\arabic{footnote}}%
\suppressfloats	


\begin{abstract}\vspace{+.01\textheight}
    The next-to-leading order (NLO) Standard Model Effective Field Theory (SMEFT) renormalization group equations are needed to account for phenomenologically relevant operator mixing and ensure renormalization scale independence in NLO calculations of observables. For the first time, we present the renormalization group equations of the baryon-number-conserving sector of the dimension-six SMEFT up to two-loop order. Our calculations have been performed using functional methods with an anticommuting $ \gamma_5 $-scheme. A variety of strategies are employed to mitigate the reading-point ambiguities inherent to this scheme choice. We also describe how a local version of the $ \boldsymbol{R}^\ast $-method is adapted to handle the evanescent operators arising in dimensional regularization. The results are provided in various supplementary files to make them accessible for both human inspection and numerical implementation.
\end{abstract}

\newpage
\section*{Table of Contents}
\toc
\newpage


\section{Introduction}

``Where is new physics hiding?'' is one of the most pressing questions in the field of particle physics. Evidence from neutrino masses, dark matter, and the observed matter-antimatter asymmetry, alongside several unresolved theoretical problems, demands the existence of new physics (NP) beyond the Standard Model (SM). However, its precise nature and energy scale remain open questions subject to ongoing and vigorous debate. In this context, the Standard Model Effective Field Theory (SMEFT)~\cite{Buchmuller:1985jz,Grzadkowski:2010es,Brivio:2017vri,Isidori:2023pyp} has emerged as the leading framework for systematically parameterizing potential deviations from the SM induced by yet-unknown heavy NP.

The flow and mixing of SMEFT operators under renormalization group (RG) evolution can significantly modify phenomenological predictions compared to tree-level expectations, but most SMEFT analyses and global fits have so far had to rely on the one-loop---that is, leading order (LO)---renormalization group equations (RGEs)~\cite{Jenkins:2013zja,Jenkins:2013wua,Alonso:2013hga,Alonso:2014zka} (see also~\cite{Machado:2022ozb,Aebischer:2025qhh}). Rather than stopping at LO in the perturbative expansion, the community is advancing toward incorporating next-to-leading order (NLO) effects in, for instance, Effective Field Theory (EFT) matching~\cite{Dekens:2019ept,Carmona:2021xtq,Fuentes-Martin:2022jrf} and the calculation of electroweak precision and Higgs observables~\cite{Hartmann:2016pil,Dawson:2019clf,Dawson:2022bxd,Biekotter:2025nln,Bellafronte:2026mhp,Biekotter:2026dlb}. A cornerstone of the NLO SMEFT program is the set of two-loop RGEs, since they impact every single observable. While two-loop contributions are often expected to provide only small corrections, there are notable counterexamples such as the sizable effects from top-quark operators mixing into electroweak precision~\cite{Allwicher:2023aql,Haisch:2024wnw,Stefanek:2024kds} and Higgs observables~\cite{DiNoi:2025tka} or semileptonic four-fermion operators inducing lepton-flavor-violating dipole transitions~\cite{Ardu:2021koz}. Moreover, the inclusion of two-loop RGEs is essential to achieving scheme independence in NLO calculations~\cite{Floratos:1977au,Bardeen:1978yd,Ciuchini:1993ks,Ciuchini:1993fk}. In this work, we provide the complete two-loop RGEs for the dimension-six SMEFT in the baryon-number-conserving sector, determining an essential, missing piece for future NLO precision studies.

Various methods and theoretical frameworks have been employed in recent EFT RG and matching calculations. These include on-shell unitarity methods~\cite{Bern:2020ikv,Machado:2022ozb,Aebischer:2025zxg,Chala:2024llp,LopezMiras:2025gar}, determinations in generic EFTs~\cite{Fonseca:2025zjb,Misiak:2025xzq,Aebischer:2025zxg,Fonseca:2025cls,Aebischer:2025ddl,Guedes:2025sax}, geometric formulations~\cite{Helset:2022pde,Jenkins:2023bls,Assi:2025fsm}, techniques based on dimensional reduction~\cite{Chala:2025crd}, and SMEFT studies at dimension eight~\cite{Chala:2021pll,DasBakshi:2022mwk,DasBakshi:2023htx,Chala:2023xjy,Bakshi:2024wzz,Liao:2024xel,Wu:2025qto}. In parallel, efforts to extend RGEs to two-loop order in NP EFTs have gathered momentum:\footnote{And even multi-loop order in general $ \phi^4$ theories~\cite{Henriksson:2025vyi}.} Two-loop RGEs for the dimension-five SMEFT and $\nu$SMEFT have been computed in~\cite{Ibarra:2024tpt,Zhang:2025ywe} while the first---albeit partial---results for the dimension-six SMEFT appeared in~\cite{Aebischer:2022anv,Born:2024mgz,DiNoi:2024ajj,Duhr:2025zqw,DiNoi:2025arz,Haisch:2025lvd,Haisch:2025vqj,Duhr:2025yor,DiNoi:2025tka} with the full baryon-number-violating sector~\cite{Banik:2025wpi} appearing as we were finalizing this work. In the Low-Energy Effective Theory (LEFT)~\cite{Jenkins:2017jig}, defined below the electroweak scale, the complete two-loop RGEs have been obtained in~\cite{Naterop:2023dek,Naterop:2024cfx,Naterop:2025lzc,Naterop:2025cwg} using the Breitenlohner--Maison--'t~Hooft--Veltman (BMHV) scheme~\cite{tHooft:1972tcz,Breitenlohner:1977hr}, with a complementary calculation of the four-fermion operators in an anticommuting $\gamma_5$-scheme~\cite{Aebischer:2025hsx}.

The scheme distinction between the LEFT calculations exemplifies one of the major complications in extending the SMEFT RGEs to two-loop order: the calculation depends on the scheme adopted for handling the $\gamma_5$-matrix in $d = 4 - 2\epsilon$ dimensions. This issue is particularly pressing in the SMEFT, where even the renormalizable sector is chiral. We opt for the computationally simpler route of using an anticommuting $\gamma_5$, although this introduces mathematical inconsistencies. Reproducibility of our result is guaranteed with a reading-point prescription, and most contributions can be separately shown to produce consistent results. Alternatively, the BMHV scheme ensures full mathematical consistency but requires significant computational overhead and the introduction of gauge-symmetry-restoring counterterms. The implications of $\gamma_5$-scheme dependence in SMEFT have recently been examined in~\cite{DiNoi:2023ygk,DiNoi:2025uan,DiNoi:2025arz}.

A second technical challenge in two-loop EFT RG calculations is the proper treatment of evanescent operators~\cite{Buras:1989xd,Dugan:1990df,Herrlich:1994kh}.\footnote{The relevance of evanescent operators in NLO SMEFT calculations was addressed in~\cite{Fuentes-Martin:2022vvu,Haisch:2024wnw}.} In non-integer $d$ dimensions, the Dirac algebra becomes formally infinite dimensional, giving rise to additional operators that vanish in the four-dimensional limit, the so-called evanescent operators. Although evanescent operators have vanishing physical matrix elements---in suitable renormalization schemes---they play a non-trivial role in the renormalization procedure: they necessitate the introduction of additional counterterms to subtract subdivergences in multi-loop calculation and modify the formulas relating the RGEs of the physical couplings to counterterms. In our calculation, we perform local subtractions using the local $\rstar$-method~\cite{Herzog:2017bjx,Born:2024mgz}; however, as emphasized in~\cite{Naterop:2024cfx}, this procedure is sensitive to the specific implementation of the evanescent scheme. We take special care to detail these subtleties as part of the setup of our calculations. 

Our approach to the calculation of the two-loop SMEFT RGEs is based on an implementation of functional methods, which have been (re-)popularized in recent years for EFT matching calculations~\cite{Henning:2014wua,Drozd:2015rsp,delAguila:2016zcb,Henning:2016lyp,Fuentes-Martin:2016uol,Zhang:2016pja,Cohen:2020fcu,Cohen:2020qvb,Fuentes-Martin:2020udw}. Alternative functional formulations, such as the derivative expansion~\cite{Iliopoulos:1974ur} and heat-kernel techniques~\cite{Jack:1982hf,Bijnens:1999hw}, have also been applied in select two-loop computations. In this work, we employ a general framework~\cite{Fuentes-Martin:2023ljp,Fuentes-Martin:2024agf}, tailored for multi-loop calculations, which maps functional contractions onto standard vacuum momentum-space loop integrals. Concretely, our calculation extends the computer implementation used to derive the RGEs in the purely bosonic SMEFT~\cite{Born:2024mgz}, which is itself based on the public \texttt{Matchete} code~\cite{Fuentes-Martin:2022jrf}.

In the remainder of the paper, we first introduce our conventions for the SMEFT, with particular emphasis on our choice of dimension-six operator basis and the rationale behind this selection. In Section~\ref{sec:reg_scheme}, we describe our regularization scheme in detail and catalog all potential issues related to the anti-commuting $ \gamma_5 $. Section~\ref{sec:Rstar} outlines our implementation of the local version of the $\rstar$-method in the presence of evanescent operators. Our main results for the SMEFT RGEs are presented in Section~\ref{sec:results}, followed by concluding remarks in Section~\ref{sec:conclusion}. Additional appendices include a detailed mapping between our preferred dimension-six operator basis and the Warsaw basis (Appendix~\ref{app:basis_transformations}); further details on evanescent schemes and basis transformations (Appendix~\ref{app:eva_schemes}); the full evanescent operator basis (Appendix~\ref{app:eva_ops}); and the anti-hermitian field redefinitions required to eliminate unphysical poles in the beta functions (Appendix~\ref{app:field_shifts}).

\section{The SMEFT and a Revised Operator Basis} 
\label{sec:SMEFT}

In this section, we review the conventions and structure of the SMEFT relevant to our analysis, with particular emphasis on the dimension-six operator basis. While most phenomenological applications adopt the Warsaw basis~\cite{Grzadkowski:2010es}, our computation benefits from a revised basis that is better suited to multi-loop calculations involving an anticommuting $\gamma_5$. Additional modifications have been implemented to improve other aspects of the Warsaw basis. In what follows, we motivate the changes and present the resulting operator basis employed for our two-loop calculations.

\subsection{The SMEFT}

The SMEFT is the EFT built from the SM fields and symmetries with an expansion organized in terms of operators of increasing canonical mass dimension. Restrictign ourselves to baryon-number-conserving terms, the dimension-six SMEFT Lagrangian consists of six parts---the SM Lagrangian, the topological $\theta$-terms, the gauge-fixing and ghost terms, the dimension-five Weinberg operator, and 59 independent dimension-six operators:
    \begin{align}
    \L_\sscript{SMEFT} = \L_\sscript{SM} + \L_\theta + \L_\mathrm{gf.} + \L_\mathrm{gh.} + \L_5 + \L_6 \, .
    \end{align}
The SM Lagrangian comprises all renormalizable operators that are not proportional to total derivatives:
    \begin{align} \label{eq:SMLag}
    \L_\sscript{SM} &= -\frac{1}{4} B^{\mu \nu} B_{\mu \nu} - \frac{1}{4} W^{I \, \mu \nu} W_{\mu \nu}^I - \frac{1}{4} G^{A \, \mu \nu} G_{\mu \nu}^A + D_\mu H^\dag D^\mu H + \mu_H^2 (H^\dag H) - \frac{\lambda}{2} (H^\dag H)^2 \nonumber\\
    &\,\quad + \sum_{\psi \in \{\ell,e,q,u,d\}} \hspace{-1.3em} i\big(\Bar{\psi} \, \gamma_\mu \, D^\mu \psi\big) - \Big[ Y_e^{pr}  (\Bar{\ell}^{\, p}_{\vphantom{i}} e^r H) + Y_u^{pr}  (\Bar{q}^p u^r \widetilde{H}) + Y_d^{pr}  (\Bar{q}^p d^r H) + \hc \Big],
    \end{align}
where $\widetilde{H}^i = \varepsilon^{ij}  H^*_j$, with $ \varepsilon^{ij} $ being the Levi-Civita tensor for the fundamental $ \SU(2)_L $ indices. The flavor indices $p, r, s, t$ run over the three fermion generations. The fermion fields $q, \ell$ are left-handed and $ u,d,e $ are right-handed.\footnote{We have $ P_\LL = \tfrac{1}{2} (\mathds{1} - \gamma_5) $ and, e.g., $ q = P_\LL q= q_\LL$.} 
Our sign convention for the covariant derivatives is defined by
    \begin{align}
    D_\mu q^{ai}= \partial_\mu q^{ai} - i g_Y Y B_\mu q^{ai} - i g_L t^I_{ij}  W_\mu^I q^{aj} - i g_s T^A_{ab} G_\mu^A q^{bi},
    \end{align}
where $t^I $ and $T^A$ denote the fundamental generators of $\SU(2)_L$ and $\SU(3)_c$, respectively. The generators are normalized such that $\mathrm{Tr}[t^I t^J] = \tfrac{1}{2} \delta^{IJ}$ and $\mathrm{Tr}[T^A T^B] = \tfrac{1}{2} \delta^{AB}$. Additionally, the renormalizable theory contains topological $\theta$-terms, given by
    \begin{align}
    \L_\theta = \frac{\theta_B}{32 \pi^2} \widetilde{B}_{\mu \nu} B^{\mu \nu} + \frac{\theta_W}{32 \pi^2} \widetilde{W}_{\mu \nu}^I W^{I \, \mu \nu} + \frac{\theta_G}{32 \pi^2} \widetilde{G}_{\mu \nu}^A G^{A \, \mu \nu} .
    \end{align}
Being total derivatives, these terms have no bearing on perturbation theory. Nonetheless, their RG running can be determined using perturbative methods. We adopt the convention $\widetilde{F}^{\mu \nu} = \frac{1}{2} \varepsilon^{\mu \nu \rho \sigma} F_{\rho \sigma}$ for the dual field-strength tensors with $\varepsilon^{0123} = -\varepsilon_{0123} = +1$.

Perturbative calculations require gauge fixing of the Lagrangian, which can be implemented in various ways; however, the specific choice does not affect the running of gauge-invariant couplings~\cite{Caswell:1974cj}---such as those appearing in $\L_{\sscript{SM}, \theta, 5, 6}$---which are the focus of our analysis. We employ the background-field gauge~\cite{DeWitt:1967ub,Abbott:1980hw} for our calculations, as it preserves background-gauge invariance and facilitates background-gauge-invariant functional methods at higher-loop orders~\cite{Fuentes-Martin:2024agf}. The gauge-fixing Lagrangian, e.g., for the $ \SU(3)_c $ group, is given by  
    \begin{align}\label{eq:gf_lag}
    \L_\text{gf.} \supset -\frac{1}{2\xi} \lzm \partial^\mu G_\mu^A + g_s f_{ABC} \widehat{G}^{B \mu} G^C_\mu \dzm^2 \equiv -\frac{1}{2\xi} \lzm \widehat{D}^\mu G_\mu^A \dzm^2 ,
    \end{align}
where $\widehat{D}$ is the background covariant derivative built solely from background fields---$\widehat{G}_\mu^A$ in this case---and $\xi$ is the gauge-fixing parameter. Analogous gauge-fixing terms are used for the other gauge groups. We adopt the Feynman gauge ($\xi = 1$) for simplicity, in line with standard practice for functional calculations. The corresponding ghost Lagrangian reads
    \begin{align}
    \L_\text{gh.} \supset - \Bar{c}^A \widehat{D}^\mu \big(\widehat{D}_\mu c^A + g_s f_{ABC} G^B_\mu c^C \big),
    \end{align}
with ($\bar c^A$) $c^A$ denoting the (anti-)ghost fields. 

We add a tower of higher-dimensional effective operators on top of the renormalizable Lagrangian. At dimension five, the only term is the Weinberg operator,  
\begin{align}
    \mathcal{L}_5 = C_{\ell \ell H H}^{pr} \big( \overline{\ell^\cc}\phantom{}^{\, p} \widetilde{H}^* \widetilde{H}^\dagger \ell^r \big) + \hc ,
\end{align}
where $\psi^{\cc} = C\, \Bar{\psi}\transpose $ denotes the charge-conjugated $\psi$ fermion. At dimension six, a much larger set of operators appears, whose organization and operator basis require more deliberate considerations.

\subsection{The Mainz basis}
\label{sec:SMEFTbasis}

Building on the renormalizable SMEFT Lagrangian supplemented by the dimension-five Weinberg operator discussed above, the next order in the heavy-mass expansion introduces dimension-six operators. The dimension-six piece of the SMEFT Lagrangian is given by 
\begin{equation}
    \L_6 = \sum_{X\in \mathrm{real}} C_X O_X + \!\! \sum_{X\in \mathrm{complex}} \!\!\big(C_X O_X + \hc \big),
\end{equation}
where $ O_X $ are the operators and $ C_X $ the corresponding Wilson coefficients (couplings). The conjugates of the inherently complex operators---with complex Wilson coefficients---are added to the Lagrangian to ensure reality. 

\begin{table}
    \centering
    \def\arraystretch{1.75} 
    \resizebox{\textwidth}{!}{%
    \begin{tabular}{|c|c|c|c|c|c|c|c|}
      \hhline{|--~--~--|}
      \multicolumn{2}{|c|}{\boldmath \cellcolor{blue!30} $X^3$}
      &
      & \multicolumn{2}{c|}{\boldmath \cellcolor{blue!30} $H^4 D^2$ \textbf{and} $H^6$}
      &
      & \multicolumn{2}{c|}{\boldmath \cellcolor{blue!30} $\psi^2 H^3$} \\
      \cline{1-2} \cline{4-5} \cline{7-8}
      $O_{W}$ & $g_L^3 f^{I J K} \, W_\mu^{I \, \nu} \, W_{\nu}^{J \, \rho} \, W^{K \, \mu}_\rho$ & & 
      $O_{H}$ & $\lzm H^\dag H \dzm^3$ & & 
      $O_{eH}^{pr}$ & $\lzm H^\dag H \dzm \lzm \Bar{\ell}^{\, p}_{\vphantom{i}} e^r H \dzm$ \\
      \cline{1-2} \arrayrulecolor{red} \noalign{\global\arrayrulewidth=1.5pt} \cline{4-5} \arrayrulecolor{black} \noalign{\global\arrayrulewidth=0.4pt} \cline{7-8}
      $O_{G}$ & $g_s^3 f^{A B C} \, G_\mu^{A \, \nu} \, G_{\nu}^{B \, \rho} \, G^{C \, \mu}_\rho$ & & 
      \multicolumn{1}{!{\color{red}\vrule width 1.5pt}c|}{$O_{HD}^{\scriptscriptstyle(\|)}$} & \multicolumn{1}{c!{\color{red}\vrule width 1.5pt}}{$\big(H^\dag H\big)  \lzm D_\mu H^\dag D^\mu H \dzm$} & & 
      $O_{uH}^{pr}$ & $\lzm H^\dag H \dzm \lzm \Bar{q}^p u^r \widetilde{H} \dzm$ \\
      \cline{1-2} \arrayrulecolor{red} \noalign{\global\arrayrulewidth=1.5pt} \cline{4-5} \arrayrulecolor{black} \noalign{\global\arrayrulewidth=0.4pt} \cline{7-8}
      $O_{\widetilde{W}}$ & $g_L^3 f^{I J K} \, \widetilde{W}_\mu^{I \, \nu} \, W_{\nu}^{J \, \rho} \, W^{K \, \mu}_\rho$ & & 
      $O_{HD}^{\scriptscriptstyle(\times)}$ & $H^*_i H^j\,  D_\mu H^*_j D^\mu H^i$ & & 
      $O_{dH}^{pr}$ & $\lzm H^\dag H \dzm \lzm \Bar{q}^p d^r H \dzm$ \\
      \cline{1-2} \cline{4-5} \cline{7-8}
      $O_{\widetilde{G}}$ & $g_s^3 f^{A B C} \, \widetilde{G}_\mu^{A \, \nu} \, G_{\nu}^{B \, \rho} \, G^{C \, \mu}_\rho$ & \multicolumn{1}{c}{} & 
      \multicolumn{2}{c}{} & 
      \multicolumn{1}{c}{} & 
      \multicolumn{2}{c}{} \\
      \cline{1-2} 
      \multicolumn{8}{c}{} \\
      \hhline{|--~--~--|}
      \multicolumn{2}{|c|}{\boldmath \cellcolor{blue!30} $X^2 H^2$}
      &
      & \multicolumn{2}{c|}{\boldmath \cellcolor{blue!30} $\psi^2 X H$}
      &
      & \multicolumn{2}{c|}{\boldmath \cellcolor{blue!30} $\psi^2 H^2 D$} \\
      \cline{1-2} \arrayrulecolor{yellow} \noalign{\global\arrayrulewidth=1.5pt} \cline{4-5} \noalign{\global\arrayrulewidth=0.4pt} \arrayrulecolor{black} \cline{7-8}
      $O_{HG}$ & $g_s^2 (H^\dag H) \, G^{A \, \mu \nu} \, G_{\mu \nu}^A$ & &
      \multicolumn{1}{!{\color{yellow}\vrule width 1.5pt}c|}{$O_{eW}^{pr}$} & \multicolumn{1}{c!{\color{yellow}\vrule width 1.5pt}}{$g_L\lzm \Bar{\ell}^{\, p} \sigma^{\mu \nu} e^r t^I H \dzm W_{\mu \nu}^I $} & &
      $O_{H\ell}^{{\scriptscriptstyle(\|)} \, pr}$ & $i \lzm H^\dag \lrderiv{\mu} H \dzm \lzm \Bar{\ell}^{\, p}_{\vphantom{i}} \gamma^\mu \, \ell^r \, \dzm $ \\
      \cline{1-2} \arrayrulecolor{yellow} \noalign{\global\arrayrulewidth=1.5pt} \cline{4-5} \noalign{\global\arrayrulewidth=0.4pt} \arrayrulecolor{black} \arrayrulecolor{green} \noalign{\global\arrayrulewidth=1.5pt} \cline{7-8} \noalign{\global\arrayrulewidth=0.4pt} \arrayrulecolor{black}
      $O_{H\widetilde{G}}$ & $g_s^2(H^\dag H) \, G^{A \, \mu \nu} \, \widetilde{G}_{\mu \nu}^A$ & & 
      $O_{eB}^{pr}$ & $g_Y\lzm \Bar{\ell}^{\, p} \, \sigma^{\mu \nu} e^r H \dzm B_{\mu \nu} $ & & 
      \multicolumn{1}{!{\color{green}\vrule width 1.5pt}c|}{$O_{H\ell}^{{\scriptscriptstyle(\times)} \, pr}$} & \multicolumn{1}{c!{\color{green}\vrule width 1.5pt}}{$i \lzm H^*_i \lrderiv{\mu} H^j \dzm \lzm \Bar{\ell}^{\, p}_j \, \gamma^\mu \, \ell^{ir} \, \dzm $} \\
      \cline{1-2} \cline{4-5} \arrayrulecolor{green} \noalign{\global\arrayrulewidth=1.5pt} \cline{7-8} \noalign{\global\arrayrulewidth=0.4pt} \arrayrulecolor{black}
      $O_{HW}$ & $g_L^2 (H^\dag H) \, W^{I \, \mu \nu} \, W_{\mu \nu}^I$ & & 
      $O_{uG}^{pr}$ & $g_s\lzm \Bar{q}^p \, \sigma^{\mu \nu} \, T^A \, u^r \widetilde{H} \dzm G_{\mu \nu}^A $ & & 
      $O_{He}^{pr}$ & $i \lzm H^\dag \lrderiv{\mu} H \dzm \lzm \Bar{e}^p \, \gamma^\mu \, e^r \dzm $ \\
      \cline{1-2} \arrayrulecolor{yellow} \noalign{\global\arrayrulewidth=1.5pt} \cline{4-5} \noalign{\global\arrayrulewidth=0.4pt} \arrayrulecolor{black} \cline{7-8}
      $O_{H\widetilde{W}}$ & $g_L^2 (H^\dag H) \, W^{I \, \mu \nu} \, \widetilde{W}_{\mu \nu}^I$ & &
      \multicolumn{1}{!{\color{yellow}\vrule width 1.5pt}c|}{$O_{uW}^{pr}$} & \multicolumn{1}{c!{\color{yellow}\vrule width 1.5pt}}{$g_L \lzm \Bar{q}^p \, \sigma^{\mu \nu} u^r t^I \widetilde{H}\dzm  W_{\mu \nu}^I $} & & 
      $O_{Hq}^{{\scriptscriptstyle(\|)} \, pr}$ & $i \lzm H^\dag \lrderiv{\mu} H \dzm \lzm \Bar{q}^p \, \gamma^\mu \, q^r \dzm $ \\
      \cline{1-2} \arrayrulecolor{yellow} \noalign{\global\arrayrulewidth=1.5pt} \cline{4-5} \noalign{\global\arrayrulewidth=0.4pt} \arrayrulecolor{black} \arrayrulecolor{green} \noalign{\global\arrayrulewidth=1.5pt} \cline{7-8} \noalign{\global\arrayrulewidth=0.4pt} \arrayrulecolor{black}
      $O_{HB}$ & $g_Y^2 (H^\dag H) \, B^{\mu \nu} \, B_{\mu \nu}$ & & 
      $O_{uB}^{pr}$ & $g_Y \lzm \Bar{q}^p \, \sigma^{\mu \nu} \, u^r \widetilde{H} \dzm B_{\mu \nu} $ & & 
      \multicolumn{1}{!{\color{green}\vrule width 1.5pt}c|}{$O_{Hq}^{{\scriptscriptstyle(\times)} \, pr}$} & \multicolumn{1}{c!{\color{green}\vrule width 1.5pt}}{$i \lzm H^*_i \lrderiv{\mu} H^j \dzm \lzm \Bar{q}^p_j \, \gamma^\mu \, q^{ir} \dzm $} \\
      \cline{1-2} \cline{4-5} \arrayrulecolor{green} \noalign{\global\arrayrulewidth=1.5pt} \cline{7-8} \noalign{\global\arrayrulewidth=0.4pt} \arrayrulecolor{black}
      $O_{H\widetilde{B}}$ & $g_Y^2 (H^\dag H) \, B^{\mu \nu} \, \widetilde{B}_{\mu \nu}$ & & 
      $O_{dG}^{pr}$ & $g_s\lzm \Bar{q}^p \, \sigma^{\mu \nu} \, T^A d^r H \dzm G_{\mu \nu}^A $ & & 
      $O_{Hu}^{pr}$ & $i \lzm H^\dag \lrderiv{\mu} H \dzm \lzm \Bar{u}^p \, \gamma^\mu \, u^r \dzm $ \\
      \arrayrulecolor{yellow} \noalign{\global\arrayrulewidth=1.5pt} \cline{1-2}  \cline{4-5} \noalign{\global\arrayrulewidth=0.4pt} \arrayrulecolor{black} \cline{7-8}
      \multicolumn{1}{!{\color{yellow}\vrule width 1.5pt}c|}{$O_{HWB}$} & \multicolumn{1}{c!{\color{yellow}\vrule width 1.5pt}}{$g_Y g_L \big( H^\dag t^I H \big) B^{\mu \nu} \, W_{\mu \nu}^I$} & &
      \multicolumn{1}{!{\color{yellow}\vrule width 1.5pt}c|}{$O_{dW}^{pr}$} & \multicolumn{1}{c!{\color{yellow}\vrule width 1.5pt}}{$g_L\lzm \Bar{q}^p \, \sigma^{\mu \nu} \, d^r t^I H\dzm  W_{\mu \nu}^I $} & &
      $O_{Hd}^{pr}$ & $i \lzm H^\dag \lrderiv{\mu} H \dzm \lzm \Bar{d}^p \, \gamma^\mu \, d^r \dzm $ \\
      \arrayrulecolor{yellow} \noalign{\global\arrayrulewidth=1.5pt} \cline{1-2} \cline{4-5} \noalign{\global\arrayrulewidth=0.4pt} \arrayrulecolor{black} \cline{7-8}
      \multicolumn{1}{!{\color{yellow}\vrule width 1.5pt}c|}{$O_{H\widetilde{W}B}$} & \multicolumn{1}{c!{\color{yellow}\vrule width 1.5pt}}{$g_Y g_L \big(H^\dag t^I H \big) B^{\mu \nu} \, \widetilde{W}_{\mu \nu}^I$} & &
      $O_{dB}^{pr}$ & $g_Y\lzm \Bar{q}^p \, \sigma^{\mu \nu} \, d^r H \dzm B_{\mu \nu} $ & &
      $O_{Hud}^{pr}$ & $i \lzm \widetilde{H}^\dag D_{\mu} H \dzm \lzm \Bar{u}^p \, \gamma^\mu \, d^r \dzm $ \\
      \arrayrulecolor{yellow} \noalign{\global\arrayrulewidth=1.5pt} \cline{1-2} \noalign{\global\arrayrulewidth=0.4pt} \arrayrulecolor{black} \cline{4-5} \cline{7-8}
    \end{tabular}
    }
    \caption{Dimension-six operators containing zero or two fermions. The operator $O_{HD}^{\scriptscriptstyle(\|)}$ (framed in red) replaces $Q_{H\Box}$ of the Warsaw basis, and the original $Q_{HD}$ is renamed $O_{HD}^{\scriptscriptstyle(\times)}$ to reflect this change. Operators in the classes $X^3$, $X^2 H^2$, and $\psi^2 X H$ explicitly include appropriate gauge coupling factors. Pauli matrices have been replaced by $\SU(2)_L$ generators in all operators framed in yellow. Operators framed in green are related to Warsaw operators through $\SU(2)_L$ Fierz identities. Explicit indices are shown only for non-trivial contractions, with $\{i,j\}$ being $\SU(2)_L$ fundamental indices.}

    \label{tab:dim6OpsNoFourFermi}
\end{table}

\begin{table}
    \centering
    \def\arraystretch{1.75} 
    \begin{tabular}{|c|c|c|c|c|}
        \hhline{|--~--|}
        \multicolumn{2}{|c|}{\boldmath \cellcolor{blue!30} $(\Bar{L}L)(\Bar{R}R)$}
        &
        & \multicolumn{2}{c|}{\boldmath \cellcolor{blue!30} $(\Bar{R}R)(\Bar{R}R)$} \\
        \cline{1-2} \cline{4-5}
        $O_{\ell e}^{prst}$ & $\lzm \Bar{\ell}^{\, p}_{\vphantom{i}} \, \gamma_\mu \, \ell^r \dzm \lzm \Bar{e}^s \, \gamma^\mu \, e^t \dzm$ & & 
        $O_{ee}^{prst}$ & $\lzm \Bar{e}^p \, \gamma_\mu \, e^r \dzm \lzm \Bar{e}^s \, \gamma^\mu \, e^t \dzm$ \\
        \cline{1-2} \cline{4-5}
        $O_{\ell u}^{prst}$ & $\lzm \Bar{\ell}^{\, p}_{\vphantom{i}} \, \gamma_\mu \, \ell^r \dzm \lzm \Bar{u}^s \, \gamma^\mu \, u^t \dzm$ & & 
        $O_{uu}^{prst}$ & $\lzm \Bar{u}^p \, \gamma_\mu \, u^r \dzm \lzm \Bar{u}^s \, \gamma^\mu \, u^t \dzm$ \\
        \cline{1-2} \cline{4-5}
        $O_{\ell d}^{prst}$ & $\lzm \Bar{\ell}^{\, p}_{\vphantom{i}} \, \gamma_\mu \, \ell^r \dzm \lzm \Bar{d}^s \, \gamma^\mu \, d^t \dzm$ & & 
        $O_{dd}^{prst}$ & $\lzm \Bar{d}^p \, \gamma_\mu \, d^r \dzm \lzm \Bar{d}^s \, \gamma^\mu \, d^t \dzm$ \\
        \cline{1-2} \cline{4-5}
        $O_{qe}^{prst}$ & $\lzm \Bar{q}^p \, \gamma_\mu \, q^r \dzm \lzm \Bar{e}^s \, \gamma^\mu \, e^t \dzm$ & & 
        $O_{eu}^{prst}$ & $\lzm \Bar{e}^p \, \gamma_\mu \, e^r \dzm \lzm \Bar{u}^s \, \gamma^\mu \, u^t \dzm$ \\
        \cline{1-2} \cline{4-5}
        $O_{qu}^{{\scriptscriptstyle(\|)} \, prst}$ & $\lzm \Bar{q}^p \, \gamma_\mu \, q^r \dzm \lzm \Bar{u}^s \, \gamma^\mu \, u^t \dzm$ & & 
        $O_{ed}^{prst}$ & $\lzm \Bar{e}^p \, \gamma_\mu \, e^r \dzm \lzm \Bar{d}^s \, \gamma^\mu \, d^t \dzm$ \\
        \arrayrulecolor{green} \noalign{\global\arrayrulewidth=1.5pt} \cline{1-2} \noalign{\global\arrayrulewidth=0.4pt} \arrayrulecolor{black} \cline{4-5}
        \multicolumn{1}{!{\color{green}\vrule width 1.5pt}c|}{$O_{qu}^{{\scriptscriptstyle(\times)} \, prst}$} & \multicolumn{1}{c!{\color{green}\vrule width 1.5pt}}{$\lzm \Bar{q}^p_a \, \gamma_\mu \, q^{br} \dzm \lzm \Bar{u}^s_b \, \gamma^\mu \, u^{at} \dzm$} & & 
        $O_{ud}^{{\scriptscriptstyle(\|)} \, prst}$ & $\lzm \Bar{u}^p \, \gamma_\mu \, u^r \dzm \lzm \Bar{d}^s \, \gamma^\mu \, d^t \dzm$ \\
        \arrayrulecolor{green} \noalign{\global\arrayrulewidth=1.5pt} \cline{1-2} \cline{4-5} \noalign{\global\arrayrulewidth=0.4pt} \arrayrulecolor{black}
        $O_{qd}^{{\scriptscriptstyle(\|)} \, prst}$ & $\lzm \Bar{q}^p \, \gamma_\mu \, q^r \dzm \lzm \Bar{d}^s \, \gamma^\mu \, d^t \dzm$ & &
        \multicolumn{1}{!{\color{green}\vrule width 1.5pt}c|}{$O_{ud}^{{\scriptscriptstyle(\times)} \, prst}$} & \multicolumn{1}{c!{\color{green}\vrule width 1.5pt}}{$\lzm \Bar{u}^p_a \, \gamma_\mu \, u^{br} \dzm \lzm \Bar{d}^s_b \, \gamma^\mu \, d^{at} \dzm$} \\
        \arrayrulecolor{green} \noalign{\global\arrayrulewidth=1.5pt} \cline{1-2} \cline{4-5} \noalign{\global\arrayrulewidth=0.4pt} \arrayrulecolor{black}
        \multicolumn{1}{!{\color{green}\vrule width 1.5pt}c|}{$O_{qd}^{{\scriptscriptstyle(\times)} \, prst}$} & \multicolumn{1}{c!{\color{green}\vrule width 1.5pt}}{$\lzm \Bar{q}^p_a \, \gamma_\mu \, q^{br} \dzm \lzm \Bar{d}^s_b \, \gamma^\mu \, d^{at} \dzm$} & \multicolumn{1}{c}{} & 
        \multicolumn{2}{c}{} \\
        \arrayrulecolor{green} \noalign{\global\arrayrulewidth=1.5pt} \cline{1-2} \noalign{\global\arrayrulewidth=0.4pt} \arrayrulecolor{black}
        \multicolumn{5}{c}{} \\
        \hhline{|--~--|}
        \multicolumn{2}{|c|}{\boldmath \cellcolor{blue!30} $(\Bar{L}R)(\Bar{R}L)$ \textbf{and} $(\Bar{L}R)(\Bar{L}R)$}
        &
        & \multicolumn{2}{c|}{\boldmath \cellcolor{blue!30} $(\Bar{L}L)(\Bar{L}L)$} \\
        \cline{1-2} \cline{4-5}
        $O_{\ell edq}^{prst}$ & $\lzm \Bar{\ell}^{\, p}_j \, e^r \dzm \lzm \Bar{d}^{s} \, q^t_j \dzm$ & & 
        $O_{\ell \ell}^{prst}$ & $\lzm \Bar{\ell}^{\, p}_{\vphantom{i}} \, \gamma_\mu \, \ell^r \dzm \lzm \Bar{\ell}^{\, s}_{\vphantom{i}} \, \gamma^\mu \, \ell^t \dzm$ \\
        \cline{1-2} \cline{4-5}
        $O_{quqd}^{{\scriptscriptstyle(\|)} \, prst}$ & $\lzm \Bar{q}^{p}_{i} \, u^r \dzm \varepsilon^{ij} \lzm \Bar{q}^s_{j} \, d^t \dzm$ & & 
        $O_{qq}^{{\scriptscriptstyle(\|)} \, prst}$ & $\lzm \Bar{q}^p \, \gamma_\mu \, q^r \dzm \lzm \Bar{q}^s \, \gamma^\mu \, q^t \dzm$ \\
        \arrayrulecolor{green} \noalign{\global\arrayrulewidth=1.5pt} \cline{1-2} \cline{4-5} \noalign{\global\arrayrulewidth=0.4pt} \arrayrulecolor{black}
        \multicolumn{1}{!{\color{green}\vrule width 1.5pt}c|}{$O_{quqd}^{{\scriptscriptstyle(\times)} \, prst}$} & \multicolumn{1}{c!{\color{green}\vrule width 1.5pt}}{$\lzm \Bar{q}^{p}_{ai} \, u^{br} \dzm \varepsilon^{ij} \lzm \Bar{q}^s_{bj} \, d^{at} \dzm$} & & 
        \multicolumn{1}{!{\color{green}\vrule width 1.5pt}c|}{$O_{qq}^{{\scriptscriptstyle(\times)} \, prst}$} & \multicolumn{1}{c!{\color{green}\vrule width 1.5pt}}{$\lzm \Bar{q}^p_i \, \gamma_\mu \, q^{jr} \dzm \lzm \Bar{q}^s_j \, \gamma^\mu \, q^{it} \dzm$} \\
        \arrayrulecolor{green} \noalign{\global\arrayrulewidth=1.5pt} \cline{1-2} \cline{4-5} \noalign{\global\arrayrulewidth=0.4pt} \arrayrulecolor{black}
        $O_{\ell equ}^{prst}$ & $\lzm \Bar{\ell}^{\, p}_i \, e^r \dzm \varepsilon^{ij} \lzm \Bar{q}^s_j \, u^t \dzm$ & & 
        $O_{\ell q}^{{\scriptscriptstyle(\|)} \, prst}$ & $\lzm \Bar{\ell}^{\, p}_{\vphantom{i}} \, \gamma_\mu \, \ell^r \dzm \lzm \Bar{q}^s \, \gamma^\mu \, q^t \dzm$ \\
        \arrayrulecolor{red} \noalign{\global\arrayrulewidth=1.5pt} \cline{1-2} \arrayrulecolor{green} \cline{4-5} \noalign{\global\arrayrulewidth=0.4pt} \arrayrulecolor{black}
        \multicolumn{1}{!{\color{red}\vrule width 1.5pt}c|}{$O_{\ell uqe}^{prst}$} & \multicolumn{1}{c!{\color{red}\vrule width 1.5pt}}{$\lzm \Bar{\ell}^{\, p}_i \, u^r \dzm \varepsilon^{ij} \lzm \Bar{q}^s_j \, e^t \dzm$} & & 
        \multicolumn{1}{!{\color{green}\vrule width 1.5pt}c|}{$O_{\ell q}^{{\scriptscriptstyle(\times)} \, prst}$} & \multicolumn{1}{c!{\color{green}\vrule width 1.5pt}}{$\lzm \Bar{\ell}^{\,p}_i \, \gamma_\mu \, \ell^{jr} \dzm \lzm \Bar{q}^{s}_j \, \gamma^\mu \, q^{it} \dzm$} \\
        \arrayrulecolor{red} \noalign{\global\arrayrulewidth=1.5pt} \cline{1-2} \arrayrulecolor{green} \cline{4-5} \noalign{\global\arrayrulewidth=0.4pt} \arrayrulecolor{black}
    \end{tabular}
    \caption{Dimension-six four-fermion operators. The operator $O_{\ell uqe}^{prst}$ (framed in red) replaces $Q_{\ell equ}^{(3) \, prst}$ of the Warsaw basis. Operators framed in green are related to Warsaw basis operators through $\SU(N)$ Fierz identities. Explicit indices are shown only for non-trivial contractions, with $\{i,j\}$ and $\{a,b\}$ being fundamental $\SU(2)_L$ and $\SU(3)_c$ indices, respectively. }
    \label{tab:dim6OpsFourFermi} 
\end{table}

Although the Warsaw basis~\cite{Grzadkowski:2010es} is the most widely adopted choice for the operator basis, we find it necessary to exchange the operator $Q^{(3)}_{\ell equ}=\lzm\bar \ell_i\sigma_{\mu\nu}e\dzm\varepsilon^{ij}\lzm\bar q_j\sigma^{\mu\nu}u\dzm$ to alleviate the reading-point ambiguities arising from our use of an anticommuting $\gamma_5$ (see Section~\ref{sec:gamma5} for details). This change, in turn, motivates making further adjustments to improve other aspects of the operator basis. The complete set of revised operators is listed in Tables~\ref{tab:dim6OpsNoFourFermi} and~\ref{tab:dim6OpsFourFermi}.\footnote{While outside the scope of the present work, the complete operator basis also includes dimension-six baryon-number-violating operators, which are taken to coincide with those in the Warsaw basis.} In what follows, we refer to this operator basis as the \emph{Mainz basis}.\footnote{The primary modifications were originally proposed at the \href{https://indico.mitp.uni-mainz.de/event/395/}{SMEFT-Tools 2025 Workshop} in Mainz, lending the basis its name. The suggested changes are open for discussion at~\url{https://github.com/NewSMEFTBasis/basis-proposal} and have been the subject of debate within the \href{https://indico.cern.ch/event/1601781/}{LHC EFT Working Group}. As the basis definition remains under discussion, we aim to update future versions of this document should the community consensus evolve.} The modifications relative to the Warsaw basis are detailed below:
\begin{itemize}
    \item As anticipated, we replace the only tensor-current four-fermion operator in the Warsaw basis, $Q^{(3)}_{\ell equ}$, with the scalar-current operator
    \begin{align}
    O_{\ell uqe} &= \lzm \bar{\ell}_i u \dzm \varepsilon^{ij} \lzm \bar{q}_j e \dzm .
    \end{align}
    The sole SM extension that generates $Q^{(3)}_{\ell equ}$ directly (before Fierz identities) at tree-level involves a rank-2 antisymmetric Lorentz tensor, for which, to the best of our knowledge, no renormalizable Lagrangian can be formulated. By contrast, $O_{\ell uqe}$ is generated at tree-level by the $R_2$ scalar leptoquark (see e.g.~\cite{Dorsner:2016wpm}). One could alternatively use $O_{\ell q^c e^c u}= \lzm\bar \ell_i q_j^\cc\dzm \epsilon^{ij} \lzm \overline{e^\cc} u\dzm$, which is generated at tree-level by the $S_1$ scalar leptoquark. However, $O_{\ell uqe}$ is preferred due to the absence of charge-conjugated fermions in the operator. When Fierzing either $O_{\ell uqe}$ or $O_{\ell q^c e^c u}$ into $Q^{(3)}_{\ell equ}$, one introduces evanescent shifts that are ambiguous in an anticommuting $\gamma_5$ scheme already at one-loop order~\cite{Fuentes-Martin:2022vvu}. This ambiguity propagates to matrix-element computations involving the $Q^{(3)}_{\ell equ}$ operator. These problems are avoided with the adoption of $ O_{\ell uqe} $. Of particular relevance to our calculations, the scalar $ O_{\ell uqe} $ introduces fewer extra Dirac matrices into Dirac traces than equivalent insertions of the $Q^{(3)}_{\ell equ}$ tensor operator, removing potential reading-point ambiguities in the two-loop counterterm calculations. 
    \item The operator $Q_{H\Box} = (H^\dagger H) \Box (H^\dagger H)$ is replaced by\footnote{To distinguish the two $H^4D^2$-type operators and adhere to the overall naming convention, we use $ O_{HD}^{\scriptscriptstyle(\times)}$ for the operator known as $ Q_{HD} $ in the Warsaw basis.}
    \begin{align}
    O_{HD}^{\scriptscriptstyle(\|)} = \big(H^\dagger H\big) \lzm D_\mu H^\dag D^\mu H \dzm.
    \end{align}
    While the operator $Q_{H\Box}$ may yield slightly simpler Feynman rules, it departs from common design principles in EFT basis constructions (such as those used in many computer tools) in that it contains derivatives acting on products of fields rather than on individual fields. Expanding the derivatives into the fields generates terms proportional to $D^2 H$, which are typically removed from any basis using field redefinitions. Moreover, the $ O_{HD}^{\scriptscriptstyle(\|)} $ operator aligns better with geometric constructions (e.g.,~\cite{Helset:2020yio}), as it can be interpreted as a modification of the metric governing the Higgs kinetic term.
    \item We have included an explicit gauge coupling $g_X$ with every field-strength tensor $X_{\mu\nu}$ in the operator definitions. This modification affects all operators in the $X^3$, $X^2 H^2$, and $\psi^2 XH$ classes. Gauge invariance ensures that any matching contribution to such operators, as well as any RG mixing from other operators into them, is always proportional to (at least) the appropriate power of the gauge coupling. Without this normalization, one encounters terms in the \befs and amplitudes proportional to, e.g., $\sqrt{\alpha_s}$, obscuring the perturbative counting. Furthermore, it is not possible to absorb all gauge couplings into the normalization of the gauge kinetic term (via the field redefinition $X_\mu \to g_X^{\eminus1} X_\mu$) without this change.\footnote{This choice provides a clean and consistent framework, allowing each gauge coupling to be associated with a single, gauge-invariant operator, in line with all other couplings. It is also the convention adopted in \texttt{Matchete} as of \texttt{v0.3}, although this is not our motivation for implementing the basis change here.}
    \item The Warsaw basis contains a variety of singlet and adjoint---triplet for $\SU(2)_L$ and octet for $\SU(3)_c$---currents, reflecting different contractions of gauge indices in the spinor lines in four-fermion operators. While this neatly aligns with some tree-level BSM contributions, singlet and adjoint operators generally mix under RG running, reducing the relevance of such distinctions. Using the Fierz identities
    \begin{align} \label{eq:group_identities}
    t^I_{ij} \, t^I_{k\ell} &= \frac{1}{2} \delta_{i\ell} \, \delta_{jk} - \frac{1}{4} \delta_{ij} \, \delta_{k\ell} , &
    T^A_{ab} \, T^A_{cd} &= \frac{1}{2} \delta_{ad} \, \delta_{bc} - \frac{1}{6} \delta_{ab} \, \delta_{cd} ,
    \end{align}
    one finds that the change of basis from singlet/adjoint currents to fundamental-index contractions is linear and involves no evanescent operators. We remove the adjoint currents in favor of contracting the fundamental gauge indices between different spinor lines, which slightly simplifies the $\SU(N)$ algebra and reduces the number of objects in the operator definitions. 
    \item In the Warsaw basis, the operators $Q_{HWB}$, $Q_{H\widetilde{W}B}$, $Q_{eW}$, $Q_{uW}$ and $Q_{dW}$ use Pauli matrices to contract the $\SU(2)_L$ indices. We replace the Pauli matrices with the $\SU(2)_L$ generators, in analogy with the gluonic operators, which employ the $\SU(3)_c$ generators rather than the Gell-Mann matrices. This changes the normalizations of these operators by a factor $ \tfrac{1}{2} $ relative to the Warsaw basis.
\end{itemize}
Having introduced the Mainz basis, it is useful to clarify the mapping between the Mainz and Warsaw bases and their RGEs. We refer the reader to Appendix~\ref{app:basis_transformations}, where the basis transformation and its impact on the RGEs are explicated (to the extent that it is possible given the $ \gamma_5$ ambiguity). That being said, we hope that the new basis choice also proves advantageous in other SMEFT calculations and phenomenological analyses.

In our calculations, we further assume that all Wilson coefficients respect the flavor symmetries exhibited by the associated operators in the four-dimensional limit. For example, the coefficient $C_{ee}^{prst}$ is taken to be symmetric not only under the exchange of flavor indices $pr \leftrightarrow st$, but also under $r \leftrightarrow t$. The latter symmetry follows from a Fierz identity, which is strictly valid exclusively in four dimensions. Coefficients not satisfying these symmetries ether manifestly vanish at the Lagrangian level or are classified as part of the evanescent rather than the physical couplings.

\section{Dimensional Continuation and Scheme Conventions} 
\label{sec:reg_scheme}

The two-loop RGEs are generally scheme dependent, contrary to their one-loop counterparts. We therefore begin with a detailed description of our regularization conventions and renormalization scheme.

\subsection{Dirac algebra in $ d $ dimensions}
\label{sec:gamma5}

When using dimensional regularization, as is standard practice in perturbative quantum field theory calculations, one has to continue four-dimensional objects to non-integer $ d=4 - 2 \epsilon $ dimensions in such a way that the correct four-dimensional limit can be recovered. 
For the $ d $-dimensional Dirac algebra, this leaves the question of how to continue intrinsically four-dimensional objects such as the infamous $ \gamma_5 $. We employ an anticommuting $ \gamma_5 $---a version of naive dimensional regularization (NDR)---such that the algebra is defined by 
    \begin{align}
    \{ \gamma_\mu,\, \gamma_\nu\} = 2 g_{\mu\nu}, \qquad 
    \{ \gamma_\mu,\, \gamma_5 \} = 0, \qquad 
    \gamma_5^2 = \mathds{1},
    \end{align}
which is known to produce a self-contradiction when enforcing the four-dimensional limit
    \begin{equation}
    \Tr\!\big[ \gamma_\mu \gamma_\nu \gamma_\rho \gamma_\sigma \gamma_5 \big] \xrightarrow[d\to 4]{} - 4 i\varepsilon_{\mu \nu \rho \sigma}.
    \end{equation}
This limit is inconsistent with cyclicity of the Dirac trace (see e.g.,~\cite{Jegerlehner:2000dz}). A resolution is to proceed heedless of the mathematical inconsistency by abandoning trace cyclicity. The price to pay for this approach is a high one: results will change depending on what reading point is chosen for $ \gamma_5$-odd traces. There is no general argument that a consistent and unambiguous choice can be made in all cases, but this solution is, nevertheless, often used at low loop order, as it greatly simplifies calculations compared to the mathematically rigorous BMHV scheme~\cite{tHooft:1972tcz,Breitenlohner:1977hr}. One has to attempt to ensure consistency of the calculation on a case-by-case basis, as we have done here by specifying a reading-point prescription for the ambiguous diagrams. 

Traces with an even number of $ \gamma_5 $-matrices evaluate through ordinary manipulations, commuting $ \gamma $-matrices around, using $\gamma_5^2 = \mathds{1} $ to remove all instances of $ \gamma_5 $, and then applying cyclicity to produce a result in terms of $ d $-dimensional metrics. For $ \gamma_5 $-odd traces, we eliminate all but a single $\gamma_5 $-matrix. We then follow~\cite{Korner:1991sx} to assign a value to the trace of the ordinary $ \gamma $-matrices with the remaining $ \gamma_5 $. In $ d $ dimensions, the antisymmetrized products 
\begin{equation}
    \Gamma_{\mu_1 \ldots \mu_n } \equiv \gamma_{[\mu_1} \cdots \gamma_{\mu_n]},
\end{equation}
(including $ \mathds{1}$ [$n=0$]) form a basis of the Dirac algebra, which is orthogonal under the Dirac trace. Any object of the Dirac algebra can be decomposed as a sum of such elements. In four dimensions, $ \gamma_5 = - \dfrac{i}{24} \varepsilon^{\mu \nu \rho \sigma} \Gamma_{\mu \nu \rho \sigma} $. The $ \gamma_5 $-odd traces are then defined in $ d $~dimensions by
    \begin{equation} \label{eq:tr5_def}
    \Tr\!\big[ \Gamma_{\mu_1 \ldots \mu_n } \gamma_5 \big] \equiv 
    \begin{dcases}
        -4 i \varepsilon_{\mu_1 \ldots \mu_4} & \text{for}\; n= 4 \\
        0 & \text{for}\; n\neq 4
    \end{dcases},
    \end{equation}
obviously reproducing the four-dimensional limit.
\texttt{Matchete} implements methods described in~\cite{Kuusela:2019iok} to efficiently decompose products of $ \gamma $-matrices in terms of the antisymmetric basis elements. Non-cyclicity of the traces is evident, e.g., from the observation that 
\begin{align}
\begin{aligned}
    \gamma_\alpha \gamma_\mu \gamma_\nu \gamma_\rho \gamma_\sigma \gamma^\alpha &= (d-8) \Gamma_{\mu\nu\rho\sigma } +\ldots, \\ 
    -\gamma^\alpha \gamma_\alpha \gamma_\mu \gamma_\nu \gamma_\rho \gamma_\sigma &= -d \,\Gamma_{\mu\nu\rho\sigma } +\ldots, 
\end{aligned}
\end{align}
where we have omitted terms proportional to metric tensors times $ \mathds{1} $ and $ \Gamma_{\alpha\beta} $ for $ \alpha,\beta \in \{ \mu,\nu,\rho, \sigma\}$. Tracing these expressions along with a $ \gamma_5 $-matrix clearly demonstrates the non-cyclicity of the definition~\eqref{eq:tr5_def}. This non-cyclicity arises specifically in $\gamma_5$-odd traces involving at least six ordinary $\gamma$-matrices.

The reading-point ambiguity in such $\gamma_5$-odd traces is $\ord{\epsilon}$.\footnote{The ambiguity refers to the difference between traces obtained using different reading points. The finite part of $\gamma_5$-odd traces always remains unambiguous.} Consequently, the one-loop RGEs remain unaffected, being determined solely by the simple $\epsilon$-poles of the counterterms. At two-loop order, however, we do not have this luxury; the simple poles of many two-loop Feynman graphs depend explicitly on the choice of reading point. In the following section, we detail the procedure we have adopted to uniquely fix our reading-point prescription. An unfortunate consequence is that any EFT matching and observables calculations connecting to our results must use a consistent reading-point prescription. Even with all these considerations, we cannot verify that all parts of the calculation are fully consistent. Such is life with an anticommuting $\gamma_5$. 

Another limitation of NDR is that this scheme fails to capture anomalous contributions from chiral field redefinitions. As the effective action is computed off-shell, determining the on-shell counterterms requires field redefinitions of the form
\begin{align}
\psi_{\LL,\RR}(x) \to \psi_{\LL,\RR}^\prime(x) &= \big(1+i\alpha(x)\big)\,\psi_{\LL,\RR}(x),
\end{align}
where $\psi(x)$ is a generic fermion field and $\alpha(x)$ a generic real-valued product of couplings and fields. These transformations are known to yield additional one-loop contributions because of the chiral anomaly. Therefore, we must explicitly account for anomalous contributions to the two-loop RGEs generated during the on-shell reduction of the one-loop counterterms. We determine these contributions using Fujikawa's method~\cite{Fujikawa:1979ay,Fujikawa:1980eg}.

\subsection{Reading-point prescription and $\gamma_5$ ambiguities} 
\label{sec:PrescriptionsAndTricks}

We employ three separate strategies to deal with ambiguous $ \gamma_5 $ traces. The most reliable and unambiguous of these is to impose reality on the obtained counterterms.
Since $\gamma_5$-odd traces acquire an extra factor of $i$ relative to $\gamma_5$-even traces, many ambiguous traces produce purely imaginary counterterms and thus violate the reality of the Lagrangian. We simply remove these imaginary contributions from our results. Even accepting imaginary counterterms, the resulting imaginary running of a real WC would be senseless. We consider this class of ambiguous traces to be entirely unproblematic given their explicitly unphysical manifestation. 

Unfortunately, not all ambiguous traces violate the reality of the counterterm Lagrangian. Our next strategy applies to $\gamma_5$-odd traces in which an EFT vertex appears only once. In this case, our prescription for reading the trace is to start at the operator insertion and proceed along the direction of the fermion flow, indicated by the fermion arrow in Feynman graphs. This reading-point prescription is compatible with~\cite{Korner:1991sx}, which noted that $ \gamma_5 $-anomalies are located in the operator starting the trace, i.e., at our reading point. Having the chiral SMEFT operators as sources of anomalies ensures that anomalous contributions are not associated with external gauge bosons. Crucially, we have been able to verify the validity of this prescription for all four-fermion operators by crosschecking the calculation with a different unambiguous operator basis (cf. Section~\ref{sec:open_basis}). 

Finally, we establish a reading-point prescription for the cases not covered by the unique EFT operator prescription. This happens in cases where
\begin{enumerate}[i)]
    \item the effective operator is not part of the Dirac trace;
    \item all fermions of a four-fermion operator connect to form a single Dirac trace, which happens in diagrams such as the one in Figure~\ref{subfig:vector4Fermi6};
    \item there is more than one effective-operator insertion in a trace, which at dimension six occurs solely through the insertion of two dimension-five operators. In the SMEFT, the only ambiguous contribution of this kind is resolved by the reality condition. 
\end{enumerate}
In these scenarios, we define the reading point by averaging over the two three-point vertices for sunset topologies ({\Large$\boldsymbol{\ominus}$}) or the two disjoint Dirac structures from the four-point vertex for figure-8 topologies ({\LARGE ${\circ\!\circ}$}). This prescription prevents the reading point from coinciding with external gauge lines or Higgs fields. Avoiding external Higgs legs ensures compatibility with NLO broken phase calculations, where parts of such insertions manifest as resummed masses within the propagators rather than as distinct interaction vertices.

Our reading-point prescription ensures reproducibility of the result and constitutes an integral component of the NDR scheme definition, suitable for use in NLO SMEFT calculations. We caution, however, that while we have verified its consistency in numerous specific instances, this prescription does not guarantee validity across all possible computations. In what follows, we systematically identify all classes of ambiguous diagrams encountered in the two-loop SMEFT calculation.

\subsubsection{Identifying ambiguous diagrams} 
\label{sec:IdentifyingAmbiguousDiagrams}

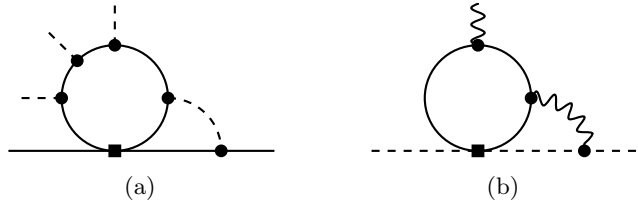
\begin{figure}
    \centering
    \def\radius{0.7} 
    \def\length{1.8} 
    \begin{subfigure}{0.3\textwidth}
        \centering
        \begin{tikzpicture}
            \coordinate (origin) at (0,0);
            \coordinate(phantomHiggs) at (0,-1.8*\radius);
            \coordinate (x1) at ({cos(0)*\radius},{sin(0)*\radius});
            \coordinate (x2) at ({cos(90)*\radius},{sin(90)*\radius});
            \coordinate (x2out) at ({\length*cos(90)*\radius},{\length*sin(90)*\radius});
            \coordinate (x3) at ({cos(135)*\radius},{sin(135)*\radius});
            \coordinate (x3out) at ({\length*cos(135)*\radius},{\length*sin(135)*\radius});
            \coordinate (x4) at ({cos(180)*\radius},{sin(180)*\radius});
            \coordinate (x4out) at ({\length*cos(180)*\radius},{\length*sin(180)*\radius});
            \coordinate (xvert) at ({cos(270)*\radius},{sin(270)*\radius});
            \coordinate (fermionLine) at (2*\radius,-\radius);
            \filldraw[black] ([xshift=-2pt,yshift=-2pt]xvert) rectangle ++(4pt,4pt);
            \filldraw[black] (x1) circle (2pt);
            \filldraw[black] (x2) circle (2pt);
            \filldraw[black] (x3) circle (2pt);
            \filldraw[black] (x4) circle (2pt);
            \filldraw[black] (fermionLine) circle (2pt);
            \draw[] (origin) circle [radius=\radius];
            \draw[dashed] (x1) to[bend left=45] (fermionLine);
            \draw[dashed] (x2) -- (x2out);
            \draw[dashed] (x3) -- (x3out);
            \draw[dashed] (x4) -- (x4out);
            \draw[] (-2*\radius,-\radius) -- (3*\radius,-\radius);
        \end{tikzpicture}
        \caption{}
        \label{subfig:unambiguous1}
    \end{subfigure}
    \begin{subfigure}{0.3\textwidth}
        \centering
        \begin{tikzpicture}
            \coordinate (origin) at (0,0);
            \coordinate (x1) at ({cos(0)*\radius},{sin(0)*\radius});
            \coordinate (x2) at ({cos(90)*\radius},{sin(90)*\radius});
            \coordinate (x2out) at ({\length*cos(90)*\radius},{\length*sin(90)*\radius});
            \coordinate (x3) at ({cos(135)*\radius},{sin(135)*\radius});
            \coordinate (x3out) at ({\length*cos(135)*\radius},{\length*sin(135)*\radius});
            \coordinate (x4) at ({cos(180)*\radius},{sin(180)*\radius});
            \coordinate (x4out) at ({\length*cos(180)*\radius},{\length*sin(180)*\radius});
            \coordinate (xvert) at ({cos(270)*\radius},{sin(270)*\radius});
            \coordinate (fermionLine) at (2*\radius,-\radius);
            \filldraw[black] ([xshift=-2pt,yshift=-2pt]xvert) rectangle ++(4pt,4pt);
            \filldraw[black] (x1) circle (2pt);
            \filldraw[black] (x2) circle (2pt);
            \filldraw[black] (fermionLine) circle (2pt);
            \draw[] (origin) circle [radius=\radius];
            \draw[decorate,decoration={snake,aspect=0,segment length=0.22cm}] (x1) to[bend left=45] (fermionLine);
            \draw[decorate,decoration={coil,aspect=0,segment length=0.22cm}] (x2) -- (x2out);
            \draw[dashed] (-2*\radius,-\radius) -- (3*\radius,-\radius);
        \end{tikzpicture}
        \caption{}
        \label{subfig:unambiguous2}
    \end{subfigure}
    \caption{Diagrams that are unambiguous even though they contain $\gamma_5$-odd traces with six $\gamma$-matrices. The dimension-six operator is indicated with a box.}
    \label{fig:unambiguous_diags}
\end{figure}

We have gathered all relevant diagrams based on the following observations:
\begin{itemize}
    \item As discussed in Section \ref{sec:gamma5}, a $\gamma_5$-odd trace in NDR is ambiguous only if it contains at least six ordinary $\gamma$-matrices;
    \item Truncating our calculation at dimension six limits the number of external legs we can attach to a diagram;
    \item The ambiguous part of a fermion trace is proportional to a Levi-Civita tensor, which allows us to discard diagrams without a sufficient number of external Lorentz indices for contraction. An example of this occurs with a vector four-fermion operator, shown in Figure~\ref{subfig:unambiguous1}, where the two $\gamma$-matrices of the fermion line (dimension-six vertex and propagator) provide the only external Lorentz indices;
    \item Diagrams whose only external fields are a single field-strength tensor and any number of Higgs fields yield operators proportional to
    \begin{align}
        D_\mu \widetilde{F}^{\mu \nu} = 0 ,
    \end{align}
    which vanishes due to the Bianchi identity. An example graph of this class is shown in Figure~\ref{subfig:unambiguous2}, whose ambiguous part leads to an operator $D_\mu F_{\nu\rho}(H^\dagger D_\sigma H) \varepsilon^{\mu\nu\rho\sigma} = 0$.
\end{itemize}
In Figures~\ref{fig:vector4FermiDiags}--\ref{fig:ambiguousWeinbergDiag}, we list the complete set of potentially ambiguous diagrams organized by operator class. These should be understood as classes of functional graphs: each vertex encodes all possible SMEFT couplings, and each internal line denotes a dressed propagator that may generate additional covariant derivatives upon expansion. In the background-field gauge, external gauge lines represent field-strength tensors rather than individual gauge fields, thereby contributing two mass dimensions to the diagram. We proceed to examine the ambiguous traces for each operator class, discuss their impact on the SMEFT \befs, and point out the cases that can be uniquely fixed by known, unambiguous results.

\subsubsection{Four-fermion operators ($\psi^4$)} 
\label{sec:FourFermionOperators}

The four-fermion operators are classified as scalar or vector according to their Dirac structure. Our choice of SMEFT basis (cf. Section~\ref{sec:SMEFT}) ensures an absence of tensor four-fermion operators, removing many potential ambiguities.
We observe that most ambiguities occur in `closed' configurations where one spinor line of the four-fermion operator forms a trace. They are, however, resolved in alternative operator bases with Fierz-related `open' configurations, where the fermion loop closes between different lines of the four-fermion operator, eliminating the ambiguous traces. We can then verify our reading-point prescription by Fierzing the problematic operators to their `open' version to obtain the \emph{unique, consistent} NDR result for the otherwise ambiguous diagrams. At the Lagrangian level, this amounts to an operator basis change, such that `open' and `closed' \befs are related by~\eqref{eq:betaChangeOfBasis}. Hence, we can directly compare the ambiguous parts of the \befs in the `closed' basis with the unambiguous ones from the `open' basis. More technical details on this procedure are provided in Section~\ref{sec:open_basis}.

\input{Figures/vectorFourFermi}

\paragraph{Vector four-fermion operators}
Figure~\ref{fig:vector4FermiDiags} shows the six ambiguous diagram classes that contain vector four-fermion operators. For the sunset-type diagrams (Figures~\ref{subfig:vector4Fermi1}--\ref{subfig:vector4Fermi5}), we validated our reading-point prescription by Fierzing the `closed' operators $O_{ed}$---representing the $(\Bar{R}R)(\Bar{R}R)$ class and the structurally similar $(\Bar{L}L)(\Bar{L}L)$ class---and $O_{qe}$---exemplifying the $(\Bar{L}L)(\Bar{R}R)$ class---into their `open' versions $O_{deed}$ and $O_{qeeq}$, respectively [cf.~\eqref{eq:Fierz_basis_change}]. The sunset diagrams in Figure~\ref{fig:vector4FermiDiags} are unambiguous in the `open' basis. Relating the results back to the `closed' basis through \eqref{eq:betaChangeOfBasis}, we find full agreement with our \befs. This comparison verifies our reading-point prescription and ensures the validity of these results.

Moreover, we note that both diagram classes~\ref{subfig:vector4Fermi1} and~\ref{subfig:vector4Fermi2} appear in the two-loop renormalization of the LEFT. We therefore compare our methods with~\cite{Aebischer:2025hsx}, finding full agreement. In particular, the class of diagrams~\ref{subfig:vector4Fermi2} is structurally similar to the illustrious triangle anomaly, which in isolation is reproduced correctly in NDR only when the trace is started at the axial (chiral) vector operator insertion~\cite{Korner:1991sx}. This observation makes it crucial to \textit{start} all traces at our reading point.

For the figure-8 diagram~\ref{subfig:vector4Fermi6}, we notice that only operators of the class $(\Bar{L}L)(\Bar{R}R)$ are allowed in the effective vertex. Furthermore, the diagram does not exist for semileptonic operators from this class. 
The ambiguities stemming from the remaining five operators in this class (namely, $O_{\ell e}$, $O_{qu}^\sscript{(\|,\times)} $, and $O_{qd}^\sscript{(\|,\times)}$) are purely imaginary and therefore vanish when imposing reality on the counterterm Lagrangian.

\begin{figure}
    \centering
    \def\radius{0.7} 
    \def\length{1.8} 
    \begin{subfigure}{0.45\textwidth}
        \centering
        \begin{tikzpicture}
            \coordinate (origin) at (0,0);
            \coordinate (x1) at ({cos(0)*\radius},{sin(0)*\radius});
            \coordinate (x2) at ({cos(90)*\radius},{sin(90)*\radius});
            \coordinate (x2out) at ({\length*cos(90)*\radius},{\length*sin(90)*\radius});
            \coordinate (x3) at ({cos(135)*\radius},{sin(135)*\radius});
            \coordinate (x3out) at ({\length*cos(135)*\radius},{\length*sin(135)*\radius});
            \coordinate (x4) at ({cos(180)*\radius},{sin(180)*\radius});
            \coordinate (x4out) at ({\length*cos(180)*\radius},{\length*sin(180)*\radius});
            \coordinate (xvert) at ({cos(270)*\radius},{sin(270)*\radius});
            \coordinate (fermionLine) at (2*\radius,-\radius);
            \filldraw[black] ([xshift=-2pt,yshift=-2pt]xvert) rectangle ++(4pt,4pt);
            \filldraw[black] (x1) circle (2pt);
            \filldraw[black] (x2) circle (2pt);
            \filldraw[black] (x4) circle (2pt);
            \filldraw[black] (fermionLine) circle (2pt);
            \draw[] (origin) circle [radius=\radius];
            \draw[decorate,decoration={snake,aspect=0,segment length=0.22cm}] (x1) to[bend left=45] (fermionLine);
            \draw[dashed] (x2) -- (x2out);
            \draw[decorate,decoration={coil,aspect=0,segment length=0.2cm}] (x4) -- (x4out);
            \draw[] (-2*\radius,-\radius) -- (3*\radius,-\radius);
        \end{tikzpicture}
        \caption{}
        \label{subfig:scalar4Fermi1}
    \end{subfigure}
    \begin{subfigure}{0.45\textwidth}
        \centering
        \begin{tikzpicture}
            \coordinate (origin) at (0,0);
            \coordinate (centerLeft) at (-\radius,0);
            \coordinate (centerRight) at (\radius,0);
            \path (centerLeft) + (135:\radius) coordinate (x1);
            \path (centerLeft) + (135:{2*\radius}) coordinate (x1out);
            \path (centerLeft) + (225:\radius) coordinate (x2);
            \path (centerLeft) + (225:{2*\radius}) coordinate (x2out);
            \path (centerRight) + (45:\radius) coordinate (x3);
            \path (centerRight) + (45:{2*\radius}) coordinate (x3out);
            \path (centerRight) + (-45:\radius) coordinate (x4);
            \path (centerRight) + (-45:{2*\radius}) coordinate (x4out);
            \filldraw[black] ([xshift=-2pt,yshift=-2pt]origin) rectangle ++(4pt,4pt);
            \filldraw[black] (x1) circle (2pt);
            \filldraw[black] (x2) circle (2pt);
            \filldraw[black] (x3) circle (2pt);
            \filldraw[black] (x4) circle (2pt);
            \draw[] (centerLeft) circle [radius=\radius];
            \draw[] (centerRight) circle [radius=\radius];
            \draw[decorate,decoration={snake,aspect=0,segment length=0.22cm}] (x1) -- (x1out);
            \draw[dashed] (x2) -- (x2out);
            \draw[decorate,decoration={snake,aspect=0,segment length=0.22cm}] (x3) -- (x3out);
            \draw[dashed] (x4) -- (x4out);
        \end{tikzpicture}
        \caption{}
        \label{subfig:scalar4Fermi2}
    \end{subfigure}
    \caption{Ambiguous diagrams from a scalar four-fermion insertion.}
    \label{fig:scalar4FermiDiags}
\end{figure}
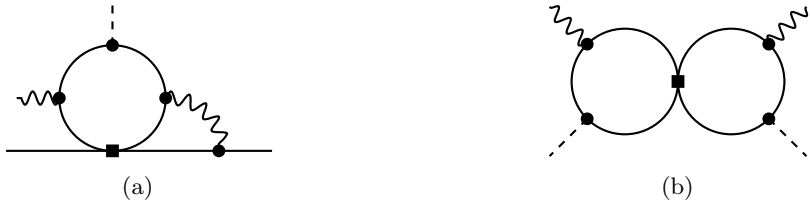

\paragraph{Scalar four-fermion operators}
There is only one class of ambiguous sunset diagrams in the SMEFT containing a scalar four-fermion operator, shown in Figure~\ref{subfig:scalar4Fermi1}, which contributes to the running of dimension-six dipoles. We have explicitly verified that our reading-point prescription reproduces the unambiguous result that can be obtained by Fierzing $O_{\ell edq}$ to $ O_{\ell qde}$---belonging to the operator class $(\Bar{L}R)(\Bar{R}L)$; cf.~\eqref{eq:Fierz_basis_change}---and $O_{quqd}^{{\scriptscriptstyle (\times)}}$ to $ O_{qqdu}$---in the operator class $(\Bar{L}R)(\Bar{L}R)$; cf.~\eqref{eq:Fierz_basis_change_LRLR}---which removes the trace and therefore the ambiguity from Figure~\ref{subfig:scalar4Fermi1}. Using Eq.~\eqref{eq:betaChangeOfBasis}, we find full agreement between the terms generated by this diagram in the `open' and `closed' basis. 

A second source of trace ambiguities arises in the figure-8 diagram shown in Figure~\ref{subfig:scalar4Fermi2}. Among the five scalar four-fermion operators of the SMEFT (listed in Table~\ref{tab:dim6OpsFourFermi}), only $O_{quqd}^{\scriptscriptstyle (\|, \times)}$ and $O_{\ell uqe}$ produce ambiguous counterterms (to the $H^2 X^2$ operator class). Remarkably, under our reading-point prescription, these diagrams generate a pure double-pole divergence with no simple pole. This behavior---reminiscent of what one would expect from factorizable Feynman diagrams~\cite{Jenkins:2023rtg}---implies that there are no contributions to the $H^2 X^2$ \befs, for either the CP-even or CP-odd operators. For the remaining two $(\Bar{L}R)(\Bar{L}R)$ operators, $O_{\ell equ}$ and $O_{\ell edq}$, the diagram factors into two separate, unambiguous traces involving leptons and quarks, respectively.

\subsubsection{Higgs current operators ($\psi^2 H^2 D$)} 
\label{sec:HiggsCurrentOperators}

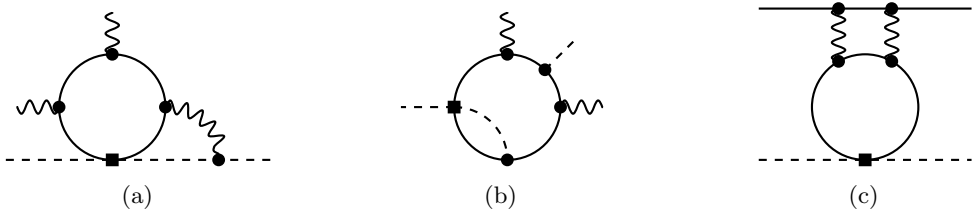
\begin{figure}
    \centering
    \def\radius{0.7} 
    \def\length{1.8} 
    \begin{subfigure}{0.3\textwidth}
        \centering
        \begin{tikzpicture}
            \coordinate (origin) at (0,0);
            \coordinate (x1) at ({cos(0)*\radius},{sin(0)*\radius});
            \coordinate (x2) at ({cos(90)*\radius},{sin(90)*\radius});
            \coordinate (x2out) at ({\length*cos(90)*\radius},{\length*sin(90)*\radius});
            \coordinate (x3) at ({cos(135)*\radius},{sin(135)*\radius});
            \coordinate (x3out) at ({\length*cos(135)*\radius},{\length*sin(135)*\radius});
            \coordinate (x4) at ({cos(180)*\radius},{sin(180)*\radius});
            \coordinate (x4out) at ({\length*cos(180)*\radius},{\length*sin(180)*\radius});
            \coordinate (xvert) at ({cos(270)*\radius},{sin(270)*\radius});
            \coordinate (fermionLine) at (2*\radius,-\radius);
            \filldraw[black] ([xshift=-2pt,yshift=-2pt]xvert) rectangle ++(4pt,4pt);
            \filldraw[black] (x1) circle (2pt);
            \filldraw[black] (x2) circle (2pt);
            \filldraw[black] (x4) circle (2pt);
            \filldraw[black] (fermionLine) circle (2pt);
            \draw[] (origin) circle [radius=\radius];
            \draw[decorate,decoration={snake,aspect=0,segment length=0.2cm}] (x1) to[bend left=45] (fermionLine);
            \draw[decorate,decoration={coil,aspect=0,segment length=0.22cm}] (x2) -- (x2out);
            \draw[decorate,decoration={coil,aspect=0,segment length=0.22cm}] (x4) -- (x4out);
            \draw[dashed] (-2*\radius,-\radius) -- (3*\radius,-\radius);
        \end{tikzpicture}
        \caption{}
        \label{subfig:HiggsCurrent1}
    \end{subfigure}
    \begin{subfigure}{0.3\textwidth}
        \centering
        \begin{tikzpicture}
            \coordinate (origin) at (0,0);
            \coordinate (x1) at ({cos(0)*\radius},{sin(0)*\radius});
            \coordinate (x1out) at ({\length*cos(0)*\radius},{\length*sin(0)*\radius});
            \coordinate (x2) at ({cos(45)*\radius},{sin(45)*\radius});
            \coordinate (x2out) at ({\length*cos(45)*\radius},{\length*sin(45)*\radius});
            \coordinate (x3) at ({cos(90)*\radius},{sin(90)*\radius});
            \coordinate (x3out) at ({\length*cos(90)*\radius},{\length*sin(90)*\radius});
            \coordinate (x4) at ({cos(135)*\radius},{sin(135)*\radius});
            \coordinate (x4out) at ({\length*cos(135)*\radius},{\length*sin(135)*\radius});
            \coordinate (xvert) at ({cos(180)*\radius},{sin(180)*\radius});
            \coordinate (xvertOut) at ({cos(180)*\radius-\radius},{sin(180)*\radius});
            \coordinate (x5) at ({cos(270)*\radius},{sin(270)*\radius});
            \coordinate (x5out) at ({\length*cos(270)*\radius},{\length*sin(270)*\radius});
            \filldraw[black] ([xshift=-2pt,yshift=-2pt]xvert) rectangle ++(4pt,4pt);
            \filldraw[black] (x1) circle (2pt);
            \filldraw[black] (x2) circle (2pt);
            \filldraw[black] (x3) circle (2pt);
            \filldraw[black] (x5) circle (2pt);
            \draw[] (origin) circle [radius=\radius];
            \draw[dashed] (xvert) to[bend left=45] (x5);
            \draw[decorate,decoration={coil,aspect=0,segment length=0.22cm}] (x1) -- (x1out);
            \draw[dashed] (x2) -- (x2out);
            \draw[decorate,decoration={coil,aspect=0,segment length=0.22cm}] (x3) -- (x3out);
            \draw[dashed] (xvert) -- (xvertOut);
        \end{tikzpicture}
        \caption{}
        \label{subfig:HiggsCurrent2}
    \end{subfigure}
    \begin{subfigure}{0.3\textwidth}
        \centering
        \begin{tikzpicture}
            \coordinate (origin) at (0,0);
            \coordinate (x1) at ({cos(60)*\radius},{sin(60)*\radius});
            \coordinate (x1out) at ($(x1)+(0,\radius)$);
            \coordinate (x2) at ({cos(120)*\radius},{sin(120)*\radius});
            \coordinate (x2out) at ($(x2)+(0,\radius)$);
            \coordinate (upperLineMiddle) at (0,{sin(60)*\radius+\radius});
            \coordinate (xvert) at (0,-\radius);
            \filldraw[black] ([xshift=-2pt,yshift=-2pt]xvert) rectangle ++(4pt,4pt);
            \filldraw[black] (x1) circle (2pt);
            \filldraw[black] (x2) circle (2pt);
            \filldraw[black] (x1out) circle (2pt);
            \filldraw[black] (x2out) circle (2pt);
            \draw[] (origin) circle [radius=\radius];
            \draw[decorate,decoration={coil,aspect=0,segment length=0.2cm}] (x1) -- (x1out);
            \draw[decorate,decoration={coil,aspect=0,segment length=0.2cm}] (x2) -- (x2out);
            \draw[dashed] (-2*\radius,-\radius) -- (2*\radius,-\radius);
            \draw[] ($(upperLineMiddle)+(-2*\radius,0)$) -- ($(upperLineMiddle)+(2*\radius,0)$);
        \end{tikzpicture}
        \caption{}
        \label{subfig:HiggsCurrent3}
    \end{subfigure}
    \caption{Ambiguous diagrams from a Higgs current insertion.}
    \label{fig:HiggsCurrentDiags}
\end{figure}

We find no uncontrolled ambiguities arising from Higgs current operators. At first glance, the diagram classes~\ref{subfig:HiggsCurrent1} and~\ref{subfig:HiggsCurrent2} contribute ambiguously to the \bef of CP-odd $X^2 H^2$ operators; however, these contributions are purely imaginary and vanish when we impose reality on the counterterm Lagrangian. Diagram class~\ref{subfig:HiggsCurrent3}, contributing to the running of flavor-universal Higgs current operators, involves another ambiguous trace. It is, however, structurally equivalent to~\ref{subfig:vector4Fermi3}, where we found agreement with both the unambiguous result of~\cite{Aebischer:2025hsx} and our `open' basis results. We conclude that the reading-point prescription produces a consistent result.\footnote{The diagram is equivalent to a LEFT contribution if we substitute the external Higgs current for a non-interacting neutrino contribution.}

\subsubsection{Dimension-six Yukawas ($\psi^2 H^3$)} 
\label{sec:DimensionSixYukawas}

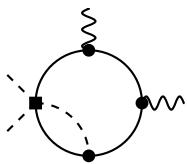
\begin{figure}
    \centering
    \def\radius{0.7} 
    \def\length{1.8} 
    \begin{subfigure}{0.3\textwidth}
        \centering
        \begin{tikzpicture}
            \coordinate (origin) at (0,0);
            \coordinate (x1) at ({cos(0)*\radius},{sin(0)*\radius});
            \coordinate (x1out) at ({\length*cos(0)*\radius},{\length*sin(0)*\radius});
            \coordinate (x2) at ({cos(45)*\radius},{sin(45)*\radius});
            \coordinate (x2out) at ({\length*cos(45)*\radius},{\length*sin(45)*\radius});
            \coordinate (x3) at ({cos(90)*\radius},{sin(90)*\radius});
            \coordinate (x3out) at ({\length*cos(90)*\radius},{\length*sin(90)*\radius});
            \coordinate (x4) at ({cos(135)*\radius},{sin(135)*\radius});
            \coordinate (x4out) at ({\length*cos(135)*\radius},{\length*sin(135)*\radius});
            \coordinate (xvert) at ({cos(180)*\radius},{sin(180)*\radius});
            \coordinate (xvertOut1) at ({-1.6*\radius},{-0.6*\radius});
            \coordinate (xvertOut2) at ({-1.6*\radius},{0.6*\radius});
            \coordinate (x5) at ({cos(270)*\radius},{sin(270)*\radius});
            \coordinate (x5out) at ({\length*cos(270)*\radius},{\length*sin(270)*\radius});
            \filldraw[black] ([xshift=-2pt,yshift=-2pt]xvert) rectangle ++(4pt,4pt);
            \filldraw[black] (x1) circle (2pt);
            \filldraw[black] (x3) circle (2pt);
            \filldraw[black] (x5) circle (2pt);
            \draw[] (origin) circle [radius=\radius];
            \draw[dashed] (xvert) to[bend left=45] (x5);
            \draw[decorate,decoration={coil,aspect=0,segment length=0.22cm}] (x1) -- (x1out);
            \draw[decorate,decoration={coil,aspect=0,segment length=0.22cm}] (x3) -- (x3out);
            \draw[dashed] (xvert) -- (xvertOut1);
            \draw[dashed] (xvert) -- (xvertOut2);
        \end{tikzpicture}
        \label{subfig:dim6Yukawa1}
    \end{subfigure}
    \caption{Ambiguous diagram from a dimension-six Yukawa insertion.}
    \label{fig:dim6Yukawa}
\end{figure}

These operators generate a single ambiguous diagram class as depicted in Figure~\ref{fig:dim6Yukawa}. It contributes to the running of the CP-odd $X^2 H^2$ operators. Unfortunately, in contrast to the four-fermion operators, we cannot check our reading-point prescription in this case by a change of basis. Nor have we been able to determine any consistency checks for this class of diagrams in the literature. Therefore, the validity of our prescription and results for this case should be subjected to additional scrutiny. 

\input{Figures/dim6Dipoles}

\subsubsection{Dimension-six dipoles ($\psi^2 H X$)} 
\label{sec:Dipoles}

\begin{figure}

    \def\radius{0.9} 
    \def\length{1.} 
    \centering
    
    \begin{tikzpicture}[baseline=-0.1cm]
    
        \coordinate (x1) at (0,0);
        \path (x1) +(135:{\length*0.6*\radius}) coordinate (e1);
        \path (x1) +(45:{\length*0.6*\radius}) coordinate (e2);
        \coordinate (x3) at 
        ({-cos(60)*\radius},{-sin(60)*\radius});
        \coordinate (x4) at 
        ({cos(60)*\radius},{-sin(60)*\radius});        
        \path (x3) +(-135:{\length*0.6*\radius}) coordinate (e3);
        \path (x4) +(-45:{\length*0.6*\radius}) coordinate (e4);
        
        \draw (-\radius,0) arc[start angle=180, end angle=0, radius=\radius];
        \draw (\radius,0) arc[start angle=180, end angle=0, radius=-\radius];

        \draw[] (x1) -- (e1);
        \draw[] (x1) -- (e2);
        \draw[dashed] (x3) -- (e3);
        \draw[decorate,decoration={coil, aspect=0, segment length=0.35cm}] (x4) -- (e4);
        
        \draw[decorate,decoration={coil, aspect=0, segment length=0.3cm}] 
        (-\radius,0) -- (x1);
        \draw[dashed] (x1) -- (\radius, 0);
        
        \filldraw ([xshift=-3pt,yshift=-3pt]x1) rectangle ++(6pt,6pt);
        \filldraw[black] (-\radius,0) circle (2pt); 
        \filldraw[black] (\radius,0) circle (2pt); 
        \filldraw[black] (x3) circle (2pt);
        \filldraw[black] (x4) circle (2pt);
        
    \end{tikzpicture} 
    \caption{Diagram containing a dipole not inserted on the ambiguous $ \gamma_5$-odd trace. Our prescription is to average over the two reading points starting at the two three-point vertices.}
    \label{fig:DipoleGamma5Ambiguity}
\end{figure}
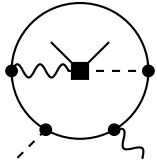

Since the Dirac tensor of the dipole operators contributes two $\gamma$-matrices to Dirac traces, they feature by far the most ambiguous diagrams, namely the eleven classes shown in Figure~\ref{fig:DipoleDiags} with the dipole operators inserted on the fermion line. We have not been able to verify the consistency of our reading-point prescription for any of them, leaving the running of dimension-six dipoles to other operators through these diagrams unchecked. The affected classes of \befs are the dipoles (diagram~\ref{subfig:Dipole1}), the CP-odd $X^2H^2$ operators (diagrams~\ref{subfig:Dipole2}--\ref{subfig:Dipole10}), and the CP-odd $X^3$ operators (diagram~\ref{subfig:Dipole11}). As the dipole coefficients are generic complex $ 3\times 3$ matrices, there is no hope that enforcing reality can cure any of these ambiguities.

Moreover, there is a single diagram class in Figure~\ref{fig:DipoleGamma5Ambiguity}, 
where the dipole insertion is not part of the Dirac trace. These diagrams contribute to the self-running of the dipoles. Following our prescription, we average the Dirac traces read from the two three-point vertices. We cannot crosscheck the result of this class either.

\subsubsection{Bosonic operators}
\label{sec:CPoddH2X2}

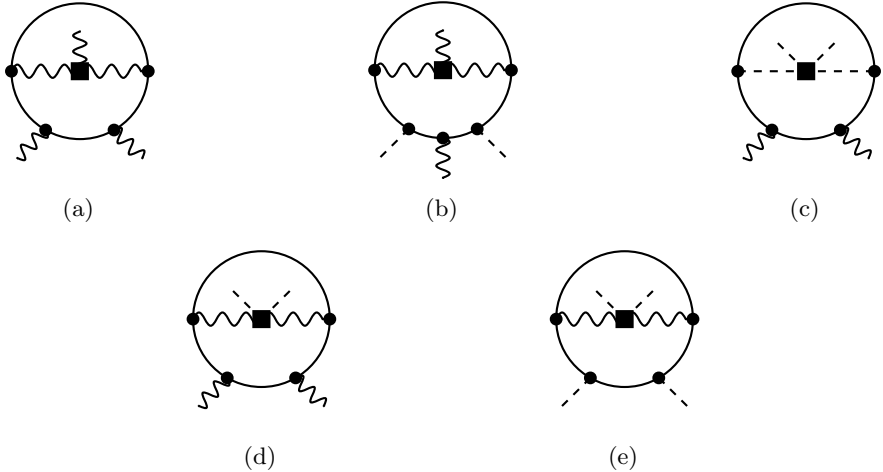
\begin{figure}

    \def\radius{0.9} 
    \def\length{1.} 
    \centering

    \begin{subfigure}{0.3\textwidth}
        \centering
        
        \begin{tikzpicture}[baseline=-0.1cm]  
        
        \coordinate (x1) at (0,0);
        \path (x1) +(90:{\length*0.6*\radius}) coordinate (e1);
        \path (x1) +(270:{\radius+\length*0.6*\radius}) coordinate (e2);
        \coordinate (x2) at (0,-\radius);
        \coordinate (x3) at 
        ({-cos(60)*\radius},{-sin(60)*\radius});
        \coordinate (x4) at 
        ({cos(60)*\radius},{-sin(60)*\radius});        
        \path (x3) +(-135:{\length*0.6*\radius}) coordinate (e3);
        \path (x4) +(-45:{\length*0.6*\radius}) coordinate (e4);
        
        \draw (-\radius,0) arc[start angle=180, end angle=0, radius=\radius];
        \draw (\radius,0) arc[start angle=180, end angle=0, radius=-\radius];

        \draw[decorate,decoration={coil,aspect=0,segment length=0.21cm}] (x1) -- (e1);
        \path (x2) -- (e2);
        \draw[decorate,decoration={coil,aspect=0,segment length=0.21cm}] (x3) -- (e3);
        \draw[decorate,decoration={coil,aspect=0,segment length=0.21cm}] (x4) -- (e4);
        
        \draw[decorate,decoration={coil,aspect=0,segment length=0.31cm}] 
        (-\radius,0) -- (\radius,0);
        
        \filldraw ([xshift=-3pt,yshift=-3pt]x1) rectangle ++(6pt,6pt);
        \filldraw[black] (-\radius,0) circle (2pt); 
        \filldraw[black] (\radius,0) circle (2pt); 
        \filldraw[black] (x3) circle (2pt);
        \filldraw[black] (x4) circle (2pt);
    
        \end{tikzpicture}
        \caption{}
        \label{fig:TripleGaugeAmbiguity1}
        
    \end{subfigure}
    \begin{subfigure}{0.3\textwidth}
        \centering
        
        \begin{tikzpicture}[baseline=-0.1cm]  
        
        \coordinate (x1) at (0,0);
        \path (x1) +(90:{\length*0.6*\radius}) coordinate (e1);
        \path (x1) +(270:{\radius+\length*0.6*\radius}) coordinate (e2);
        \coordinate (x2) at (0,-\radius);
        \coordinate (x3) at 
        ({-cos(60)*\radius},{-sin(60)*\radius});
        \coordinate (x4) at 
        ({cos(60)*\radius},{-sin(60)*\radius});        
        \path (x3) +(-135:{\length*0.6*\radius}) coordinate (e3);
        \path (x4) +(-45:{\length*0.6*\radius}) coordinate (e4);
        
        \draw (-\radius,0) arc[start angle=180, end angle=0, radius=\radius];
        \draw (\radius,0) arc[start angle=180, end angle=0, radius=-\radius];

        \draw[decorate,decoration={coil,aspect=0,segment length=0.21cm}] (x1) -- (e1);
        \draw[decorate,decoration={coil,aspect=0,segment length=0.21cm}] (x2) -- (e2);
        \draw[dashed] (x3) -- (e3);
        \draw[dashed] (x4) -- (e4);
        
        \draw[decorate,decoration={coil,aspect=0,segment length=0.31cm}] 
        (-\radius,0) -- (\radius,0);
        
        \filldraw ([xshift=-3pt,yshift=-3pt]x1) rectangle ++(6pt,6pt);
        \filldraw[black] (-\radius,0) circle (2pt); 
        \filldraw[black] (\radius,0) circle (2pt); 
        \filldraw[black] (x2) circle (2pt);
        \filldraw[black] (x3) circle (2pt);
        \filldraw[black] (x4) circle (2pt);
    
        \end{tikzpicture}
        \caption{}
        \label{fig:TripleGaugeAmbiguity2}
        
    \end{subfigure}
    \begin{subfigure}{0.3\textwidth}
        \centering
        
        \begin{tikzpicture}[baseline=-0.1cm]  
        
        \coordinate (x1) at (0,0);
        \path (x1) +(135:{\length*0.6*\radius}) coordinate (e1);
        \path (x1) +(45:{\length*0.6*\radius}) coordinate (e2);
        \path (x1) +(270:{\radius+\length*0.6*\radius}) coordinate (eph);
        \coordinate (x2) at (0,-\radius);
        \coordinate (x3) at 
        ({-cos(60)*\radius},{-sin(60)*\radius});
        \coordinate (x4) at 
        ({cos(60)*\radius},{-sin(60)*\radius});        
        \path (x3) +(-135:{\length*0.6*\radius}) coordinate (e3);
        \path (x4) +(-45:{\length*0.6*\radius}) coordinate (e4);
        
        \draw (-\radius,0) arc[start angle=180, end angle=0, radius=\radius];
        \draw (\radius,0) arc[start angle=180, end angle=0, radius=-\radius];

        \draw[dashed] (x1) -- (e1);
        \draw[dashed] (x1) -- (e2);
        \path (x2) -- (eph);
        \draw[decorate,decoration={coil,aspect=0,segment length=0.21cm}] (x3) -- (e3);
        \draw[decorate,decoration={coil,aspect=0,segment length=0.21cm}] (x4) -- (e4);
        
        \draw[dashed] 
        (-\radius,0) -- (\radius,0);
        
        \filldraw ([xshift=-3pt,yshift=-3pt]x1) rectangle ++(6pt,6pt);
        \filldraw[black] (-\radius,0) circle (2pt); 
        \filldraw[black] (\radius,0) circle (2pt); 
        \filldraw[black] (x3) circle (2pt);
        \filldraw[black] (x4) circle (2pt);
    
        \end{tikzpicture}
        \caption{}
        \label{fig:CHDAmbiguity}
        
    \end{subfigure}

    \vspace{0.3cm}

    \begin{subfigure}{0.3\textwidth}
        \centering
        
        \begin{tikzpicture}[baseline=-0.1cm]  
        
        \coordinate (x1) at (0,0);
        \path (x1) +(135:{\length*0.6*\radius}) coordinate (e1);
        \path (x1) +(45:{\length*0.6*\radius}) coordinate (e2);
        \path (x1) +(270:{\radius+\length*0.6*\radius}) coordinate (eph);
        \coordinate (x2) at (0,-\radius);
        \coordinate (x3) at 
        ({-cos(60)*\radius},{-sin(60)*\radius});
        \coordinate (x4) at 
        ({cos(60)*\radius},{-sin(60)*\radius});        
        \path (x3) +(-135:{\length*0.6*\radius}) coordinate (e3);
        \path (x4) +(-45:{\length*0.6*\radius}) coordinate (e4);
        
        \draw (-\radius,0) arc[start angle=180, end angle=0, radius=\radius];
        \draw (\radius,0) arc[start angle=180, end angle=0, radius=-\radius];

        \draw[dashed] (x1) -- (e1);
        \draw[dashed] (x1) -- (e2);
        \path (x2) -- (eph);
        \draw[decorate,decoration={coil,aspect=0,segment length=0.21cm}] (x3) -- (e3);
        \draw[decorate,decoration={coil,aspect=0,segment length=0.21cm}] (x4) -- (e4);
        
        \draw[decorate,decoration={coil,aspect=0,segment length=0.31cm}] 
        (-\radius,0) -- (\radius,0);
        
        \filldraw ([xshift=-3pt,yshift=-3pt]x1) rectangle ++(6pt,6pt);
        \filldraw[black] (-\radius,0) circle (2pt); 
        \filldraw[black] (\radius,0) circle (2pt); 
        \filldraw[black] (x3) circle (2pt);
        \filldraw[black] (x4) circle (2pt);
    
        \end{tikzpicture}
        \caption{}
        \label{fig:CPOddX2H22}
        
    \end{subfigure}
    \begin{subfigure}{0.3\textwidth}
        \centering
        
        \begin{tikzpicture}[baseline=-0.1cm]  
        
        \coordinate (x1) at (0,0);
        \path (x1) +(135:{\length*0.6*\radius}) coordinate (e1);
        \path (x1) +(45:{\length*0.6*\radius}) coordinate (e2);
        \path (x1) +(270:{\radius+\length*0.6*\radius}) coordinate (eph);
        \coordinate (x2) at (0,-\radius);
        \coordinate (x3) at 
        ({-cos(60)*\radius},{-sin(60)*\radius});
        \coordinate (x4) at 
        ({cos(60)*\radius},{-sin(60)*\radius});        
        \path (x3) +(-135:{\length*0.6*\radius}) coordinate (e3);
        \path (x4) +(-45:{\length*0.6*\radius}) coordinate (e4);
        
        \draw (-\radius,0) arc[start angle=180, end angle=0, radius=\radius];
        \draw (\radius,0) arc[start angle=180, end angle=0, radius=-\radius];

        \draw[dashed] (x1) -- (e1);
        \draw[dashed] (x1) -- (e2);
        \path (x2) -- (eph);
        \draw[dashed] (x3) -- (e3);
        \draw[dashed] (x4) -- (e4);
        
        \draw[decorate,decoration={coil,aspect=0,segment length=0.31cm}] 
        (-\radius,0) -- (\radius,0);
        
        \filldraw ([xshift=-3pt,yshift=-3pt]x1) rectangle ++(6pt,6pt);
        \filldraw[black] (-\radius,0) circle (2pt); 
        \filldraw[black] (\radius,0) circle (2pt); 
        \filldraw[black] (x3) circle (2pt);
        \filldraw[black] (x4) circle (2pt);
    
        \end{tikzpicture}
        \caption{}
        \label{fig:CPOddX2H2}
        
    \end{subfigure}
    \caption{Diagrams with a bosonic operator insertion and a $ \gamma_5 $-odd trace with pure SM interactions. The reading point prescription is to average over starting points at the two three-point vertices.}
    \label{fig:BosonicGamma5Ambiguity}
    
\end{figure}

The bosonic SMEFT operators are never part of Dirac traces. The prescription for the $ \gamma_5$-odd traces in the diagram classes in Figure~\ref{fig:BosonicGamma5Ambiguity} is to average the reading point over the two three-point vertices. Conveniently, the diagrams in Figures~\ref{fig:TripleGaugeAmbiguity1}--\ref{fig:CPOddX2H22} generate purely imaginary ambiguities, which are resolved by imposing reality on the Lagrangian. However, the class depicted in Figure~\ref{fig:CPOddX2H2}, for which the square vertex represents an insertion of any of the CP-odd $X^2 H^2$ operators ($ C_{H\widetilde{B}} $, $ C_{H\widetilde{W}} $, $ C_{H\widetilde{G}} $ or $ C_{H\widetilde{W}B} $), contains an ambiguity, which cannot be removed. They generate a contribution to the counterterm of the off-shell operator
\begin{equation} \label{eq:ambiguous_operator}
    i \big( H^\dagger H \big) \big( H^\dagger D^2 H - D^2 H^\dagger H\big ) .
\end{equation}
In this case, the Levi-Civita tensor from the dual field-strength tensors $ \widetilde{X}_{\mu\nu} = \tfrac{1}{2} \varepsilon^{\mu\nu\rho \sigma} X_{\mu\nu} $ reduces to metrics when combined with the Levi-Civita tensor from the $ \gamma_5 $-odd Dirac trace [cf.~\eqref{eq:lc_lc_reduction}], thus producing non-vanishing contributions. Upon using field redefinitions for the Higgs field to remove the off-shell operator~\eqref{eq:ambiguous_operator}, the ambiguity is shifted to the $ C_{eH} $, $C_{dH}$, and $ C_{uH}$ Wilson coefficients (the dimension-six Yukawa couplings).

\subsubsection{Weinberg operator} 
\label{sec:WeinbergOperator}
\begin{figure}
    \centering
    \def\radius{0.9} 
    \def\length{1.7} 
    \begin{subfigure}{0.45\textwidth}
        \centering
        \begin{tikzpicture}
            \coordinate (origin) at (0,0);
            \coordinate (xvert1) at ({cos(0)*\radius},{sin(0)*\radius});
            \coordinate (xvert1Out) at ({\length*cos(0)*\radius},{\length*sin(0)*\radius});
            \coordinate (x2) at ({cos(45)*\radius},{sin(45)*\radius});
            \coordinate (x2out) at ({\length*cos(45)*\radius},{\length*sin(45)*\radius});
            \coordinate (x3) at ({cos(90)*\radius},{sin(90)*\radius});
            \coordinate (x3out) at ({\length*cos(90)*\radius},{\length*sin(90)*\radius});
            \coordinate (x4) at ({cos(135)*\radius},{sin(135)*\radius});
            \coordinate (x4out) at ({\length*cos(135)*\radius},{\length*sin(135)*\radius});
            \coordinate (xvert2) at ({cos(180)*\radius},{sin(180)*\radius});
            \coordinate (xvert2Out) at ({\length*cos(180)*\radius},{\length*sin(180)*\radius});
            \coordinate (x5) at ({cos(270)*\radius},{sin(270)*\radius});
            \coordinate (x5out) at ({\length*cos(270)*\radius},{\length*sin(270)*\radius});
            \filldraw[black] ([xshift=-2pt,yshift=-2pt]xvert1) rectangle ++(4pt,4pt);
            \filldraw[black] ([xshift=-2pt,yshift=-2pt]xvert2) rectangle ++(4pt,4pt);
            \filldraw[black] (x2) circle (2pt);
            \filldraw[black] (x4) circle (2pt);
            \draw[] (origin) circle [radius=\radius];
            \draw[dashed] (xvert1) -- (xvert2);
            \draw[decorate,decoration={coil,aspect=0,segment length=0.25cm}] (x2) -- (x2out);
            \draw[decorate,decoration={coil,aspect=0,segment length=0.25cm}] (x4) -- (x4out);
            \draw[dashed] (xvert1) -- (xvert1Out);
            \draw[dashed] (xvert2) -- (xvert2Out);
        \end{tikzpicture}
        \label{subfig:ambiguousWeinbergDiag1}
    \end{subfigure}
    \caption{Potentially ambiguous diagram from two Weinberg operator insertions. The ambiguity vanishes when imposing the reality condition.}
    \label{fig:ambiguousWeinbergDiag}
\end{figure}
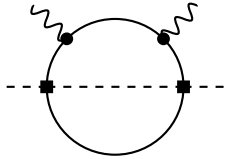

Since the Weinberg operator has dimension five, two insertions may appear in the running of dimension-six operators. If both operators appear in a single fermion trace, the reading point is no longer fixed at the EFT operator insertion. It turns out that the only ambiguous trace involving the Weinberg operator is the one shown in Figure~\ref{fig:ambiguousWeinbergDiag}. Fortunately, the reality condition alone is sufficient to ensure the absence of both contributions and ambiguities into the CP-odd $X^2H^2$ operator class. 

Interestingly, when calculating the diagram in Figure~\ref{fig:ambiguousWeinbergDiag} with averaging of reading points at the two three-point vertices, no contribution to the CP-odd $X^2H^2$ operator class is generated at all. However, it is important to emphasize that this is a prescription dependent statement: using another choice of reading point might give rise to non-zero counterterms, which need to be removed by the reality condition. 
This serves as an instructive example that certain prescriptions directly lead to vanishing ambiguous terms without imposing external conditions.

\subsubsection{Summary}

\begin{table}
    \centering
    \begin{tabularx}{\textwidth}{|c|c|Y|}
         \hline \rowcolor{blue!30} \textbf{Operator class} & \textbf{Problematic diagrams} & \textbf{Affected \boldsymbol{$\beta$}-functions} \\ \hline
         \rowcolor{blue!10} \multicolumn{3}{|c|}{Reading point at the dimension-six vertex} \\ \hline
         Scalar $\psi^4$ & --- & --- \\ \hline
         Vector $\psi^4$ & --- & --- \\ \hline
         $\psi^2 H^2 D$ & --- & --- \\ \hline
         $\psi^2 H^3$ & Figure~\ref{fig:dim6Yukawa} & CP-odd $X^2 H^2$ \\ \hline
         $\psi^2 X H$ & Figures~\ref{subfig:Dipole1}--\ref{subfig:Dipole11} & $ \psi^2 X H$, CP-odd $X^2 H^2$, and CP-odd $X^3$ \\ \hline
         \rowcolor{blue!10} \multicolumn{3}{|c|}{Averaging over vertex reading points} \\ \hline
         Scalar $\psi^4$ & Figure~\ref{subfig:scalar4Fermi2} & CP-odd $ X^2H^2$ \\ \hline
         $\psi^2 X H$ & Figure~\ref{fig:DipoleGamma5Ambiguity} & $\psi^2 X H$ \\ \hline
         $H^2 X^2$ & Figure~\ref{fig:CPOddX2H2} & $\psi^2 H^3$ \\ \hline
         $ X^3$ and $ H^4 D^2$ & --- & --- \\ \hline
         Weinberg operator & --- & --- \\ \hline
    \end{tabularx}
    \caption{Summary of potentially inconsistent contribution to the SMEFT \befs due to ambiguous $\gamma_5$-odd traces.}
    \label{tab:gamma5_problems}
\end{table}

We summarize all potentially inconsistent contributions in Table~\ref{tab:gamma5_problems}. Among all the ambiguous traces, only those originating from the diagrams in Figures.~\ref{subfig:scalar4Fermi2}, \ref{fig:dim6Yukawa}, \ref{subfig:Dipole1}--\ref{subfig:Dipole11}, \ref{fig:DipoleGamma5Ambiguity}, and~\ref{fig:CPOddX2H2} cannot be verified. Fortunately, they only contribute to the running of the dimension-six Yukawas, dipoles, and CP-odd $H^2X^2$ operators, all of which play a relatively minor role (or are loop suppressed) in phenomenology. While we cannot fully guarantee the consistency of these results, we conversely have no indication that our reading-point prescription fails to reproduce the correct single-pole structure for these diagram classes.

\subsection{Treatment of evanescent operators} 
\label{sec:eva_precsription}

Evanescent operators were first identified in the four-fermion sector of the Weak Effective Hamiltonian (WEH)~\cite{Buras:1989xd,Dugan:1990df,Herrlich:1994kh}, where it was observed that extending the Dirac algebra to $d$ dimensions invalidates several identities that hold in strictly four dimensions. Consequently, the reduction of certain $d$-dimensional operators onto a four-dimensional basis generates remnants---\emph{evanescent operators}---which vanish in the $d \to 4$ limit but cannot be neglected in dimensional regularization. In the two-loop SMEFT calculation, further complications arise from CP-odd bosonic operators, which induce $d$-dimensional structures containing Levi-Civita tensors. In the following, we describe how these evanescent contributions are isolated, while their impact on the RG functions is analyzed in the next sections.

For a generic $d$-dimensional operator $O^i_d$, one can separate physical components $Q^a$ from evanescent ones $E^\alpha$. This separation depends on the chosen \emph{evanescent prescription}, which can be encoded in terms of a linear projector $\mathcal{P}$ onto the physical subspace:
    \begin{equation} 
    \mathcal{P} O_d^i = \mathcal{P}_{ia}(\epsilon) Q^a, \qquad
    \mathcal{P}_{ia}(\epsilon) = \sum_{n=0}^\infty \mathcal{P}_{ia,n}\,\epsilon^n ,
    \end{equation}
where the $\epsilon$-dependence of $\mathcal{P}_{ia}$ is kept explicit, as it plays a role in the determination of the two-loop RG functions. The complementary projector $\mathcal{E} \equiv \mathds{1} - \mathcal{P}$ selects the evanescent subspace, and any $d$-dimensional operator decomposes as
    \begin{equation} \label{eq:d_to_P_basis}
    O^i_d = \mathcal{P}_{ia} Q^a + \mathcal{E}_{i\alpha} E^\alpha .
    \end{equation}
The evanescent basis is typically chosen such that $\mathcal{E}$ is independent of $\epsilon$ with any $\ord{\epsilon}$ terms absorbed into the definition of $E^\alpha$. The evanescent operators are formally of rank~$\epsilon$ and vanish in the four-dimensional limit.

The evanescent prescription adopted in our two-loop SMEFT calculation follows~\cite{Fuentes-Martin:2022vvu}, and is consistent with~\cite{Aebischer:2025hsx,Dekens:2019ept},\footnote{Alternative prescriptions for four-fermion evanescent operators have been proposed in, e.g.,~\cite{Aebischer:2024xnf}.} while extending these approaches to structures involving Levi--Civita tensors contracted with Dirac matrices. In a first step, we reduce such contractions by replacing the Levi--Civita tensor with an additional $\gamma_5$:
    \begin{equation} \label{eq:lc_gamma_reduction}
    \varepsilon_{\mu\nu\rho\sigma} \gamma^\sigma \xrightarrow{\;\mathcal{P}\;} - i \Gamma_{\mu\nu\rho} \gamma_5,
    \end{equation}
which is exact only in four dimensions. The introduction of this extra $ \gamma_5 $ does not cause additional complications, as the theory is already fully chiral. If a Levi--Civita tensor contracts several Dirac matrices, the result of~\eqref{eq:lc_gamma_reduction} may depend on the chosen contraction; we fix the prescription by averaging over all possible choices. For example, the following combination that appears at two loops projects to zero with our prescription:\footnote{The projection $ \mathcal{P} $ is kept after the first step, as the two structures (the left–right tensor operators) are still redundant individually and would ordinarily be further transformed to yield a physical structure. In this case, they cancel exactly.} 
    \begin{equation}
    \mathcal{P} \Big[ \varepsilon^{\mu\nu\rho \sigma} (\sigma_{\mu\nu} P_\LL \otimes \sigma_{\rho\sigma} P_\RR) \Big] = \mathcal{P} \Big[ - i (\sigma^{\rho\sigma} \gamma_5  P_\LL \otimes \sigma_{\rho\sigma} P_\RR) - i (\sigma_{\mu\nu} P_\LL \otimes \sigma^{\mu\nu} \gamma_5 P_\RR) \Big] = 0.
    \end{equation}
Here `$ \otimes $' indicates that the Dirac structures are inserted into two ordered but otherwise unspecified spinor lines. Multiple Levi--Civita tensors are also reduced with the four-dimensional identity,
    \begin{equation} \label{eq:lc_lc_reduction}
    \varepsilon_{\alpha_1 \alpha_2 \alpha_3 \alpha_4 } \varepsilon^{\beta_1 \beta_2 \beta_3 \beta_4} \xrightarrow{\;\mathcal{P}\;} -\delta\du{\alpha_1}{[\beta_1} \delta\du{\alpha_2}{\beta_2} \delta\du{\alpha_3}{\beta_3} \delta\du{\alpha_4}{\beta_4]}.
    \end{equation}

\begin{table}[]
    \centering
    \begin{tabular}{|b|l|}
    \hline
    {\bfseries Bilinear structures} & $ P_\LL \otimes P_\LL,\; P_\LL \otimes P_\RR,\; P_\RR \otimes P_\LL,\; P_\RR \otimes P_\RR,$ \\
    & $\gamma_\mu P_\LL \otimes \gamma^\mu P_\LL,\; \gamma_\mu P_\LL \otimes \gamma^\mu P_\RR,\; \gamma_\mu P_\RR \otimes \gamma^\mu P_\LL,\; \gamma_\mu P_\RR \otimes \gamma^\mu P_\RR,$ \\
    & $ \sigma_{\mu\nu} P_\LL \otimes \sigma^{\mu\nu} P_\LL,\; \sigma_{\mu\nu} P_\RR \otimes \sigma^{\mu\nu} P_\RR $ \\ \hline
    {\bfseries Linear structures} & $ P_\LL,\; P_\RR,\; \gamma_\mu P_\LL,\; \gamma_\mu P_\RR,\; \sigma_{\mu\nu} P_\LL,\; \sigma_{\mu\nu} P_\RR $ \\ \hline
    \end{tabular}
    \caption{Chiral basis of four-dimensional Dirac structures.}
    \label{tab:dirac_basis}
\end{table}

Following the removal of Levi-Civita tensors, we reduce any bilinear Dirac structures not contained in our four-dimensional Dirac basis. Reflecting the chiral nature of the SMEFT, we adopt the chiral basis shown in Table~\ref{tab:dirac_basis}. We implement this reduction via a projection using Dirac traces following~\cite[Section~3.2.2]{Fuentes-Martin:2022vvu}; all powers of $\epsilon$ are retained so the projection is invariant under Dirac-algebra manipulations prior to its application. A subtlety arises because this reduction is not invariant under the transposition of fermion currents (see also~\cite{Dekens:2019ept}). To resolve this, we arrange all currents in the direction of the fermion flow prior to reduction whenever possible; for instance, terms such as $(\overline{\psi_1^\cc} \psi_2^\cc)$ are rewritten as $(\bar\psi_2 \psi_1)$. In cases where a unique flow direction is ill-defined (e.g., due to a single charge-conjugated fermion), we symmetrize over the equivalent forms. For example, we arrange $(\overline{\psi_1^\cc} \psi_2)$ as $ \frac{1}{2} [(\overline{\psi_1^\cc} \psi_2)+ (\overline{\psi_2^\cc} \psi_1)]$ before reduction of Dirac structures.

At the end of our reduction of fermion bilinears, we then apply four-dimensional Fierz relations according to the field content of the operator to reduce redundant four-fermion structures to the Mainz basis. A special example is the $ O_{ee}^{prst} = \big(\overline{e}^p \gamma_\mu e^r\big) \big(\overline{e}^s \gamma^\mu e^t \big)$ operator. In four dimensions, Fierz identities impose the extra flavor symmetry $ O_{ee}^{prst} = O_{ee}^{ptsr} $ (in addition to the obvious $ d $-dimensional $ O_{ee}^{prst} = O_{ee}^{stpr} $ symmetry). Thus, the physical part of the operator is
    \begin{equation}
        \mathcal{P} \big[O_{ee}^{prst}\big] = \tfrac{1}{2} \big( O_{ee}^{prst} + O_{ee}^{ptsr} \big).
    \end{equation}
Finally, we employ Schouten identities, which dictate that fully antisymmetric combinations of five or more Lorentz indices vanish in four dimensions. These identities are used to eliminate certain bosonic operators, e.g.,
    \begin{equation}
    \mathcal{P} \big[ B_{\mu\nu} G^A_{\nu\rho} \widetilde{G}^A_{\mu\rho} \big] = 0,
    \end{equation}
within the $ X^3 $ operator class. 

The reduction procedure described above implicitly specifies the set of evanescent operators through the decomposition in~\eqref{eq:d_to_P_basis}. A list of reduction identities for the redundant $d$-dimensional structures arising in the two-loop SMEFT calculation---those not included in our physical operator basis---is provided in Appendix~\ref{app:eva_ops}.

\subsection{Renormalization schemes}

Contrary to their one-loop counterparts, two-loop RGEs are generally renormalization scheme dependent. We aim to extract the two-loop RGEs in a variant of the \msbar scheme, including finite counterterms that are required to properly decouple the evanescent operators under the RG~\cite{Buras:1989xd,Dugan:1990df,Herrlich:1994kh}. 
In renormalizable theories, the \msbar scheme is usually regarded as a unique prescription where the counterterms subtract only the divergences of the Green's functions.\footnote{This impression is belied by an ambiguity in choosing the square root of the wave-function renormalization~\cite{Jack:1990eb,Fortin:2012cq,Herren:2021yur,Zhang:2025ywe,Thomsen:2025kka}.} 
This notion breaks down in EFTs, as the operator basis is not unique and the $d$-dimensional Lorentz and Dirac algebra often induce explicitly $ \epsilon $-dependent transformations between bases. Thus, \msbar counterterms in one basis may acquire finite parts when expressed in another. It is clear that the `\msbar scheme' is a basis-dependent statement rather than something universal to the theory.

In setting up our calculation, we must manage multiple renormalization schemes. We, therefore, begin by outlining the basic elements that distinguish such schemes. A key element of any scheme definition is the operator basis. For a generic scheme~$\mathcal{S}$,\footnote{We restrict attention to schemes based on dimensional regularization with the same continuation of~$\gamma_5$.} we denote the associated (off-shell) operator basis by $\mathscr{B}_{\mathcal{S}} = \{ O^i_{\mathcal{S}} \}$, such that the bare action takes the form\footnote{With an abuse of notation, we use the variables associated with Lagrangian operators, such as $ O^i $ and $ Q^\alpha $, also in the context of actions. In this case, they properly refer to the integrated operators, e.g., $ \int\, \dd^d x\, O^i$. Whether the symbols refer to the Lagrangian operators or the integrated operators should be clear from context.}
\begin{equation}
    S_\mathcal{S}(\bar{\lambda}_{\mathcal{S}}) 
    = S_\mathrm{kin} + \bar{\lambda}_{\mathcal{S}, i} O_{\mathcal{S}}^i,
\end{equation}
where $S_\mathrm{kin}$ denotes the kinetic terms of the matter fields. The functional form of $ S_{\mathcal{S}} $ is shared for all schemes with the associated basis $ \mathscr{B}_{\mathcal S}$. The bare couplings decompose as
\begin{equation}
    \bar{\lambda}_{\mathcal{S}, i} 
    = \mu^{k_i \epsilon}
      \big(\lambda_{\mathcal{S},i} 
      + \delta_\mathcal{S} \lambda_i 
      + \Delta_\mathcal{S} \lambda_i \big),
\end{equation}
with $\lambda_{\mathcal{S},i}$ the finite renormalized couplings, $\delta_\mathcal{S}\lambda_i$ the divergent counterterms, and $\Delta_\mathcal{S}\lambda_i$ any finite counterterms (if present). The parameter $\mu$ is the renormalization scale, and $k_i$ is fixed such that the mass dimension of the renormalized couplings agrees with its four-dimensional counterpart. For brevity, we invariably omit the overall factor $\mu^{k_i \epsilon}$ in what follows. While any operator basis is formally infinite due to the $d$-dimensional Dirac algebra, only a finite subset is needed in practice, once suitable methods are employed. 

As a starting point, we consider a genuine $d$-dimensional operator basis $\mathscr{B}_{d}=\{O^i_{d}\}$ that does not impose any ordering on the (infinite) set of elements. Such a basis is not practical for physics, as it requires infinitely many couplings that mix under RG evolution. Nevertheless, it appears naturally at intermediate stages of calculations, in particular in our functional approach. The issue of infinitely many couplings is resolved by introducing a projected basis $\mathscr{B}_{\mathcal{P}}$, which decomposes into a finite set of physical operators $\{Q^i\}$ (at each order in the EFT expansion) and an infinite set of evanescent operators $\{E^\alpha\}$. As discussed in the previous section, this separation is uniquely specified by the choice of projector $\mathcal{P}$ (evanescent prescription) onto the physical space acting on the $d$-dimensional operators. In suitable renormalization schemes, the dynamics of the physical theory is determined entirely by the renormalized values of the physical couplings.

To summarize, a renormalization scheme is specified by an operator basis, a physical projector/evanescent prescription (if relevant), and a renormalization prescription fixing the counterterms. We employ the following schemes in what follows:
\begin{itemize}
    \item[$ \mathcal{S} $:] a generic scheme 
        \[
        \mathcal{S} = \big( \mathscr{B}_\mathcal{S},\, \mathcal{P}_\mathcal{S},\, \hat{\delta}_{\mathcal{S}}\lambda \equiv \delta_{\mathcal{S}}\lambda + \Delta_\mathcal{S} \lambda \big),
        \]
        with action
        \begin{equation}
        S_\mathcal{S}(\bar{\lambda}_\mathcal{S}) 
        = S_\mathrm{kin} + \bar{\lambda}_{\mathcal{S},i} O_\mathcal{S}^{i} 
        = S_\mathrm{kin} + \big(\lambda_{\mathcal{S},i} + \delta_\mathcal{S} \lambda_i + \Delta_\mathcal{S} \lambda_i \big) O_\mathcal{S}^{i}\, .
        \end{equation}
        We use this notation whenever a discussion or setup applies generally, without reference to a particular scheme;
    \item[$ \dbar $:] the \msbar scheme in a $ d $-dimensional (unordered) basis,
        \[
        \dbar = \big( \mathscr{B}_d,\, \mathds{1},\, \delta_{\dbar} \lambda \big),
        \]
        with action
        \begin{equation}
        S_{\dbar}(\bar{\lambda}_{\dbar}) 
        = S_\mathrm{kin} + \bar{\lambda}_{\dbar,i} O_d^{i} 
        = S_\mathrm{kin} + \big(\lambda_{\dbar,i} + \delta_{\dbar} \lambda_i \big) O_d^{i}\, .
        \end{equation}
        Since it has no evanescent prescription ($\mathcal{P}_{\dbar} = \mathds{1}$), this scheme is not viable for physical calculations but remains useful at intermediate steps;
    \item[$ \Pbar $:] the \msbar scheme with an evanescent prescription $ \mathcal{P} $,
        \[
        \Pbar = \big( \mathscr{B}_\mathcal{P},\, \mathcal{P},\, (\delta_{\Pbar} g,\, \delta_{\Pbar} \eta ) \big),
        \]
        with action
        \begin{equation}
        \begin{split}
        S_{\Pbar}(\bar{g}_{\Pbar},\, \bar{\eta}_{\Pbar}) 
        &= S_\mathrm{kin} + \bar{g}_{\Pbar,a} Q_{\mathcal{P}}^{a} + \bar{\eta}_{\Pbar,\alpha} E_{\mathcal{P}}^{\alpha} \\
        &= S_\mathrm{kin} + \big( g_{\Pbar, a} + \delta_{\Pbar} g_a \big) Q_{\mathcal{P}}^{a} 
        + \big( \eta_{\Pbar, \alpha} + \delta_{\Pbar} \eta_\alpha \big) E_{\mathcal{P}}^{\alpha}\, .
        \end{split}
        \end{equation}
        This scheme separates physical and evanescent operators, but the two sectors mix uncontrollably under RG evolution. It is used as a stepping stone toward physical schemes rather than a good scheme in itself;
    \item[$ \Pf $:] the \emph{finitely-compensated} evanescent scheme,
        \[
        \Pf = \big( \mathscr{B}_\mathcal{P},\, \mathcal{P},\, (\delta_{\Pf} g + \Delta_{\Pf} g,\, \delta_{\Pf} \eta ) \big),
        \]
        with action
        \begin{equation}
        \begin{split}
        S_{\Pf}( \bar{g}_{\Pf},\, \bar{\eta}_{\Pf} ) 
        &= S_\mathrm{kin} + \bar{g}_{\Pf, a} Q_{\mathcal{P}}^{a} + \bar{\eta}_{\Pf, \alpha} E_{\mathcal{P}}^{\alpha} \\
        &= S_\mathrm{kin} + \big( g_{\Pf, a} + \delta_{\Pf} g_a + \Delta_{\Pf} g_a \big) Q_{\mathcal{P}}^{a} 
        + \big( \eta_{\Pf, \alpha} + \delta_{\Pf} \eta_\alpha \big) E_{\mathcal{P}}^{\alpha}\, .
        \end{split}
        \end{equation}
        Here, finite renormalization constants for the physical operators cancel the dependence on evanescent couplings ensuring that physical processes are fully defined by the physical couplings in the four-dimensional limit~\cite{Buras:1989xd,Dugan:1990df,Herrlich:1994kh}. In the absence of evanescent couplings, the physical couplings renormalize as in \msbar, i.e.,~by minimal subtraction. This scheme is equivalent (though philosophically distinct) to the \emph{evanescent-subtracted} scheme~\cite{Fuentes-Martin:2022vvu}, where the evanescent couplings are removed via a finite shift of the physical couplings. We refer the reader to Appendix~\ref{app:eva_schemes} for further details.
\end{itemize}
Whenever the operator bases coincide, as is the case in the $\Pf$- and $\Pbar$-schemes, the one-loop divergent counterterms are the same, i.e., $\delta_{\Pf}^{(1)}g_a = \delta_{\Pbar}^{(1)}g_a$ and $\delta_{\Pf}^{(1)}\eta_\alpha = \delta_{\Pbar}^{(1)}\eta_\alpha$, if the renormalized couplings coincide. As we discuss below, this implies scheme independence of the one-loop \befs. It is well known that the scheme independence no longer holds at two-loop order.

\subsection{RG functions from counterterms} 
\label{sec:RG_formula}

Having established the relevant renormalization schemes, we now turn to the extraction of the RG functions. As anticipated, our aim is to determine the two-loop SMEFT \befs in the finitely-compensated scheme~$\Pf$, where physical amplitudes are independent of evanescent couplings, and, of equal importance, the flow of the physical couplings decouples from the evanescent ones. In practice, however, counterterms are most conveniently computed in the minimal-subtraction scheme with an evanescent prescription, $\Pbar$; the two-loop \befs in the $\Pf$-scheme can be reconstructed directly from $\Pbar$ counterterms by including a one-loop shift induced by evanescent terms. We summarize the essential ingredients below and refer to Appendix~\ref{app:eva_schemes} for further details.

The $\beta$ functions for the physical couplings in the $\Pf$-scheme are defined in the usual manner,
\begin{equation}
    \beta_{\Pf, a}(g) \equiv \frac{\dd g_{\Pf, a}}{\dd \ln \mu}.
\end{equation}
A convenient relation to the $\Pbar$-scheme arises from the fact that at $\eta_\alpha=0$, the bare couplings coincide in the two schemes:
\begin{equation}
    \bar{g}_{\Pf}(g, \eta=0 ) = \bar{g}_{\Pbar}(g, \eta=0), 
    \qquad 
    \bar{\eta}_{\Pf}(g, \eta=0) = \bar{\eta}_{\Pbar}(g, \eta=0).
\end{equation}
At one-loop order, only the divergent counterterms of the physical couplings contribute, and the result coincides with that of any scheme:
\begin{equation} \label{eq:one_loop_beta}
    \beta_{\Pf, a}^{(1)} (g)  
    = 2 \, \delta^{(1)}_{\Pbar, 1} g_{a}(g, 0) ,
\end{equation}
where $\delta^{(\ell)}_{\Pbar,1}\lambda$ denotes the coefficient of the $1/\epsilon$ pole of the $\lambda$ counterterm at loop order~$\ell$.\footnote{In general, we write 
    \[
    \delta_{\mathcal S} \lambda_i = \sum_{n=1}^\infty \dfrac{\delta_{\mathcal S,n} \lambda_i}{\epsilon^n},
    \]
for the expansion of the counterterms in powers of $\epsilon$.}

\begin{figure}[t]
    \centering
    \includegraphics[width=.8\textwidth]{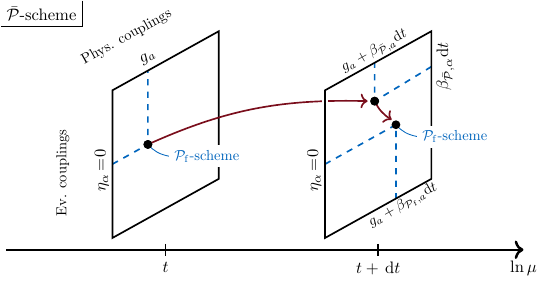}
    \caption{Schematic depiction of the RG flow in the $\Pbar$-scheme. After RG evolution, the evanescent couplings can be absorbed into a shift of physical couplings (without changing physics). At vanishing evanescent couplings, the $ \Pbar $- and $ \Pf $-scheme coincides.}
    \label{fig:RG_schemes}
\end{figure}

At two-loop order, the relation becomes more intricate. In the $\Pbar$-scheme, the evanescent couplings mix into the physical sector, whereas in $\Pf$ this flow is absorbed by shifting the physical couplings, as illustrated in Figure~\ref{fig:RG_schemes}. The resulting \befs can be expressed in terms of $\Pbar$ counterterms as
\begin{equation} \label{eq:two_loop_beta}
    \beta_{\Pf, a}^{(2)} (g)  
    = 4 \, \delta^{(2)}_{\Pbar, 1} g_{a}(g, 0) 
    + 2 \, \delta^{(1)}_{\Pbar,1} \, \eta_{\alpha}(g, 0) 
    \frac{\partial \Delta^{(1)}_{\Pbar\to \Pf} g_a }{\partial \eta_{\Pbar,\alpha}}(g, 0),
\end{equation}
with all counterterms evaluated at $\eta=0$. The first term is the standard minimal-subtraction contribution, determined by the $1/\epsilon$ poles of the two-loop counterterms of the physical couplings. The second term compensates for the mixing into the evanescent sector. It depends both on the divergent one-loop counterterms of evanescent couplings and on a finite shift that relates the physical couplings in the two schemes at one loop order:
\begin{equation} 
    \Delta_{\Pbar\to \Pf}^{(1)} g_a(g, \eta) \, Q_{\mathcal{P}}^{a} 
    = -\Delta_{\Pf}^{(1)}\! g_a(g, \eta) Q_{\mathcal{P}}^{a} 
    = \mathcal{P} \rir \! \big[\Gamma^{(1)}_{\mathcal{P}}(g, \eta) - \Gamma^{(1)}_{\mathcal{P}}(g, 0) \big] \Big|_{\epsilon=0}.
\end{equation}
Here, $\Gamma_\mathcal{P}^{(1)}$ denotes the one-loop effective action, with the subindex $\mathcal{P}$ indicating that the $\mathscr{B}_\mathcal{P}$-basis is employed, and $\rir$ an infrared (IR) regulator (cf. Section~\ref{sec:rstar_method}), ensuring that divergences arise exclusively from the ultraviolet (UV). This finite shift encodes the contribution of evanescent couplings in the $\Pbar$-scheme to physical Green’s functions. The $\ord{\epsilon^0}$ piece arises when evanescent contributions multiply the $1/\epsilon$ ultraviolet poles of divergent one-loop integrals, making it local. Consequently, the second term in~\eqref{eq:two_loop_beta} can be computed entirely from the product of two one-loop calculations.

\section[The Local \texorpdfstring{$ \boldsymbol{R}^{\ast} $}{R*}-Method in the Presence of Evanescent Operators]{The Local \texorpdfstring{$ R^{\ast} $}{R*}-Method in the Presence of Evanescent Operators} \label{sec:Rstar}

The $\rstar$-method~\cite{Chetyrkin:1982nn,Chetyrkin:1984xa,Chetyrkin:2017ppe,Herzog:2017bjx} provides a systematic prescription for subtracting all subdivergences from Feynman graphs via the \rstarop to isolate their overall divergence. In our previous work~\cite{Born:2024mgz}, we adapted the local version of the \rstarop from~\cite{Herzog:2017bjx} to functional methods. With the inclusion of fermionic operators in the full dimension-six SMEFT, however, evanescent operators become unavoidable, which complicates the direct link between the counterterm Lagrangian and the UV counterterms for the subdivergences from the original formulation. As emphasized in~\cite{Naterop:2024cfx}, both the definition of evanescent operators and the choice of renormalization scheme affect the action of the \rstarop. 

In the local approach of~\cite{Born:2024mgz}, the \rstarop acts solely on \emph{generalized loop integrals}. These generalized integrals carry all Lorentz and Dirac structures, momentum dependence, and vertex/propagator assignments, while remaining agnostic to specific field identities, flavor indices, and gauge group representations. The UV counterterms introduced by the local $\rstar$-operation to cancel subdivergences are therefore drawn from the full $d$-dimensional basis $ \mathscr{B}_d$ of off-shell operators (spanning all Lorentz and Dirac invariants, independent of the fields), rather than from the physical/evanescent operator basis of the SMEFT Lagrangian. One must carefully account for the $\mathcal{O}(\epsilon)$ differences between the counterterms that should be inserted according to the global, Lagrangian prescription and those generated by the local \rstarop. 

The following discussion first describes the local $\rstar$-method in general terms, then illustrates the renormalization of subdivergences in both global and local pictures in the presence of evanescent structures using an example, and finally presents a general strategy for incorporating evanescent operators into the local framework.

\subsection[The \texorpdfstring{$ \boldsymbol{R}^{\ast} $}{R*}-method]{The \texorpdfstring{\mathversion{boldsans}$ R^{\ast} $}{R*}-method} 
\label{sec:rstar_method}

We introduce the shorthand notation
    \begin{equation}
    \delta_\mathcal{S} S \equiv \delta_{\mathcal{S}} \lambda_i \, O_{\mathcal{S}}^i \,,
    \qquad 
    \hat\delta_\mathcal{S} S \equiv \hat\delta_{\mathcal{S}} \lambda_i \, O_{\mathcal{S}}^i \,,
    \qquad 
    \Gamma_\mathcal{S}(\lambda_\mathcal{S}) \equiv \Gamma \big[ S_\mathcal{S}(\lambda_\mathcal{S}) \big] \,,
    \end{equation}
where $\delta_\mathcal{S} S$ denotes the divergent counterterm action, $\hat{\delta}_\mathcal{S} S$ the full (divergent $+$ finite) counterterm action, and $\Gamma_\mathcal{S}$ the quantum effective action calculated from the action $ S_\mathcal{S} $ associated with the $ \mathscr{B}_{\mathcal{S}} $ operator basis, employed in the arbitrary $ \mathcal{S} $-scheme. The effective action defined like this is renormalized only if the bare couplings are used as arguments---as in $ \Gamma_\mathcal{S}(\bar\lambda_\mathcal{S})$---otherwise no counterterms are included. The renormalized effective action can also be written in terms of the recursive $ \boldsymbol{R} $-operation (see, e.g.,~\cite{Chetyrkin:2017ppe}), which subtracts all nested UV divergences from each graph it is applied to. The $ \boldsymbol{R}_{\mathcal{S}} $ operator is scheme dependent, as it inserts UV counterterms in the appropriate scheme. Thus, we have\footnote{The reader may be unfamiliar with the notion of applying the $ \boldsymbol{R} $-operation to the entire effective action, rather than to individual graphs. We take $ \boldsymbol{R}_{\mathcal{S}} $ to act linearly as it is applied to the sum of all 1PI Green's functions, which make up the effective action.}
    \begin{equation}
    \Gamma_\mathcal{S}(\bar{\lambda}_\mathcal{S}) = \boldsymbol{R}_\mathcal{S} \Gamma_\mathcal{S}(\lambda_\mathcal{S}) = \hat{\delta}_\mathcal{S} S(\lambda_{\mathcal{S}} ) + \bar{\boldsymbol{R}}_\mathcal{S} \Gamma_\mathcal{S}(\lambda_\mathcal{S}),
    \end{equation}
Here, $ \bar{\boldsymbol{R}}_{\mathcal{S}} $ acts as $ \boldsymbol{R}_{\mathcal{S}} $ except that it does not insert an overall counterterm for a graph (if present). 

The \rstarop~\cite{Chetyrkin:1982nn,Chetyrkin:1984xa} is a close cousin of $ \boldsymbol{R} $, but it also subtracts infrared divergences to render graphs fully finite. This is particularly handy when calculating counterterms, as IR-rearrangement is known to introduce spurious IR divergences, which must be carefully accounted for to isolate the UV divergence of a graph. We employ a local version of the \rstarop adapted to functional methods~\cite{Herzog:2017bjx,Born:2024mgz}.
The scheme-dependent \rstarop renders the effective action finite and is defined by
    \begin{equation} \label{eq:Rstar_def}
    \boldsymbol{R}^\ast_{\mathcal{S}} \Gamma_{\mathcal{S}}(\lambda_{\mathcal{S}}) = \ctop_{\mathcal{S}} \Gamma_{\mathcal{S}}(\lambda_{\mathcal{S}})+ \bar{\boldsymbol{R}}_{\mathcal{S}}^\ast \Gamma_{\mathcal{S}}(\lambda_{\mathcal{S}}),
    \end{equation}
where $ \ctop_{\mathcal{S}} $ is the counterterm operation in the $ \mathcal{S} $ scheme, which produces the overall UV counterterm associated to a graph. It follows that 
    \begin{equation}
    \ctop_{\mathcal{S}} \Gamma_{\mathcal{S}}(\lambda_{\mathcal{S}}) = \hat{\delta}_\mathcal{S} S(\lambda_{\mathcal{S}} )
    \end{equation}
by construction. We will shortly see how to determine $ \ctop_{\mathcal{S}} $ recursively in minimal subtraction schemes. 

The second term in~\eqref{eq:Rstar_def} applies the $\rstarbar_{\mathcal{S}}$-operator, which removes all IR and UV subdivergences from a graph (or a generalized loop integral) by inserting the corresponding counterterms. Its action on a Feynman graph $G$ is defined as
    \begin{equation} \label{eq:Rstar_on_graph}
    \rstarbar_{\mathcal{S}}(G) = \sum_{\substack{\gamma \in \overline{W}_{\!\UV}(G)\\ \gamma'\in W_{\!\sscript{IR}}(G) \\ \gamma \cap \gamma' = \emptyset}} \!\!\!\!\! \boldsymbol{\Delta}_{\sscript{IR}}(\gamma') \ast \ctop_{\mathcal{S}} (\gamma) \ast G/ \gamma \setminus \gamma'.
    \end{equation}
Here the sum runs over all UV-divergent proper subgraphs $\gamma$ (unions of disjoint 1PI subgraphs including the empty set), denoted by $\overline{W}_{\!\UV}(G)$, together with non-overlapping IR-divergent subgraphs $\gamma'\in W_{\!\sscript{IR}}(G)$. Schematically, the UV counterterm of $\gamma$ is inserted (denoted by `$\ast$') into the reduced graph $G/\gamma$, obtained by shrinking each connected component of $\gamma$ to a point. The IR counterterms $\ctop_{\sscript{IR}}$, whose explicit form we do not need here, are independent of the scheme $\mathcal{S}$, which is left implicit. We refer the reader to~\cite{Herzog:2017bjx} for further details on the \rstarop. When acting on a generalized loop integral, $\rstar_{\mathcal{S}}$ operates in complete analogy with its action on the underlying graph, producing a finite result.

The \rstarop provides a systematic way to obtain counterterms in minimal subtraction schemes, where no finite counterterms are present; that is, schemes where $\hat{\delta}_\mathcal{S} S = \delta_\mathcal{S} S$. To isolate divergences, we introduce the pole operator $\kop_\mathcal{S}$, which extracts the pole part of the $\epsilon$ expansion. For $ \epsilon $-independent coefficients $C_{n,i}$, its action is defined by
\begin{equation}
   \kop_\mathcal{S} \! \sum_{n= \eminus \infty}^{\infty}\!  C_{n,i} \,\epsilon^n O_{\mathcal{S}}^i 
   =   \sum_{n= \eminus \infty}^{\eminus 1} \! C_{n,i} \,\epsilon^n O_{\mathcal{S}}^i .
\end{equation}
The operator $\kop_\mathcal{S}$ is tied to a specific operator basis $\mathscr{B}_{\mathcal{S}}$ and evanescent prescription $ \mathcal{P}_{\mathcal{S}}$, since transformations between different bases [e.g.,~\eqref{eq:d_to_P_basis}] generally depend on $\epsilon$. Applying $\kop_\mathcal{S}$ to~\eqref{eq:Rstar_def}, and recalling that $\rstar_\mathcal{S}$ renders graphs finite, we can express the divergent part of the $\mathcal{S}$-scheme counterterms as
    \begin{equation} \label{eq:div_ct_Sscheme}
    \delta_{\mathcal{S}} S(\lambda_\mathcal{S}) = \kop_{\mathcal{S}} \ctop_{\mathcal{S}} \Gamma_{\mathcal{S}}(\lambda_{\mathcal{S}}) = - \kop_{\mathcal{S}} \bar{\boldsymbol{R}}_{\mathcal{S}}^\ast \Gamma_{\mathcal{S}}(\lambda_{\mathcal{S}}).
    \end{equation}
This equation provides a recursive procedure for calculating counterterms in minimal subtraction schemes: the entire counterterm is determined by the pole part of $\rstarbar_\mathcal{S}$ acting on the effective action. The recursion arises because the action of $\rstarbar_\mathcal{S}$ on the right-hand side involves insertions of lower-loop $\mathcal{S}$-scheme counterterms, themselves obtained from~\eqref{eq:div_ct_Sscheme}. The recursion terminates at one-loop order, where the right-hand side contains no UV subdivergences.

Two additional observations make the $ \rstar $-method particularly convenient for the calculation of counterterms. The counterterms, being local operators, are polynomial in fields, their derivatives (external momenta), and couplings. The Taylor expansion operator $ \boldsymbol{T} $, which expands an expression treating all these objects as small compared to loop momenta, therefore, acts trivially on the l.h.s. of~\eqref{eq:div_ct_Sscheme}. When applied to a Feynman graph/generalized loop integral (inside the momentum integrals) $ \boldsymbol{T} $ can give rise to spurious IR divergences, but these are compensated for by the IR subtraction of $ \rstarbar_{\mathcal{S}} $. Consequently, the two operations commute~\cite{Herzog:2017bjx}, and we obtain
    \begin{equation} \label{eq:S_scheme_ct_action}
    \delta_{\mathcal{S}} S(\lambda_{\mathcal{S}})  = - \kop_{\mathcal{S}} \rstarbar_{\mathcal{S}} \boldsymbol{T} \, \Gamma_{\mathcal{S}} (\lambda_{\mathcal{S}}). 
    \end{equation}
The application of $ \boldsymbol{T} $ leaves scaleless loop integrals on the r.h.s. (they do not vanish due to the application of $ \rstarbar_{\mathcal{S}} $), of which only those with logarithmic superficial divergence contribute to the counterterms. At this point, we can freely perform IR rearrangement on the generalized loop integrals inside $ \rstarbar_{\mathcal{S}} $ without changing the UV divergences. This approach allows, for instance, for the introduction of an auxiliary mass parameter in the propagators to regulate all IR divergences. A detailed, concrete example of this procedure, as we apply, it is given in~\cite{Born:2024mgz}.

For our present purposes, we focus on one- and two-loop counterterms. At one-loop order, we have
    \begin{equation} \label{eq:RstarS_Gamma_S_1}
     \rstarbar_\mathcal{S} \Gamma^{(1)}_\mathcal{S}(\lambda_\mathcal{S}) = \rir \Gamma^{(1)}_\mathcal{S}(\lambda_\mathcal{S}),
    \end{equation}
where $\rir$ denotes the IR regulator operation within the $\rstarbar$-operation [corresponding to the terms with $\gamma = \emptyset$ in~\eqref{eq:Rstar_on_graph}]. This follows from the one-loop effective action not containing any UV subdivergences. Consequently, $\rstarbar_{\mathcal{S}}$ acts identically on one-loop graphs in \emph{any} scheme $ \mathcal S $. At two-loop order, however, the action of the $\rstarbar$-operation takes the form
    \begin{equation} \label{eq:RstarS_Gamma_S}
    \rstarbar_\mathcal{S} \Gamma^{(2)}_\mathcal{S}(\lambda_\mathcal{S}) =  
    \hat{\delta}_\mathcal{S}^{(1)} \! \lambda_{i}(\lambda_\mathcal{S}) \dfrac{\partial \rir \Gamma^{(1)}_\mathcal{S}}{\partial \lambda_{\mathcal{S},i}} (\lambda_\mathcal{S}) 
    + \rir \Gamma^{(2)}_\mathcal{S}(\lambda_\mathcal{S}),
    \end{equation}
which is scheme dependent. In addition to the pure IR counterterms, $ \rstarbar_{\mathcal{S}} (\rstar_{\mathcal{S}}) $ also inserts one-loop $ \mathcal{S}$-scheme counterterms in all possible ways across the one-loop graphs contained in $ \Gamma^{(1)}_\mathcal{S} $. Eq.~\eqref{eq:RstarS_Gamma_S} constitutes a more \emph{global} perspective on the \rstarop,\footnote{In the sense that it is applied at the level of the action, with one-loop counterterms canceling subdivergences across multiple two-loop graphs. In contrast, the \emph{local} version of the \rstarop acts on individual graphs, recursively determining the required counterterms to cancel their specific subdivergences.} which, while conceptually transparent, is not suitable for our calculations. Still, it provides valuable insight into how scheme changes can be implemented. Before presenting the general formalism for basis transformations within the \rstarop, we illustrate the procedure with a concrete example to highlight the differences between global and local implementations of the \rstarop.

\subsection{Practical example} 
\label{sec:example}

As a simple toy model, we consider an effective theory with a left-handed fermion $\psi$, carrying unit charge under a $\U(1)$ gauge group with coupling $ g $:
    \begin{align}\label{eq:Lagrangian_toy-model}
    \mathcal{L} = -\frac{1}{4} A_{\mu \nu}^2 + i \, \Bar{\psi} \slashed{D} \psi + C \, Q + \ldots \, , \qquad 
    Q = (\Bar{\psi} \gamma_\mu P_\LL \psi)^2\, .
    \end{align}
In this model, we seek to determine the $ \Pbar$-scheme counterterms, namely the \msbar counterterms defined with respect to the evanescent-ordered basis $\mathscr{B}_\mathcal{P}$: this would allow us to determine the $ \Pf $-scheme \bef per~\eqref{eq:two_loop_beta}, similar to our approach to the SMEFT \befs.

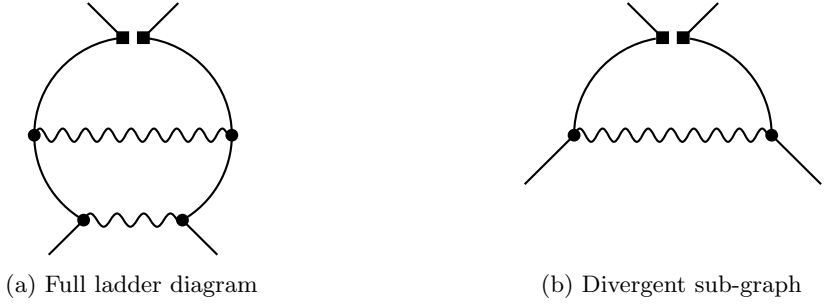
\begin{figure}
    \def\radius{1.3} 
    \begin{subfigure}[t]{0.45\textwidth}
        \centering
        \begin{tikzpicture}[baseline=-0.1cm]
            
            \coordinate (x1) at (96:\radius);
            \coordinate (x2) at (84:\radius);
            \path (x1) +(135:{0.5*\radius}) coordinate (e1);
            \path (x2) +(45:{0.5*\radius}) coordinate (e2);
            \coordinate (x3) at 
            ({-cos(60)*\radius},{-sin(60)*\radius});
            \coordinate (x4) at 
            ({cos(60)*\radius},{-sin(60)*\radius});        
            \path (x3) +(-135:{0.5*\radius}) coordinate (e3);
            \path (x4) +(-45:{0.5*\radius}) coordinate (e4);
            
            \draw (-\radius,0) arc[start angle=180, end angle=96, radius=\radius];
            \draw (\radius,0) arc[start angle=0, end angle=84, radius=\radius];
            \draw (\radius,0) arc[start angle=0, end angle=-60, radius=\radius];
            \draw (-\radius,0) arc[start angle=0, end angle=60, radius=-\radius];
    
            \draw (x1) -- (e1);
            \draw (x2) -- (e2);
            \draw (x3) -- (e3);
            \draw (x4) -- (e4);
            
            \draw[decorate,decoration={coil,aspect=0,segment length=0.3cm}] 
            (-\radius,0) -- (\radius,0);
            \draw[decorate,decoration={coil,aspect=0}] (x3) -- (x4);
            
            \filldraw[black] (-\radius,0) circle (2pt);
            \filldraw[black] (\radius,0) circle (2pt); 
            \filldraw ([xshift=-2pt,yshift=-2pt]x1) rectangle ++(4pt,4pt);
            \filldraw ([xshift=-2pt,yshift=-2pt]x2) rectangle ++(4pt,4pt);
            \filldraw[black] (x3) circle (2pt);
            \filldraw[black] (x4) circle (2pt);
        
        \end{tikzpicture} 
    \caption{Full ladder diagram}
    \label{fig:ExampleFullSunset}
    \end{subfigure}
    %
    \begin{subfigure}[t]{0.45\textwidth}
        \centering
        \begin{tikzpicture}[baseline=-0.1cm]
            
            \coordinate (x1) at (96:\radius);
            \coordinate (x2) at (84:\radius);
            \path (x1) +(135:{0.5*\radius}) coordinate (e1);
            \path (x2) +(45:{0.5*\radius}) coordinate (e2);
            
            \draw (-\radius,0) arc[start angle=180, end angle=100, radius=\radius];
            \draw (\radius,0) arc[start angle=0, end angle=80, radius=\radius];
    
            \draw (-\radius,0) -- (-1.5*\radius,-0.5*\radius);
            \draw (\radius,0) -- (1.5*\radius,-0.5*\radius);
            \draw (x1) -- (e1);
            \draw (x2) -- (e2);
            
            \draw[decorate,decoration={coil,aspect=0,segment length=0.3cm,segment length=0.3cm}] 
            (-\radius,0) -- (\radius,0);
            
            \filldraw[black] (-\radius,0) circle (2pt); 
            \filldraw[black] (\radius,0) circle (2pt);
            \filldraw ([xshift=-2pt,yshift=-2pt]x1) rectangle ++(4pt,4pt);
            \filldraw ([xshift=-2pt,yshift=-2pt]x2) rectangle ++(4pt,4pt);

            \path (-\radius,0) arc[start angle=0, end angle=180, radius=-\radius];
            \coordinate (x3) at 
            ({-cos(60)*\radius},{-sin(60)*\radius});       
            \draw (x3) +(-135:{0.5*\radius}) coordinate (e3);
        
        \end{tikzpicture} 
    \caption{Divergent sub-graph}
    \label{fig:ExampleSubGraph}
    \end{subfigure}

    \caption{Ladder diagram and its divergent subgraph. The four-fermion vertex at the top is to be understood as a single contact interaction and is  separated here only to clarify the spinor contractions.}
    \label{fig:ExampleFigure}
\end{figure}

It is instructive to look at the ladder diagram depicted in Figure~\ref{fig:ExampleFullSunset}, which arises as part of the two-loop counterterm calculation for the effective four-fermion operator $ Q $. To construct the overall counterterm of this diagram, we have to add the counterterm for the one-loop subdivergence shown in Figure~\ref{fig:ExampleSubGraph} to the full two-loop diagram. Schematically, we write this procedure under the $\rstar$-method as
    \begin{align} \label{eq:ladder_example_counterterm}
    \ctop_{\Pbar} \! \left\{ G \right\} 
    = -\kop_{\mathcal{P}} \! \left\{  
    \raisebox{-0.375\height}{\includegraphics[trim={0.15cm 0.15cm 0.15cm 0.15cm},clip,height=2.5em]{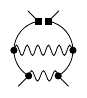}} 
    +
    \ctop_{\Pbar} \!
    \left\{
    \raisebox{-0.375\height}{\includegraphics[trim={0.15cm 0.15cm 0.15cm 0.15cm},clip,height=2.5em]{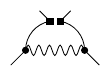}}
    \right\}
    \ast
    \raisebox{-0.375\height}{\includegraphics[trim={0.1cm 0.1cm 0.1cm 0.1cm},clip,height=2.5em]{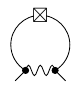}}
    \right\},
    \end{align}
in the $ \Pbar $-scheme. The graph $G$ corresponds to the full two-loop graph (without Taylor expansion) displayed in Figure~\ref{fig:ExampleFullSunset}, and the operator $\kop_{\mathcal{P}}$ extracts the $\epsilon$-poles in the operator basis  $\mathscr{B}_\mathcal{P}$. 
The counterterm formula~\eqref{eq:ladder_example_counterterm} is obtained via application of \eqref{eq:S_scheme_ct_action}, and the diagrams on the r.h.s. represent the superficially logarithmically divergent parts of the Taylor expansion, which has then been IR rearranged to remove IR divergences. To extract two-loop RG-functions, we need only the contribution to the counterterms of the physical operators, so we can freely apply the physical projector $ \mathcal P $ to~\eqref{eq:ladder_example_counterterm} to isolate the objects of interest.
 
Ideally, the \rstarop will insert the counterterm for the one-loop subgraph in the appropriate scheme, $ \Pbar $; however, this would require the (computer) implementation of the $ \rstar $-operator to be able to map the subdivergence of the graph into an operator of the $ \mathscr{B}_{\mathcal{P}}$ basis. Generally, this requires some information about the external fields of the subgraph; that is, we would need to provide the counterterm operator with additional `global' information, not contained in the generalized loop integrals, and to transform the Lorentz and Dirac structures of the subdiagram to match that of the Feynman rule from the appropriate $ \mathscr{B}_{\mathcal{P}} $ operator. Instead, we find it much more convenient to implement a `local' version of the $ \rstar $-operator, which can determine the counterterm directly from the local information contained in the divergent subgraph and, even, the generalized loop integral. In practice, this implies that we are employing counterterms from the $ d $-dimensional operator basis $ \mathscr{B}_d $. Consequently, we end up using $ \rstar_{\dbar} $ for our computer implementation. The minimal subtraction of the subdivergences ensures that the counterterm subtraction can be done fully locally, in fact, even without deriving Feynman rules for the counterterms---their structure is exactly that resulting from the subdivergence of the full two-loop diagram. Using $ \rstar_{\dbar} $ rather than $ \rstar_{\Pbar} $ to regulate the subdivergences will, evidently, change the calculated two-loop counterterm, as we will show with our example here. However, it does so in a controlled manner, which we can easily rectify {\itshape a posteriori}.

We focus now on the one-loop counterterm insertion, as this is where the scheme sensitivity of the \rstarop enters. In a generic scheme $\mathcal{S}$, it takes the form\footnote{Momentum integrals are abbreviated \[
    \int_k \;\; \equiv \int \dfrac{\dd^d k}{(2\pi)^d}\,.
\]} 
    \begin{align} \label{eq:ExampleOneLoopRes}
    \ctop_\mathcal{S} \! 
    \left\{
    \raisebox{-0.375\height}{\includegraphics[trim={0.15cm 0.15cm 0.15cm 0.15cm},clip,height=2.5em]{Figures/one_loop.pdf}}
    \right\}
    &= -g^2 C \kop_\mathcal{S} \! \left\{ \frac{1}{d} \big( \gamma_\mu \gamma_\nu \gamma_\rho P_\LL \otimes \gamma^\rho \gamma^\nu \gamma^\mu P_\LL \big) \int_k \, \frac{1}{(k^2-a)^2} \right\} ,
    \end{align}
where we introduced the auxiliary mass-squared $a$ to make the integral IR safe.\footnote{IR rearrangement lets us change the IR behavior of logarithmically divergent integrals without changing the overall UV divergence.} 
At intermediate stages, we work in a $d$-dimensional basis of fully antisymmetrized $\gamma$-matrices---following our approach in the full SMEFT calculation---and rewrite~\eqref{eq:ExampleOneLoopRes} as\footnote{Our evanescent prescription is compatible with all $d$-dimensional Dirac algebra manipulations, so this intermediate manipulation is necessary in the $\dbar$-scheme and allowed in the $\Pbar$-scheme.}
    \begin{align} \label{eq:ExampleOneLoopResAntiSym}
    \ctop_\mathcal{S} \!
    \left\{
    \raisebox{-0.375\height}{\includegraphics[trim={0.15cm 0.15cm 0.15cm 0.15cm},clip,height=2.5em]{Figures/one_loop.pdf}}
    \right\} = - g^2 C \kop_\mathcal{S} \! \left\{ \frac{1}{d} \Big[ (3d-2) \big( \gamma_\mu P_\LL \big)^{\otimes 2} - \big(\Gamma_{\mu\nu\rho} P_\LL \big)^{\otimes 2} \Big] \int_k \, \frac{1}{(k^2-a)^2} \right\} .
    \end{align}
Short-hand notation such as $\big( \gamma_\mu P_\LL \big)^{\otimes 2} \equiv \big( \gamma_\mu P_\LL \otimes \gamma^\mu P_\LL \big)$ is employed to improve the readability of the expressions. Depending on whether one adopts a global or a local approach to the counterterms for the subdivergences, the calculation diverges from here.

\subsubsection{Global renormalization}

In a traditional, `global' approach to renormalization, one-loop counterterm insertions correspond to the Feynman rules of the one-loop counterterm Lagrangian in the $\Pbar$-scheme. In the language of the $\rstar$-method, this procedure is equivalent to ensuring that $\mathcal{S}=\Pbar$ in~\eqref{eq:RstarS_Gamma_S}. 
If we were to apply this global approach to the example at hand, we would include the external fields and therefore interpret the second term of~\eqref{eq:ExampleOneLoopResAntiSym} as the one-loop counterterm to the $d$-dimensional operator
    \begin{align} 
    O_d = \big(\Bar{\psi} \, \Gamma_{\mu\nu\rho} P_\LL \, \psi\big)^2 \, .
    \end{align}
To obtain the $\Pbar$-scheme counterterm of our original physical operator $Q$, we follow our evanescent prescription and decompose
    \begin{align} \label{eq:DiracRed3Gamma}
     O_d = (6 + 2 \epsilon) \, Q + E \, ,    
    \end{align}
which implicitly defines the evanescent operator $E$. It can quickly be verified that $E $ is indeed evanescent (it vanishes in four dimensions). 
The $\Pbar$-scheme counterterm of the one-loop subgraph corresponds exactly to the one-loop counterterm renormalizing the Lagrangian. We find for~\eqref{eq:ExampleOneLoopResAntiSym} that
    \begin{align} \label{eq:evExGlobalOneLoop}
    \begin{split}
    \ctop_{\Pbar} \! 
    \left\{
    \raisebox{-0.375\height}{\includegraphics[trim={0.15cm 0.15cm 0.15cm 0.15cm},clip,height=2.5em]{Figures/one_loop.pdf}}
    \right\} 
    &= -g^2 C \kop_{\mathcal P} \! \left\{ \frac{1}{d} \left[ \big(4-8\epsilon\big) Q - E \right] \int_k \, \frac{1}{(k^2-a)^2} \right\} \\
    &= - \frac{g^2 C}{16 \pi^2 \epsilon} \, Q + \frac{g^2 C}{64 \pi^2 \epsilon} \,  E \, .
    \end{split}
    \end{align}
Observe that the $ \ord{\epsilon} $ coefficient of the evanescent reduction \eqref{eq:DiracRed3Gamma} drops out; it does not contribute to the counterterm. This a salient difference from what we will find in the $ \dbar$-scheme.

As a next step, one constructs Feynman rules for the one-loop counterterms of both physical and evanescent operators, which are then inserted into a one-loop diagram to properly renormalize the theory at two-loop order. 
One finds that the divergence of the one-loop diagram with the one-loop evanescent counterterm is purely evanescent and, so, does not contribute to the two-loop counterterms for $ Q $.  
After expanding around large loop momentum $k$, we find that the total contribution to the physical counterterm for the diagram in Figure~\ref{fig:ExampleFullSunset} is
\begin{align} 
    \kop_{\mathcal P} \mathcal P  \!
    \left\{
    \ctop_{\Pbar} \!
    \left\{
    \raisebox{-0.375\height}{\includegraphics[trim={0.15cm 0.15cm 0.15cm 0.15cm},clip,height=2.5em]{Figures/one_loop.pdf}}
    \right\}
    \ast
    \raisebox{-0.375\height}{\includegraphics[trim={0.15cm 0.1cm 0.15cm 0.1cm},clip,height=2.5em]{Figures/ct_insertion.pdf}}
    \right\}
    &= -\frac{g^4 C}{16\pi^2} \kop_{\mathcal P} \mathcal P \! \left\{ \frac{1}{d \epsilon} \big( \Bar{\psi} \gamma_\mu \gamma_\nu \gamma_\rho P_\LL \psi \big) \big(\Bar{\psi} \gamma^\rho \gamma^\nu \gamma^\mu P_\LL \psi \big) \int_k \, \frac{1}{(k^2-a)^2} \right\} \nonumber \\
    &= -\frac{g^4 C}{16\pi^2} \kop_{\mathcal P} \! \left\{ \frac{1}{d \epsilon} \big(4-8\epsilon\big) Q \int_k \, \frac{1}{(k^2-a)^2} \right\} \nonumber \\
    &= - \frac{g^4 C}{\left(16\pi^2\right)^2} \left[ \frac{1}{\epsilon^2} - \frac{3}{2\epsilon} + \frac{1}{\epsilon}\log\!\left(\frac{\mu^2}{a} \right) \right] Q \, .
    \label{eq:evExGlobalResult}
\end{align}
In the second line, we used~\eqref{eq:DiracRed3Gamma} again to reduce the Dirac structure. The evanescent piece immediately vanishes under the projection $ \mathcal P$. The logarithm in~\eqref{eq:evExGlobalResult} is an artifact of viewing the counterterm contribution to~\eqref{eq:ladder_example_counterterm} in isolation; it cancels against a logarithm in the divergence of the full ladder diagram.

\subsubsection{Local $\boldsymbol{R^\ast}$ renormalization}
We can implement the $\rstar_{\dbar}\,$-operator of the $ \dbar $-scheme using only the `local' information contained in the generalized loop integral. This allows for constructing one-loop counterterm insertions automatically  at the integrand level. Unlike in the $ \Pbar $-scheme, the $\mathcal{O}(\epsilon)$ part of the evanescent reduction~\eqref{eq:DiracRed3Gamma} remains hidden here in the Dirac structure of the counterterms; it only reappears at the very end when it can multiply a double pole, thus giving a contribution to the two-loop single pole that was previously discarded due to the $\kop_{\mathcal{P}}$ operation on the subdivergence. This contribution needs to be subtracted to eventually recover the $ \Pbar $-scheme result.

Specifically in our example, the one-loop $\rstar_{\dbar}\,$-counterterm derived from~\eqref{eq:ExampleOneLoopResAntiSym} reads
\begin{align} \label{eq:eq:evExLocalOneLoop}
\begin{split}
    \ctop_{\dbar} \!
    \left\{
    \raisebox{-0.375\height}{\includegraphics[trim={0.15cm 0.15cm 0.15cm 0.15cm},clip,height=2.5em]{Figures/one_loop.pdf}}
    \right\} 
    &= - g^2 C \kop_{d} \! \left\{ \frac{1}{d} \left[ (3d-2) \big( \gamma_\mu P_\LL \big)^{\otimes 2} - \big(\Gamma_{\mu\nu\rho} P_\LL \big)^{\otimes 2} \right] \int_k \ \frac{1}{(k^2-a)^2} \right\} \\
    &= -\frac{5 g^2 C}{32\pi^2 \epsilon} \big( \gamma_\mu P_\LL \big)^{\otimes 2} + \frac{g^2 C}{64\pi^2 \epsilon} \big(\Gamma_{\mu\nu\rho} P_\LL \big)^{\otimes 2} \, .
\end{split}
\end{align}
In some sense, we get the counterterm Feynman rules directly in this `local' approach without the need to pass through a true operator basis.
When inserting these counterterms into the remaining one-loop graph, we obtain
\begin{align}
\begin{split}
    \kop_{\mathcal P} \!
    \left\{
    \ctop_{\dbar} \! 
    \left\{
    \raisebox{-0.375\height}{\includegraphics[trim={0.15cm 0.15cm 0.15cm 0.15cm},clip,height=2.5em]{Figures/one_loop.pdf}}
    \right\}
    \ast
    \raisebox{-0.375\height}{\includegraphics[trim={0.15cm 0.1cm 0.15cm 0.1cm},clip,height=2.5em]{Figures/ct_insertion.pdf}}
    \right\}
    &= -\frac{5 g^4 C}{32 \pi^2} \kop_{\mathcal P} \!
    \left\{ \frac{1}{d \epsilon} \big(\gamma_\mu \gamma_\nu \gamma_\rho P_\LL \otimes \gamma^\rho \gamma^\nu \gamma^\mu P_\LL \big) \int_k \, \frac{1}{(k^2-a)^2} \right\} \\
    &\mathrel{\phantom{=}} +\frac{g^4 C}{64 \pi^2} \kop_{\mathcal P} \!
    \left\{ \frac{1}{d \epsilon} \big(\Gamma_{\mu\nu\rho} \gamma_\sigma \gamma_\kappa P_\LL \otimes \gamma^\kappa \gamma^\sigma \Gamma^{\mu\nu\rho} P_\LL \big) \int_k \,  \frac{1}{(k^2-a)^2} \right\} \, .
\end{split}
\end{align}
for the counterterm topology. Expanding the terms in brackets to the basis of antisymmetrized $\gamma$-matrices yields
    \begin{align}
    \kop_{\mathcal P} \!
    \left\{
    \ctop_{\dbar} \! 
    \left\{
    \raisebox{-0.375\height}{\includegraphics[trim={0.15cm 0.15cm 0.15cm 0.15cm},clip,height=2.5em]{Figures/one_loop.pdf}}
    \right\}
    \ast
    \raisebox{-0.375\height}{\includegraphics[trim={0.15cm 0.1cm 0.15cm 0.1cm},clip,height=2.5em]{Figures/ct_insertion.pdf}}
    \right\}
    =\,&+ \frac{g^4 C}{32 \pi^2} \kop_{\mathcal P} \!
    \left\{ \frac{1}{d \epsilon} (4-6d-3d^2)\big(\gamma_\mu P_\LL \big)^{\otimes 2} \int_k \, \frac{1}{(k^2-a)^2} \right\} \nonumber \\
    & -\frac{g^4 C}{64 \pi^2} \kop_{\mathcal P} \!
    \left\{ \frac{1}{d \epsilon} (8-7d)\big(\Gamma_{\mu\nu\rho} P_\LL \big)^{\otimes 2} \int_k \,  \frac{1}{(k^2-a)^2} \right\} \\
    & -\frac{g^4 C}{64 \pi^2} \kop_{\mathcal P} \!
    \left\{ \frac{1}{d \epsilon} \big(\Gamma_{\mu\nu\rho\sigma\kappa} P_\LL \big)^{\otimes 2} \int_k \,  \frac{1}{(k^2-a)^2} \right\} \, . \nonumber
    \end{align}
To proceed, we make use of the decomposition
\begin{align}
    \big(\Bar{\psi} \, \Gamma_{\mu\nu\rho\sigma\kappa} P_\LL \, \psi \big)^2 &= 36 \epsilon \, Q + E' \, ,
\end{align}
defining a second evanescent operator $E'$.
Along with \eqref{eq:DiracRed3Gamma}, we determine the one-loop counterterm insertion in the local $\rstar$-method to be
\begin{align} \label{eq:evExLocalResult}
    \kop_{\mathcal P} \mathcal P \!
    \left\{
    \ctop_{\dbar} \!
    \left\{
    \raisebox{-0.375\height}{\includegraphics[trim={0.15cm 0.15cm 0.15cm 0.15cm},clip,height=2.5em]{Figures/one_loop.pdf}}
    \right\}
    \ast
    \raisebox{-0.375\height}{\includegraphics[trim={0.15cm 0.1cm 0.15cm 0.1cm},clip,height=2.5em]{Figures/ct_insertion.pdf}}
    \right\}
    &= -\frac{g^4 C}{\left(16\pi^2\right)^2} \left[ \frac{1}{\epsilon^2} - \frac{2}{\epsilon} + \frac{1}{\epsilon} \log \! \left( \frac{\mu^2}{a} \right) \right] Q \, .
\end{align}
This is the contribution to the two-loop counterterm of the physical operator from the one-loop counterterm insertion.

\subsubsection{Comparison} 

Since the UV divergence of the Taylor expanded two-loop diagram is the same in both the global and local renormalization approach, the mismatch between the two approaches arises solely through the scheme dependence of the one-loop counterterm insertions calculated above. Comparing \eqref{eq:evExLocalResult} to \eqref{eq:evExGlobalResult}, we find 
    \begin{align} \label{eq:evExDiff}
    \kop_{\mathcal P} \mathcal P \big( \ctop_{\Pbar}\{G\} - \ctop_{\dbar}\{G\} \big) = -\frac{g^4 C}{2\left(16\pi^2\right)^2 \epsilon} Q \, .
    \end{align}
As mentioned, this difference between the full two-loop counterterms comes from the $\mathcal{O}(\epsilon)$ piece of the evanescent reduction~\eqref{eq:DiracRed3Gamma}, which dropped out in~\eqref{eq:evExGlobalOneLoop} while it remained implicit in~\eqref{eq:eq:evExLocalOneLoop}. To account for this difference in our local $\rstar$-method, we need to add the very same term to~\eqref{eq:evExLocalResult}. The difference is systematically computed by inserting the finite piece that arises through the evanescent reduction in~\eqref{eq:ExampleOneLoopResAntiSym} into the one-loop graph. In our example, this finite term takes the form 
    \begin{equation}
    \ctop_{\Pbar} \! 
    \left\{
    \raisebox{-0.375\height}{\includegraphics[trim={0.15cm 0.15cm 0.15cm 0.15cm},clip,height=2.5em]{Figures/one_loop.pdf}}
    \right\}- \mathcal P \ctop_{\dbar} \! 
    \left\{
    \raisebox{-0.375\height}{\includegraphics[trim={0.15cm 0.15cm 0.15cm 0.15cm},clip,height=2.5em]{Figures/one_loop.pdf}}
    \right\} = -\frac{g^2 C}{32 \pi^2} Q\,,
    \end{equation}
which inserted into the one-loop counterterm graph yields
    \begin{align} \label{eq:exRStarShift}
     \delta_{R^\ast} = - \frac{g^4 C}{32 \pi^2} \kop_{\mathcal P} \mathcal P \! \left\{ \frac{1}{d} \big( \gamma_\mu \gamma_\nu \gamma_\rho P_\LL \otimes \gamma^\rho \gamma^\nu \gamma^\mu P_\LL \big) \int_k \, \frac{1}{(k^2-a)^2} \right\} = -\frac{g^4 C}{2\left(16\pi^2\right)^2 \epsilon} \, Q \,.
    \end{align}
This is exactly the difference in~\eqref{eq:evExDiff}, such that
\begin{align}
    \mathcal P  \ctop_{\Pbar}\{G\} = \kop_{\mathcal P} \mathcal P \ctop_{\dbar}\{G\} + \delta_{R^\ast} \, .
\end{align}

The $ \rstar $-method can be made fully local in the $\dbar$-scheme, which makes it particularly suited for computer implementations to automatically subtract the subdivergences. The takeaway from the example explored here is that one can track and account for the difference from the $\Pbar$-scheme. This makes it possible to recover the $\Pbar$-scheme result, the one needed to extract physical RGEs.

\subsection[Applying the \texorpdfstring{$ \boldsymbol{R}^{\ast} $}{R*}-operation in different schemes]{Applying the \texorpdfstring{\mathversion{boldsans}$ R^{\ast} $}{R*}-operation in different schemes}

The complications in the previous example arise from a mismatch between the operator basis produced by the local \rstarop and the counterterm basis of the chosen renormalization scheme. More generally, we now consider the situation where the \rstarop is applied in a generic $\mathcal{T}$-scheme different from our preferred $\mathcal{S}$-scheme. Working off shell, we can safely assume that the two operator spaces coincide, i.e.,
\begin{equation}
    O_{\mathcal{S}}^i = B_{\mathcal{ST}}^{ij}(\epsilon) \, O_{\mathcal{T}}^{j},
\end{equation}
where $B_{\mathcal{ST}}(\epsilon)$ is an invertible $\epsilon$-dependent transformation matrix satisfying $ B_{\mathcal{ST}} = B_{\mathcal{TS}}^{\eminus 1} $. Using this matrix, we can express the action in the $\mathcal{S}$-scheme in terms of the $\mathcal{T}$-scheme basis:
\begin{equation}
    S_{\mathcal{S}}(\lambda_\mathcal{S})  = 
    S_{\mathcal{T}}(\{\lambda_{\mathcal{S}, j} B_{\mathcal{ST}}^{ji} \} ),
\end{equation}
where $\lambda_\mathcal{S}=\{\lambda_{\mathcal{S}, i}\}$ is the set of couplings. Accordingly, the unrenormalized effective action (that is, without counterterms) satisfies
\begin{equation} \label{eq:eff_action_ST}
    \Gamma_{\mathcal{S}}(\lambda_\mathcal{S})  = 
    \Gamma_{\mathcal{T}}(\{\lambda_{\mathcal{S}, j} B_{\mathcal{ST}}^{ji} \} ).
\end{equation} 
There is nothing deep in this statement; all we have done is to write the action in a different operator basis and then used the corresponding Feynman rules in all loop diagrams. Clearly, this cannot change the perturbative result.

To determine how the \rstarop differs in the two schemes, we introduce the $\mathcal{T}$-scheme counterterms for the effective action in the $\mathcal{S}$-scheme, $\hat{\delta}_\mathcal{TS} \lambda_i$, defined by
\begin{equation} \label{eq:TS-scheme_cts_def}
    \hat{\delta}_\mathcal{TS} \lambda_{i} (\lambda_{\mathcal{S}}) \,O_{\mathcal{T}}^i \equiv \ctop_{\mathcal{T}} \Gamma_{\mathcal{S}}(\lambda_{\mathcal{S}}) = \ctop_{\mathcal{T}} \Gamma_{\mathcal{T}}(\{\lambda_{\mathcal{S}, k} B_{\mathcal{ST}}^{k\ell} \}).
\end{equation} 
At one-loop order, no UV counterterms are inserted by $ \rstarbar_{\mathcal{T}} $ ($\rstarbar$ in the $\mathcal{T}$-scheme) according to~\eqref{eq:RstarS_Gamma_S_1}, and it trivially holds that 
\begin{equation}
    \rstarbar_{\mathcal{T}} \Gamma^{(1)}_{\mathcal{S}}(\lambda_{\mathcal{S}} ) = \rir \Gamma^{(1)}_{\mathcal{S}}(\lambda_{\mathcal{S}} ) = \rstarbar_{\mathcal{S}} \Gamma^{(1)}_{\mathcal{S}}(\lambda_{\mathcal{S}} ) .
\end{equation}
At two-loop order, however, a difference appears due to scheme dependence introduced by the insertion of one-loop UV counterterms to cancel subdivergences. Combining~\eqref{eq:RstarS_Gamma_S} and~\eqref{eq:eff_action_ST}, we obtain
\begin{equation} \label{eq:RstarT_Gamma_S}
\begin{split}
    \rstarbar_\mathcal{T} \Gamma^{(2)}_\mathcal{S}(\lambda_\mathcal{S}) &=  
    \hat{\delta}_\mathcal{TS}^{(1)} \lambda_{i} (\lambda_{\mathcal{S}}) \dfrac{\partial \rir \Gamma^{(1)}_{\mathcal{T}}}{\partial \lambda_{\mathcal{T},i}} (\{\lambda_{\mathcal{S}, k} B_{\mathcal{ST}}^{k\ell} \} ) 
    + \rir \Gamma^{(2)}_\mathcal{S}(\lambda_\mathcal{S}) \\
    &= \hat{\delta}_\mathcal{TS}^{(1)} \lambda_{i} (\lambda_{\mathcal{S}}) B_{\mathcal{TS}}^{ij} \dfrac{\partial \rir \Gamma^{(1)}_{\mathcal{S}} (\lambda_\mathcal{S}) }{\partial \lambda_{\mathcal{S},j}}
    + \rir \Gamma^{(2)}_\mathcal{S}(\lambda_\mathcal{S}),
\end{split}
\end{equation}
which makes it manifest that the scheme dependence arises from the insertion of one-loop counterterms into one-loop subgraphs, i.e., from products of one-loop contributions rather than genuine two-loop terms.

A direct comparison of the expressions for the actions of $ \rstarbar $ in the $\mathcal{S}$- and $\mathcal{T}$-schemes---\eqref{eq:RstarS_Gamma_S} and \eqref{eq:RstarT_Gamma_S}, respectively---yields
    \begin{equation} \label{eq:2loopRstarST}
    \rstarbar_\mathcal{S} \Gamma^{(2)}_\mathcal{S}(\lambda_\mathcal{S}) 
    = \rstarbar_\mathcal{T} \Gamma^{(2)}_\mathcal{S}(\lambda_\mathcal{S}) - \Big[ \hat{\delta}_\mathcal{TS}^{(1)} \lambda_{j} (\lambda_{\mathcal{S}}) B_{\mathcal{TS}}^{ji} - \hat{\delta}_\mathcal{S}^{(1)}\! \lambda_{i}(\lambda_{\mathcal{S}}) \Big]\, \dfrac{\partial \rir \Gamma^{(1)}_{\mathcal{S}} (\lambda_\mathcal{S})}{\partial \lambda_{\mathcal{S},i}}.
    \end{equation}
The shift from one action to the other involves only one-loop quantities, making its evaluation considerably less demanding than a full two-loop computation. The result~\eqref{eq:2loopRstarST} generalizes the relation between $\rstar$-operations in different schemes, which to our knowledge was first put forward in~\cite{Naterop:2024cfx}. It provides the formal justification for employing a local version of $ \rstarbar $, effectively $ \rstarbar_{\dbar} $, to compute $\Pbar$-scheme counterterms at the price of introducing a suitable compensating term.

\subsection{Calculating counterterms for the physical operators}

In this section, we particularize the general framework of the previous section to determine the two-loop counterterms for the physical operators in the $\Pbar$-scheme. As we rely on a local version of $ \rstarbar $ for the counterterm extraction, we need to apply a scheme change from $\dbar$ to $\Pbar$. Importantly, since both schemes are of the minimal-subtraction variety, all counterterms can be extracted by application of the $ \kop_{\mathcal{S}} $-operator, as described in Section~\ref{sec:rstar_method}. The mapping between the operator bases $ \mathscr{B}_{d} $ and $ \mathscr{B}_{\mathcal{P}} $ is fixed by the choice of physical projection $ \mathcal{P} $. To conform with the more compact notation of the previous section, we write the renormalized couplings and basis transformation collectively as
\begin{equation}
    \lambda_{\Pbar,i} = g_{\Pbar,a} \oplus \eta_{\Pbar,\alpha}, 
    \qquad 
    B_{d \mathcal P}^{ij}(\epsilon) = \mathcal{P}_{ia}(\epsilon) \oplus \mathcal{E}_{i\alpha},
\end{equation}
with the collective index decomposing as $ \{i\} = \{a\} \cup \{\alpha\} $. 

The $ \dbar $-counterterms for the effective action in the $\mathscr{B}_\mathcal{P}$ operator basis, $\Gamma_\mathcal{P}$, take the form
\begin{equation} \label{eq:1-loop_ct_d_basis}
    \delta^{(1)}_{\dbar \Pbar} \lambda_i(\lambda_{\Pbar})\, O_d^i 
    \equiv \ctop_{\dbar} \Gamma^{(1)}_{\mathcal P}(\lambda_{\Pbar})
    = - \kop_{d} \rir \boldsymbol{T} \, \Gamma^{(1)}_{\mathcal P}(\lambda_{\Pbar})
\end{equation}
at one-loop order following from definition~\eqref{eq:TS-scheme_cts_def} and the \msbar specific~\eqref{eq:S_scheme_ct_action}. Here, we simply perform exact $ d $-dimensional simplifications of the operators without imposing any evanescent prescription.
In practice, the quantity~\eqref{eq:1-loop_ct_d_basis} is straightforward to compute; indeed, it appears as an intermediate step when deriving the one-loop counterterms in the $ \Pbar $-scheme. To obtain the counterterms in the $ \Pbar $-scheme, note that
    \begin{equation}
    \boldsymbol{K}_\mathcal{P} = \boldsymbol{K}_\mathcal{P} \boldsymbol{K}_d  + \boldsymbol{K}_\mathcal{P} (\mathds{1}- \boldsymbol{K}_d) = \boldsymbol{K}_\mathcal{P} \boldsymbol{K}_d .
    \end{equation}
The second term vanishes since $ (\mathds{1}- \boldsymbol{K}_d) $ projects onto the finite coefficients of the $ O^i_d $ operators, while $ B_{d \mathcal P}(\epsilon) $ is analytic at $ \epsilon = 0 $. Applying this identity to the counterterms in~\eqref{eq:1-loop_ct_d_basis} yields the $ \Pbar $-scheme counterterms:
\begin{equation} \label{eq:1-loop_ct_phys_basis}
    \delta^{(1)}_{\Pbar} \! \lambda_i(\lambda_{\Pbar})\, O_\mathcal{P}^i
    = -\kop_\mathcal{P} \rir \boldsymbol{T} \, \Gamma^{(1)}_{\mathcal P}(\lambda_{\Pbar}) 
    = \delta^{(1)}_{\dbar \Pbar} \lambda_i(\lambda_{\Pbar})\, B_{d \mathcal P}^{ij}(0)\, O_{\mathcal{P}}^j ,
\end{equation}
where, in the last step, we have used that the one-loop counterterm $ \delta^{(1)}_{\dbar \Pbar} \lambda_i $ is proportional to a simple $ \epsilon $-pole. The fact that the basis transformation appears as $ B_{d \mathcal P}^{ij}(0) $, stripped of its full $ \epsilon $-dependence, accounts for the mismatch 
    \begin{equation}
   \delta^{(1)}_{\Pbar} \!\lambda_i(\lambda_{\Pbar})\, O_\mathcal{P}^i 
   \;\neq\; \delta^{(1)}_{\dbar \Pbar}\lambda_i(\lambda_{\Pbar})\, O_d^i
   \quad \text{at order } \ord{\epsilon^0},
   \end{equation}
as observed in the example in Section~\ref{sec:example}. 

At two-loop order, the $ \dbar $-counterterms for the effective action in the $\mathscr{B}_\mathcal{P}$ basis read
    \begin{equation} \label{eq:delta2_dP}
    \delta^{(2)}_{\dbar \Pbar} \lambda_i(\lambda_{\Pbar})\, O_d^i \equiv \ctop_{\dbar} \Gamma^{(2)}_{\mathcal P}(\lambda_{\Pbar}) = - \kop_{d} \rstarbar_{\dbar}\boldsymbol{T} \, \Gamma^{(2)}_{\mathcal P}(\lambda_{\Pbar}).
    \end{equation}
This is sufficient genuine two-loop information to obtain the two-loop counterterms in the $ \Pbar $-scheme. Invoking~\eqref{eq:2loopRstarST} to change the scheme of the \rstarop, we get
    \begin{align} \label{eq:del2_Pbar_1}
    \delta^{(2)}_{\Pbar}\! \lambda_{i}&(\lambda_{\Pbar}) \,O^i_\mathcal{P} 
    = - \kop_{\mathcal{P}} \rstarbar_{\Pbar} \boldsymbol{T} \, \Gamma^{(2)}_{\mathcal P}(\lambda_{\Pbar}) \\
    &= - \kop_{\mathcal{P}} \rstarbar_{\dbar} \boldsymbol{T} \, \Gamma^{(2)}_{\mathcal P}(\lambda_{\Pbar}) + \kop_{\mathcal{P}} \!\left( \big[ \delta^{(1)}_{\dbar \Pbar} \lambda_{j}(\lambda_{\Pbar}) B_{d \mathcal{P}}^{ji}(\epsilon) - \delta^{(1)}_{\Pbar} \lambda_{\Pbar,i}(\lambda_{\Pbar}) \big] \dfrac{\partial \rir \boldsymbol{T} \, \Gamma^{(1)}_{\mathcal P} (\lambda_{\Pbar})}{\partial \lambda_{\Pbar, i} } \right). \nonumber
    \end{align} 
The second term is simplified using~\eqref{eq:1-loop_ct_phys_basis}, from which we have that
    \begin{equation}
    \begin{split}
    \delta^{(1)}_{\dbar \Pbar} \lambda_{j} (\lambda_{\Pbar}) B_{d \mathcal{P}}^{ji}(\epsilon) - \delta^{(1)}_{\Pbar} \lambda_{\Pbar,i}(\lambda_{\Pbar}) &= 
    \delta^{(1)}_{\dbar \Pbar} \lambda_{j}(\lambda_{\Pbar}) \big[ B_{d \mathcal{P}}^{ji}(\epsilon) - B_{d \mathcal{P}}^{ji}(0) \big] \\
    &= \Big( \delta^{(1)}_{\dbar \Pbar, 1} \lambda_{j}(\lambda_{\Pbar}) \mathcal{P}_{ja,1} \oplus 0 \Big) +\ord{\epsilon}.
    \end{split}
    \end{equation}
Only the scheme transformation involving physical operators contains $ \ord{\epsilon} $ pieces, and they produce finite contributions when selecting the simple $\epsilon$-pole of the one-loop counterterm, denoted by $ \delta^{(1)}_{\dbar \Pbar, 1} \lambda_i $. Substituting this expression back into~\eqref{eq:del2_Pbar_1}, we find 
    \begin{equation} \label{eq:RStarShift}
    \begin{split}
    \delta^{(2)}_{\Pbar}\! \lambda_{i}(\lambda_{\Pbar}) \,O^i_\mathcal{P} 
    &= - \kop_{\mathcal{P}} \rstarbar_{\dbar} \boldsymbol{T} \, \Gamma^{(2)}_{\mathcal P}(\lambda_{\Pbar}) +   \delta^{(1)}_{\dbar \Pbar, 1} \lambda_{i}(\lambda_{\Pbar})\,  \mathcal{P}_{ia,1} \dfrac{\partial\kop_{\mathcal{P}} \rir \boldsymbol{T} \, \Gamma^{(1)}_{\mathcal P}}{\partial g_{\Pbar, a} } (\lambda_{\Pbar}) \\
    &= \kop_{\mathcal{P}} \! \big[ \delta^{(2)}_{\dbar \Pbar} \lambda_{i}(\lambda_{\Pbar}) \,O^i_d \big] - \delta^{(1)}_{\dbar \Pbar, 1} \lambda_{i}(\lambda_{\Pbar})\, \mathcal{P}_{ia,1} \dfrac{\partial\delta_{\Pbar}^{(1)}\! \lambda_{j}}{\partial g_{\Pbar, a} } (\lambda_{\Pbar}) \, O_{\mathcal{P}}^{j},
    \end{split}
    \end{equation} 
for the two-loop $ \Pbar $-scheme counterterms. It follows that these can be determined with the local $\rstarbar$-operation ($\rstar_d$), as long as we compensate the result with a one-loop counterterm inserted into other one-loop counterterms. As mentioned, both the $ \delta^{(1)}_{\dbar \Pbar} \lambda_i $ and the $ \delta^{(1)}_{\Pbar} \lambda_i $ counterterms are already determined in the course of calculating the one-loop counterterms, so all necessary ingredients are readily available.

What we have established in~\eqref{eq:RStarShift} [and~\eqref{eq:2loopRstarST}] is a method to exploit the \rstarop for subtracting local subdivergences in different renormalization schemes. This is reminiscent of the calculation of two-loop RGEs, where a one-loop computation suffices to shift the RG functions between schemes. We expect that these techniques can be extended beyond two-loop order, enabling a local $ \rstar $-method for counterterm calculations in theories with evanescent operators. 

A final note on our implementation of the \rstarop relates to our reading-point prescription for $ \gamma_5 $. In short, one has to ensure that the reading point for graphs with $\rstar$-subtracted subdivergences matches what one would obtain if manually calculating one-loop diagrams with one-loop counterterm insertions.\footnote{Alternatively, one could explore the feasibility of defining a $ \gamma_5 $-scheme where counterterms are only ever inserted by the $ \rstar $-method and where the reading point is fully determined by the 2-loop diagram. Note that this entails some reading points in the middle of the divergent subgraphs, preventing simplification of the associated Dirac algebra. In such cases the counterterm of the subgraph generated by $ \rstar $ would not correspond to any Feynman rule from any operators.} We adjust the reading point of contributions from counterterm insertions subtracting divergent subgraphs so they are read as though the corresponding $\dbar$-scheme counterterm had been inserted. With these general considerations in place, we now turn to the SMEFT calculation.

\section{Calculation and Results} 
\label{sec:results}

As with most higher-loop calculations, our determination of the two-loop SMEFT \befs relies on a bespoke computer implementation that chains the individual steps of our method together.

\begin{figure}
    \centering
    \def\vdistance{-3}
    \def\hdistance{4}
    \begin{tikzpicture}[
        ellipsenode/.style={ellipse, draw=black, fill=blue!15, thick, inner sep=0.5em},
        squarenode/.style={rectangle, draw=black, fill=blue!15, thick, minimum size=0.9cm, inner sep=0.5em, rounded corners= 5pt} 
        ]
        
        \node [ellipsenode] (lag) at (0,0) {Input Lagrangian};
        
        \node [squarenode] (Xterms) at (-\hdistance,\vdistance) {$X_{i_1i_2}$};
        \node [squarenode] (CDterms) at (\hdistance,\vdistance) {$V_{i_1i_2i_3},V_{i_1i_2i_3i_4}$};
        
        \node [squarenode, align=center] (oneLoopCT) at (-\hdistance,{2*\vdistance}) {One-loop counterterms \\ in $\mathscr{B}_d$ basis: $\delta_{\dbar\Pbar}^{(1)} \lambda_i$};
        \node [squarenode, align=center] (twoLoopCT) at (\hdistance,{2*\vdistance}) {Two-loop counterterms \\ in $\mathscr{B}_d$ basis: $\delta_{\dbar\Pbar}^{(2)} \lambda_i$};

        \node [squarenode, align=center] (oneLoopPhysical) at ({-1.5*\hdistance},{3*\vdistance}) {Physical \\ counterterms: \\[0.1cm] $\mathcal{P} \big(\delta_{\dbar\Pbar}^{(1)} \lambda_i \mathcal{O}^i\big)$};
        \node [squarenode, align=center] (oneLoopEvanescent) at ({-.7*\hdistance},{3*\vdistance}) {Evanescent \\ counterterms: \\[0.1cm] $\mathcal{E} \big(\delta_{\dbar\Pbar}^{(1)} \lambda_i \mathcal{O}^i\big)$};
        \node [squarenode, align=center] (twoLoopPhysical) at ({0.7*\hdistance},{3*\vdistance}) {Physical \\ counterterms: \\[0.1cm] $\mathcal{P} \big(\delta_{\dbar\Pbar}^{(2)} \lambda_i\mathcal{O}^i\big)$};
        \node [squarenode, align=center] (twoLoopEvanescent) at ({1.5*\hdistance},{3*\vdistance}) {Evanescent \\ counterterms: \\[0.1cm] $\mathcal{E} \big(\delta_{\dbar\Pbar}^{(2)} \lambda_i\mathcal{O}^i\big)$};
        
        \node [] (RStarShift) at ({0*\hdistance},{3.*\vdistance}) {$\rstar$-shifts};
        
        \node [squarenode, align=center] (offShellCT) at (1.1*\hdistance,{4*\vdistance}) {Physical counterterm Lagrangian \\ in an off-shell $\mathscr{B}_{\mathcal{P}}$ basis: \\ $ \delta_{\Pbar} g_a^{(2)} Q^a$};
        \node[align=center, text width =4.5cm] (oneLoopShift) at (-.7*\hdistance,3.7*\vdistance) {Inserting evanescent operators into one-loop graphs};

        \node [squarenode, align=center] (evShift) at ({-.7*\hdistance},{4.3*\vdistance}) {Evanescent shifts to the \\ $\Pf$-scheme \befs};
        
        \coordinate (oneLoopPoint) at (-1.5*\hdistance, 4.5*\vdistance);

        \node [] (goingOnShell) at (.9*\hdistance,{5.5*\vdistance}) {Field redefinitions to go on-shell};
        \node [] (goingOnShell1) at (-.9*\hdistance,{5.5*\vdistance}) {Field redefinitions to go on-shell};

        \node[ellipsenode, align=center, text width =5.3cm] (befs1) at (-.9*\hdistance, {6.3*\vdistance}) {Finitely-compensated, physical, \\ on-shell \befs: $\beta^{(1)}_{\Pbar_\mathrm{f}}$};
        \node[ellipsenode, align=center, text width =5.3cm] (befs) at (.9*\hdistance, {6.3*\vdistance}) {Finitely-compensated, physical, \\ on-shell \befs: $\beta^{(2)}_{\Pbar_\mathrm{f}}$};
        
        
        \draw [->] (lag.south) to [out=270,in=90](Xterms.north);
        \draw [->] (lag.south) to [out=270,in=90] (CDterms.north);
        
        \draw [->] (Xterms.south) to (oneLoopCT.north);
        \draw [->] (CDterms.south) to (twoLoopCT.north);
        \draw [->] (Xterms.east) to [out=0,in=180] (twoLoopCT.west);
        
        \draw [->] (oneLoopCT.south) to [out=270,in=90] (oneLoopPhysical.north);
        \draw [->] (oneLoopCT.south) to [out=270,in=90] (oneLoopEvanescent.north);
        \draw [->] (twoLoopCT.south) to [out=270,in=90] (twoLoopPhysical.north);
        \draw [->] (twoLoopCT.south) to [out=270,in=90] (twoLoopEvanescent.north);

        \draw [-] (oneLoopCT.east) to [out=0, in=90] (RStarShift.north);

        \draw [->] (twoLoopPhysical) to [out=270,in=90] (offShellCT.north);
        \draw [] (oneLoopEvanescent) to (oneLoopShift);

        \draw [->] (RStarShift.south) to [out=270,in=180] (offShellCT.west);

        \draw [] (offShellCT) to[out=270,in=90] (goingOnShell);
        \draw [->] (oneLoopShift) to[out=270,in=90] (evShift);
        
        \draw [] (oneLoopPhysical) to [out=270,in=90] (oneLoopPoint);
        \draw [] (oneLoopPoint) to [out=270,in=90] (goingOnShell1);
        
        \draw [] (evShift.south) to [out=270,in=90] (goingOnShell);

        \draw [->] (goingOnShell) to [out=270,in=90] (befs);
        \draw [->] (goingOnShell1) to [out=270,in=90] (befs1);
    \end{tikzpicture}\\[1em]
    \caption{Calculation workflow to determine one- and two-loop \befs in EFTs.}
    \label{fig:workflow}
\end{figure}
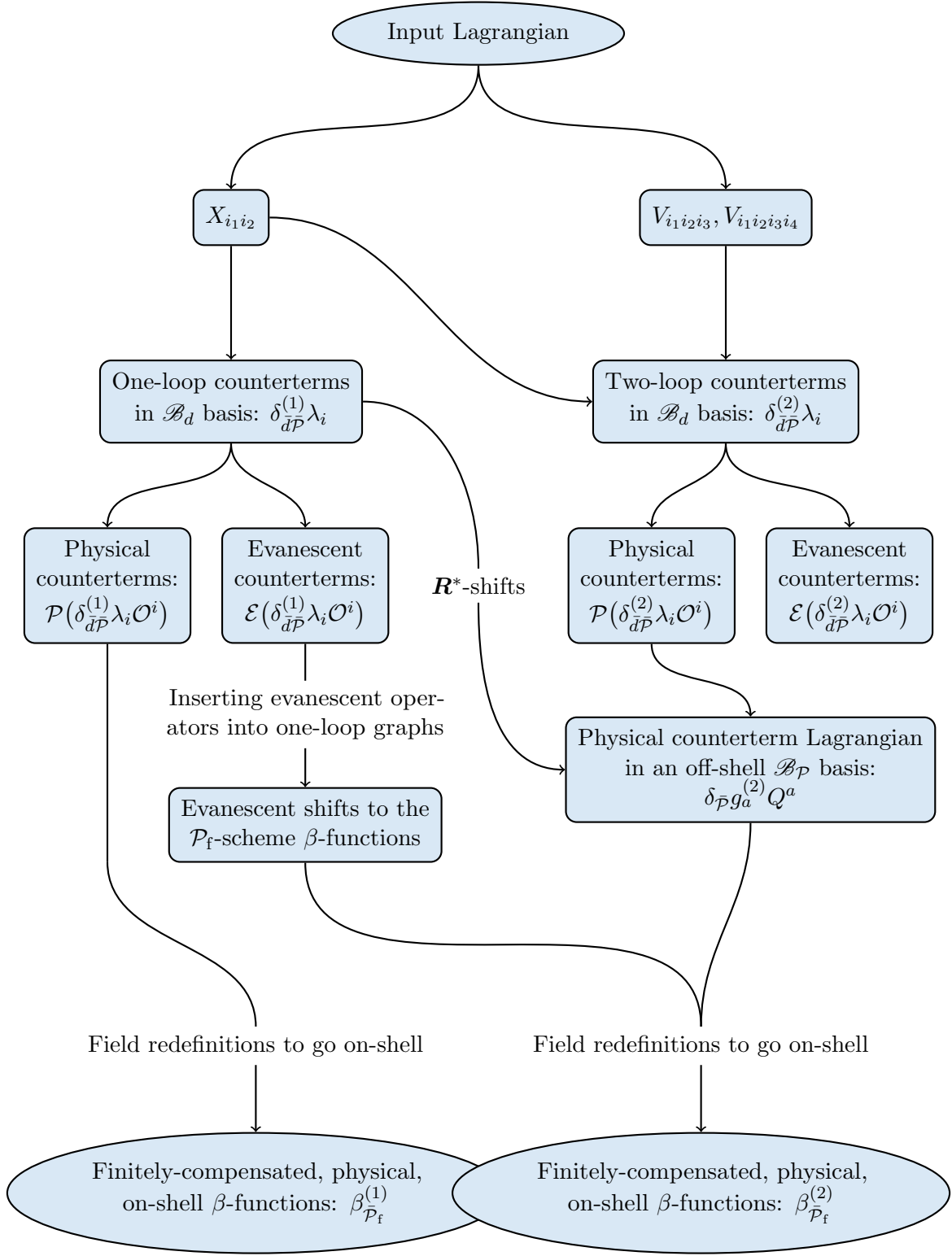

\subsection{Computer implementation}

We have continued the development of the implementation~\cite{Born:2024mgz} of the general functional methods~\cite{Fuentes-Martin:2024agf} in a code built on the \texttt{Matchete} Mathematica package~\cite{Fuentes-Martin:2022jrf} without any other external dependencies. The code employs the local $\rstar$-method in the $ \dbar $-scheme to calculate the one- and two-loop counterterms directly from the one- and two-loop functional supergraphs. We then rely on the methods new to \texttt{Matchete v0.3} to perform simplifications in the presence of evanescent operators and for calculating the one-loop finite shifts in~\eqref{eq:RStarShift}, allowing for the extraction of the two-loop counterterms in the $\Pbar$ scheme. We refer the reader to~\cite{Born:2024mgz} for the details and considerations behind the computer implementation for bosonic theories and detail the changes needed to incorporate fermions and evanescent operators here. The primary additions to the code are
\begin{enumerate}[i)]
    \item the inclusion of Dirac algebra and spinor indices to the generalized loop integrals, which are passed to the $ \rstar $ counterterm method [to calculate $ \delta^{(1,2)}_{\dbar \Pbar}\lambda_i $ of~\eqref{eq:delta2_dP}];
    \item a method to calculate the shift~\eqref{eq:RStarShift} needed to account for the discrepancy between $ \rstarbar_{\Pbar} $ and $ \rstarbar_{\dbar} $ and allow for the determination of $ \delta^{(2)}_{\Pbar} g_a $;
    \item the one-loop calculation of evanescent shifts and evanescent counterterms in~\eqref{eq:two_loop_beta} to determine $ \beta_{\Pf}^{(2)} $, the physical two-loop \befs in the finitely-compensated evanescent scheme.
\end{enumerate}
The last two points exclusively involve the evaluation of one-loop topologies and are less computationally demanding.

In Figure~\ref{fig:workflow}, we depict the workflow of the computation. From the input Lagrangian $\mathcal{L}$, we derive the relevant 2-, 3-, and 4-point vertex operators ($X_{i_1i_2}$, $V_{i_1i_2i_3}$, and $V_{i_1i_2i_3i_4}$ in the notation of~\cite{Fuentes-Martin:2024agf}) through the application of functional derivatives. Next, the contributions from sunset and figure-8 topologies are computed from these vertices using the functional master formula~\cite{Fuentes-Martin:2024agf}. Independently, the one-loop counterterms are determined and used to construct the $\rstar$-scheme shift~\eqref{eq:RStarShift}. Adding everything up, we obtain an off-shell counterterm Lagrangian in the $ \dbar $-scheme, that is, in the $d$-dimensional basis $ \mathscr{B}_d$, containing operators that are redundant in $d=4$ dimensions. In the next step, we call \texttt{Matchete}'s internal simplification routines to reduce the operators to the $ \mathscr{B}_\mathcal{P} $ basis and extract the $ \Pbar $-scheme counterterms. To this end, we employ the evanescent prescription outlined in Section~\ref{sec:eva_precsription} and Appendix~\ref{app:eva_ops}. The one-loop evanescent counterterms are then used to compute the evanescent shift~\eqref{eq:two_loop_beta} to the two-loop $\Pf$-scheme \befs. Lastly, using field redefinitions, we go to an on-shell basis and extract the on-shell $\Pf$-scheme \befs for the physical couplings.

\subsection{Validation of methods and implementation}

The predecessor code~\cite{Born:2024mgz} for computing the \befs of the bosonic SMEFT sector was already the subject of several crosschecks. Extending the framework to include fermions introduces many additional subtleties and potential complications, making further validation essential. Thanks to its model-generic design, our code can easily be applied to compute RGEs in models other than SMEFT, although direct comparisons may be difficult where different renormalization schemes are adopted. We have confirmed agreement with the following results in the literature:
\begin{itemize}
    \item all two-loop bosonic crosschecks used in~\cite{Born:2024mgz};
    \item the full two-loop SM in the unbroken phase (extracted with \texttt{RGBeta}~\cite{Thomsen:2021ncy}). Curiously, some $\gamma_5$ reading-point prescriptions can give rise to spurious running of the topological $\theta$-terms, but these vanish by the reality condition (cf. Section~\ref{sec:PrescriptionsAndTricks}); 
    \item the two-loop running of the Weinberg operator in the dimension-five SMEFT~\cite{Ibarra:2024tpt};
    \item partial crosscheck with the 2-loop LEFT \befs~\cite{Aebischer:2025hsx}. We explicitly verified against the four-fermion operators in the sectors
    \begin{enumerate}
        \item five-flavor LEFT $\Delta F = 2$ $u$-type,
        \item five-flavor LEFT $\Delta F = 2$ $d$-type,
        \item five-flavor LEFT $\Delta F = 2$ $e$-type,
        \item three-flavor LEFT $\Delta F = 1$ $ed$-type, 
        \item three-flavor LEFT $\Delta F = 1$ $\nu edu$-type,
        \item three-flavor LEFT $\Delta F^{\bar{f}f} = 1$ $\nu$-type.
    \end{enumerate}
\end{itemize}
Of these, the LEFT crosscheck was especially useful, as it allowed us to validate the consistency of the $ \gamma_5 $ reading-point prescription for the diagrams classes in Figure~\ref{subfig:vector4Fermi1} and~\ref{subfig:vector4Fermi2}.
Being the only two-loop calculation we found with a compatible evanescent prescription, Ref.~\cite{Aebischer:2025hsx} provides an essential crosscheck of our implementation of evanescent operators and the shift~\eqref{eq:RStarShift} needed to recover the correct $ \Pbar$-scheme counterterms. 

As a supplemental avenue to validation, we have been able to conduct limited consistency checks on the SMEFT result itself. We confirmed the zeros~\cite{Bern:2020ikv} of the two-loop SMEFT anomalous dimension matrix (ADM) that were predicted using on-shell methods.\footnote{We note that a few of the entries vanish only in the absence of both Yukawa \emph{and} hypercharge couplings.} 
In addition, we have checked the `t Hooft consistency conditions~\cite{tHooft:1973mfk}, namely that the double poles of the two-loop SMEFT counterterms are consistent with the one-loop counterterms.

\subsubsection{Consistency of the counterterm poles}

The `t Hooft consistency condition is usually phrased as the requirement that the \befs are finite, i.e., they do not exhibit any poles in the $ \epsilon $-expansion. However, with the introduction of flavor groups to a theory---any theory where multiple fields share quantum numbers, such as the three fermion generations of the SMEFT---the RG-functions can in fact develop unphysical poles, which generate a flow in the direction of a flavor rotation~\cite{Herren:2021yur}. Various recent EFT calculations have documented additional spurious divergences of the on-shell RG functions~\cite{Manohar:2024xbh,Zhang:2025ywe,Naterop:2025cwg}, posing further questions regarding the validity of the usual version of the consistency conditions. Ref.~\cite{Thomsen:2025kka} shows that these divergences are a consequence of omitting new source terms when shifting the path integral from an off-shell to an on-shell EFT basis: the upshot is that the \befs should depart from finiteness only through poles generating unphysical flavor rotations. The consistency check is passed if we can find a divergent flavor rotation (as in a field redefinition) to render all \befs finite.

The simple $\epsilon$-poles of the two-loop \befs are given by
    \begin{equation} \label{eq:tHooft_consistency}
    \beta_{1,a}^{(2)} \equiv  4 \, \delta^{(2)}_2 \!g_{a} - 2 \,\delta^{(1)}_1\! g_{b} \, \partial^b \delta^{(1)}_1 \!g_{a}.
    \end{equation}
As is common practice, our SMEFT calculation uses Hermitian (square roots of the) wave function renormalization factors. With this choice, we find that $ \beta_{1,a}^{(2)} $ indeed vanishes for all the Wilson coefficients and most of the renormalizable couplings, as one would expect. The \befs of the Yukawa couplings, on the other hand, exhibit non-trivial poles: 
\begin{align} \label{eq:div_ye}
    \big(\beta^{(2)}_{1,y_e}\big)^{pr} = \frac{\mu_H^2}{(16\pi^2)^2} \bigg\{&  3 \Big[ 10i \, g_Y^4 \, C_{H\widetilde{B}} + 6i \, g_L^4 \, C_{H\widetilde{W}} + \Tr \! \big(  Y_e^{\vdagph}  \, C_{eH}^\dagger - Y_e^\dagger \, C_{eH}^{\vdagph}  \big) \nonumber \\
    &+ 3 \Tr\! \big( Y_d^{\vdagph}  \, C_{dH}^\dagger - Y_d^\dagger \, C_{dH}^{\vdagph}  \big) - 3 \Tr\! \big( Y_u^{\vdagph}  \, C_{uH}^\dagger - Y_u^\dagger \, C_{uH}^{\vdagph}  \big) \Big] \, Y_e^{\vdagph pr} \nonumber \\
    &- 3 \Big[ (Y_e^{\vdagph}  \, Y_e^\dagger \, C_{eH}^{\vdagph} )^{pr} + \frac{1}{2}(C_{eH}^{\vdagph}  \, Y_e^\dagger \,  Y_e^{\vdagph} )^{pr} - \frac{3}{2}( Y_e^{\vdagph}  \, C_{eH}^\dagger \,  Y_e^{\vdagph} )^{pr} \Big] \nonumber \\
    &+ 3 \, g_Y^2 \, \Big[ ( Y_e^{\vdagph}  \, Y_e^\dagger \, C_{eB}^{\vdagph} )^{pr} - \frac{1}{2}(C_{eB}^{\vdagph}  \, Y_e^\dagger \,  Y_e^{\vdagph} )^{pr} - \frac{1}{2}( Y_e^{\vdagph}  \, C_{eB}^\dagger \,  Y_e^{\vdagph} )^{pr}  \Big] \nonumber \\
    &+ \frac{9}{2} \, g_L^2 \, \Big[ ( Y_e^{\vdagph}  \, Y_e^\dagger \, C_{eW}^{\vdagph} )^{pr} - \frac{1}{2}(C_{eW}^{\vdagph}  \, Y_e^\dagger \,  Y_e^{\vdagph} )^{pr} - \frac{1}{2}( Y_e^{\vdagph}  \, C_{eW}^\dagger \,  Y_e^{\vdagph} )^{pr} \Big] \bigg\} \, , \\
    \label{eq:div_yu}
    \big(\beta^{(2)}_{1,y_u}\big)^{pr} = \frac{\mu_H^2}{(16\pi^2)^2} \bigg\{& \frac{1}{3}\Big[ 34i \, g_Y^4 \, C_{H\widetilde{B}} + 54i \, g_L^4 \, C_{H\widetilde{W}} + 192i \, g_s^4 \, C_{H\widetilde{G}} - 9\Tr\!\big(  Y_e^{\vdagph}  \, C_{eH}^\dagger - Y_e^\dagger \, C_{eH}^{\vdagph}  \big) \nonumber \\
    &- 27 \Tr\!\big( Y_d^{\vdagph}  \, C_{dH}^\dagger - Y_d^\dagger \, C_{dH}^{\vdagph}  \big) + 27 \Tr\!\big( Y_u^{\vdagph}  \, C_{uH}^\dagger - Y_u^\dagger \, C_{uH}^{\vdagph}  \big) \Big] \, Y_d^{\vdagph pr} \nonumber \\
    &- 3 \Big[ (Y_u^{\vdagph}  \, Y_u^\dagger \, C_{uH}^{\vdagph} )^{pr} + \frac{1}{2}(C_{uH}^{\vdagph}  \, Y_u^\dagger \,  Y_u^{\vdagph} )^{pr} - \frac{3}{2}( Y_u^{\vdagph}  \, C_{uH}^\dagger \,  Y_u^{\vdagph} )^{pr} \Big] \nonumber \\
    &- 3 \, g_Y^2 \, \Big[ ( Y_u^{\vdagph}  \, Y_u^\dagger \, C_{uB}^{\vdagph} )^{pr} - \frac{1}{2}(C_{uB}^{\vdagph}  \, Y_u^\dagger \,  Y_u^{\vdagph} )^{pr} - \frac{1}{2}( Y_u^{\vdagph}  \, C_{uB}^\dagger \,  Y_u^{\vdagph} )^{pr}  \Big] \nonumber \\
    &+ \frac{9}{2} \, g_L^2 \, \Big[ ( Y_u^{\vdagph}  \, Y_u^\dagger \, C_{uW}^{\vdagph} )^{pr} - \frac{1}{2}(C_{uW}^{\vdagph}  \, Y_u^\dagger \,  Y_u^{\vdagph} )^{pr} - \frac{1}{2}( Y_u^{\vdagph}  \, C_{uW}^\dagger \,  Y_u^{\vdagph} )^{pr} \Big] \nonumber \\
    &- \frac{3}{2} \Big[(C_{dH}^{\vdagph}  \, Y_d^\dagger \,  Y_u^{\vdagph} )^{pr} - ( Y_d^{\vdagph}  \, C_{dH}^\dagger \,  Y_u^{\vdagph} )^{pr} \Big] \nonumber \\
    &+ \frac{3}{2} \, g_Y^2 \, \Big[(C_{dB}^{\vdagph}  \, Y_d^\dagger \,  Y_u^{\vdagph} )^{pr} - ( Y_d^{\vdagph}  \, C_{dB}^\dagger \,  Y_u^{\vdagph} )^{pr} \Big] \nonumber \\
    &- \frac{9}{4} \, g_L^2 \, \Big[(C_{dW}^{\vdagph}  \, Y_d^\dagger \,  Y_u^{\vdagph} )^{pr} - ( Y_d^{\vdagph}  \, C_{dW}^\dagger \,  Y_u^{\vdagph} )^{pr} \Big] \bigg\} \, , \\
    \label{eq:div_yd}
    \big(\beta^{(2)}_{1,y_d}\big)^{pr} = \frac{\mu_H^2}{(16\pi^2)^2} \bigg\{& \frac{1}{3}\Big[ 10i \, g_Y^4 \, C_{H\widetilde{B}} + 54i \, g_L^4 \, C_{H\widetilde{W}} + 192i \, g_s^4 \, C_{H\widetilde{G}} + 9\Tr\!\big(  Y_e^{\vdagph}  \, C_{eH}^\dagger - Y_e^\dagger \, C_{eH}^{\vdagph} \big) \nonumber \\
    &+ 27 \Tr\!\big( Y_d^{\vdagph}  \, C_{dH}^\dagger - Y_d^\dagger \, C_{dH}^{\vdagph}  \big) - 27 \Tr\!\big( Y_u^{\vdagph}  \, C_{uH}^\dagger - Y_u^\dagger \, C_{uH}^{\vdagph}  \big) \Big] \, Y_d^{\vdagph pr} \nonumber \\
    &- 3 \Big[ (Y_d^{\vdagph}  \, Y_d^\dagger \, C_{dH}^{\vdagph} )^{pr} + \frac{1}{2}(C_{dH}^{\vdagph}  \, Y_d^\dagger \,  Y_d^{\vdagph} )^{pr} - \frac{3}{2}( Y_d^{\vdagph}  \, C_{dH}^\dagger \,  Y_d^{\vdagph} )^{pr} \Big] \nonumber \\
    &+ 3 \, g_Y^2 \, \Big[ ( Y_d^{\vdagph}  \, Y_d^\dagger \, C_{dB}^{\vdagph} )^{pr} - \frac{1}{2}(C_{dB}^{\vdagph}  \, Y_d^\dagger \,  Y_d^{\vdagph} )^{pr} - \frac{1}{2}( Y_d^{\vdagph}  \, C_{dB}^\dagger \,  Y_d^{\vdagph} )^{pr}  \Big] \nonumber \\
    &+ \frac{9}{2} \, g_L^2 \, \Big[ ( Y_d^{\vdagph}  \, Y_d^\dagger \, C_{dW}^{\vdagph} )^{pr} - \frac{1}{2}(C_{dW}^{\vdagph}  \, Y_d^\dagger \,  Y_d^{\vdagph} )^{pr} - \frac{1}{2}( Y_d^{\vdagph}  \, C_{dW}^\dagger \,  Y_d^{\vdagph} )^{pr} \Big] \nonumber \\
    &- \frac{3}{2} \Big[(C_{uH}^{\vdagph}  \, Y_u^\dagger \,  Y_d^{\vdagph} )^{pr} - ( Y_u^{\vdagph}  \, C_{uH}^\dagger \,  Y_d^{\vdagph} )^{pr} \Big] \nonumber \\
    &+ \frac{3}{2} \, g_Y^2 \, \Big[(C_{uB}^{\vdagph}  \, Y_u^\dagger \,  Y_d^{\vdagph} )^{pr} - ( Y_u^{\vdagph}  \, C_{uB}^\dagger \,  Y_d^{\vdagph} )^{pr} \Big] \nonumber \\
    &- \frac{9}{4} \, g_L^2 \, \Big[(C_{uW}^{\vdagph}  \, Y_u^\dagger \,  Y_d^{\vdagph} )^{pr} - ( Y_u^{\vdagph}  \, C_{uW}^\dagger \,  Y_d^{\vdagph} )^{pr} \Big] \bigg\} \, .
\end{align} 
These poles have been organized so one may readily recognize imaginary scalars or anti-Hermitian matrix combinations acting on the relevant Yukawa couplings. The extra field redefinitions required to remove the poles~\eqref{eq:div_ye} to~\eqref{eq:div_yd} are collected in Appendix \ref{app:field_shifts}. They indeed take the form of divergent flavor rotations of the field and do not reintroduce any new divergences elsewhere. 

Unfortunately, this consistency check of the double poles is  not sensitive to the treatment of evanescent operators or the $ \gamma_5 $ reading-point ambiguity. Although both of these influence only the simple pole of the two-loop counterterms, the check still serves to validate the broad strokes of our calculation. To further validate our choice of $ \gamma_5 $ reading points, we have adopted a self-consistency check against the SMEFT with different four-fermion operators as previously advertised.

\subsubsection{Check of $ \gamma_5 $ reading points through basis change} 
\label{sec:open_basis}

The appearance of $ \gamma_5 $ ambiguities requires the presence of closed fermion loops. Whenever an EFT operator contains a spinor line with both a fermion and its conjugate, one risks that the insertion of such an operator in a loop-graph gives rise to an ambiguity (cf. Figures~\ref{fig:scalar4FermiDiags}--\ref{fig:DipoleDiags}). The situation is exacerbated in the SMEFT, where spinor lines with left-handed quarks or leptons with their conjugate right-handed counterparts can also be closed with the insertion of Yukawa couplings.
A clever strategy was devised in the LEFT~\cite{Buras:1992tc,Aebischer:2025hsx} to circumvent this issue for certain four-fermion operators: by Fierzing an operator, a fermion can be moved to a different spinor lines, preventing the closure of the original spinor line at the four-fermion operator. The \befs may be calculated unambiguously in the new, `open' basis and transformed into the old, `closed' basis afterwards. Crucially, this strategy relies on the evanescent contributions stemming from the basis change being unambiguous themselves. This is  fortuitously the case in the SMEFT, as long as we only Fierz between scalar and vectorial current four-fermion operators~\cite{Fuentes-Martin:2022vvu}. Although the Yukawa couplings often create new issues in the Fierzed operator, the strategy remains valid in many cases, especially for semileptonic operators, where their Fierzed versions possess spinor lines with quark-lepton pairs. 

As a cross-check, we selected representative semi-leptonic operators from the $ (\Bar{R}R)(\Bar{R}R) $, $ (\Bar{L}L)(\Bar{R}R) $, and $ (\Bar{L}R)(\Bar{R}L) $ classes,\footnote{The $ (\Bar{L} L)(\Bar{L} L)$ class is structurally similar to the $ (\Bar{R} R)(\Bar{R} R) $, except for a sign multiplying $ \gamma_5$.} which we Fierz transform to an `open' (spinor line) basis as follows
    \begin{equation} \label{eq:Fierz_basis_change}
    \begin{aligned}
    O_{ed}^{prst} &= (\Bar{e}^p \gamma^\mu e^r) (\Bar{d}^s \gamma_\mu d^t) &  
    \longrightarrow \; O_{deed}^{prst} &= (\Bar{d}^p \gamma_\mu e^r) (\Bar{e}^s \gamma^\mu d^t),\\
    O_{qe}^{prst} &= (\Bar{q}^p \gamma_\mu q^r)(\Bar{e}^s \gamma^\mu e^t) & 
    \longrightarrow\;  O_{qeeq}^{prst} &= (\Bar{q}^p e^r)(\Bar{e}^s q^t), \\
    O_{\ell edq}^{prst} &= (\Bar{\ell}^p e^r)(\Bar{d}^s q^t) & 
    \longrightarrow\; O_{\ell qde}^{prst} &= (\Bar{\ell}^p \gamma_\mu q^r)(\Bar{d}^s \gamma^\mu e^t).
    \end{aligned}
    \end{equation}
We further consider a pure-quark operator 
    \begin{equation} \label{eq:Fierz_basis_change_LRLR}
   O_{quqd}^{{\scriptscriptstyle(\times)} \, prst} = (\Bar{q}^p_{ai} u^{b r}) \varepsilon^{ij} (\Bar{q}^s_{bj} d^{a t}) \; \longrightarrow \; O_{qqdu}^{prst} = \big(\Bar{q}_{ai}^{p} q_{bj}^{\cc\, r} \big) \varepsilon^{ij} \big( \overline{d^\cc} \phantom{}^{as} u^{bt} \big).
    \end{equation}
to represent $ (\Bar{L}R)(\Bar{L}R) $ class. 
We have verified that the full two-loop SMEFT \bef calculation in the `closed' (Mainz) basis is reproduced by the calculation performed in the open basis. The comparison utilizes Eq.~\eqref{eq:betaChangeOfBasis} to map \befs between bases, accounting for contributions from one-loop evanescent shifts. This calculation confirms that the $\gamma_5$ reading-point prescription for the closed four-fermion operators in sunset diagrams correctly reproduces the unambiguous results obtained in the open basis.
Although not all four-fermion operators are semi-leptonic like the representative examples in Eq.~\eqref{eq:Fierz_basis_change}, the Dirac structures of potentially ambiguous diagrams within the same class are identical. Thus, we conclude that all sunset diagrams in Figures~\ref{fig:vector4FermiDiags} and~\ref{fig:scalar4FermiDiags} are under control with our reading-point prescription.

A separate source of ambiguity arises from the open-basis operator $ O_{\ell qde} $ within the figure-8 topology. While the closed-basis operator $ O_{\ell edq} $ generates two closed traces in Figure~\ref{subfig:scalar4Fermi2} that are too short to be ambiguous, the insertion of $ O_{\ell qde} $ yields a single ambiguous trace that affects contributions to CP-odd $ X^2 H^2 $ operators. Given this ambiguity, we adopt the closed-basis result as definitive for our SMEFT results. The issue in the Figure~\ref{subfig:scalar4Fermi2} diagrams extends to $ O_{quqd}^{{\scriptscriptstyle(\times)}} $ and $ O_{qqdu} $ (and indeed the entire $ (\Bar{L}R)(\Bar{L}R) $ class), implying that no unambiguous result can be ascribed to this diagram. However, we are encouraged to find that the \befs nevertheless agree in both bases when applying our reading-point prescription.

\subsection{Results}

Our calculations were performed without parallelization (single-threaded) on a modern laptop equipped with an AMD Ryzen 7 Pro 7840U processor. To avoid memory overflow, the code was executed separately for each effective operator. This strategy is viable because, at EFT-order six, the dimension-six couplings enter the \befs linearly. Besides, the only quadratic contributions are due to two Weinberg operator insertions, as this is the only dimension-five SMEFT operator. For most operators, the runtime ranges from 15 minutes to one hour. Notable exceptions include $O_H$, which requires only about five minutes, and $O_{\widetilde{W}}$, which exceeds nine hours. In general, the triple gauge operators are by far the most time-consuming, owing to the large number of terms in the self-interaction vertex.\footnote{Interestingly, this differs from the computational demand with on-shell methods, where triple gauge vertices are simpler and operators with many external legs, such as $O_H$, tend to be more demanding.} The full calculation was completed within approximately 60 hours.

Due to their length, the full set of two-loop SMEFT \befs is not included in this article. Instead, the results are provided in three supplementary files, each offering a different format:
\begin{enumerate}[i)]
    \item \paragraph{\texttt{SMEFT\char`_Beta\char`_Functions.pdf}}  
    A \texttt{pdf} file containing all SMEFT \befs up to two-loop order in human-readable form. For clarity, the effective operators are highlighted in red and the contributions grouped by Wilson coefficient. This version is most convenient for quickly checking or cross-referencing specific terms by hand.
    \item \paragraph{\texttt{SMEFT\char`_Beta\char`_Functions.m}}  
    A \texttt{Mathematica} file with all SMEFT \befs up to two-loop order in \texttt{Wolfram} format (with \texttt{Matchete} conventions for couplings). Results are stored as an association of the form
    \begin{verbatim}
        <| C -> Beta_C, ... |>
    \end{verbatim}
    \vspace{-20pt}
    where \verb|C| denotes any SMEFT coupling and \verb|Beta_C| its corresponding \bef. This format is best suited for further code-based processing.  
    For Wilson coefficients with parallel and cross contractions, such as $C_{Hq}^{{\scriptscriptstyle (\|)} \, pr}$ and $C_{Hq}^{{\scriptscriptstyle (\times)} \, pr}$, we use the notations \verb|cHq1| and \verb|cHq2|, respectively. These conventions are also reflected in the association keys. For instance, the \bef of $C_{Hq}^{{\scriptscriptstyle (\|)} \, pr}$ appears under the key \verb|cHq1|. Additionally, the running of the squared gauged couplings, $ g_i^2 $, can be extracted with the keys \verb|gY2|, \verb|gL2| and \verb|gs2|.
    \item \paragraph{\texttt{SMEFT\char`_ADM.cdf}}  
    An interactive \texttt{Mathematica} file providing the individual entries of the dimension-six SMEFT ADM together with the complete \befs and ADM columns. Expressions are shown both in human-readable form and in \LaTeX~format, making it a versatile tool for studying the mixing of individual operators. We define the $ \ell $-loop ADM for the dimension-six operators from the \befs by
    \begin{equation}
    \gamma^{(\ell)}_{ab} = \dfrac{\partial \beta^{(\ell)}_{\Pf,C_a}}{\partial C_b},
    \end{equation}
    where $ C_a $ takes values in all dimension-six Wilson coefficients (and, where relevant, their conjugates). 
\end{enumerate}
In all files, we use the shorthand $\hbar \equiv 1/(16\pi^2)$ to distinguish between one- and two-loop contributions.

\section{Conclusion and Outlook} 
\label{sec:conclusion}

With the calculation of two-loop \befs for the baryon-number-conserving dimension-six SMEFT coefficients, we hope to facilitate an era of NLO SMEFT analyses. Now that these results are available, the community can start adopting them for phenomenological studies and even incorporate them in automated analysis tools. Our results are provided in supplementary files, covering multiple formats to expedite the use and exploration of the \befs. The baryon-number-violating operators at dimension six decouple completely under the RG and their \befs have been calculated separately in~\cite{Banik:2025wpi}.

The SMEFT inherits all the complexity of the full SM, with its chiral gauge groups and Higgs-Yukawa interactions. This introduces obstacles absent in the LEFT, where only the EFT operators are chiral. It is perhaps unsurprising then that our calculation stretches the applicability of an anticommuting $\gamma_5$---arguably beyond its limits in some instances (cf. Section~\ref{sec:gamma5}). To avoid unnecessary ambiguities related to the four-fermion tensor operator, we even found it opportune to introduce a new SMEFT basis. While the vast majority of our results are either unambiguous or validated by independent checks, a small subset of contributions---particularly from the dipoles---remains fixed only by an ad-hoc trace prescription (cf. Table~\ref{tab:gamma5_problems}). These could not be crosschecked by available results in the literature. Fortunately, their numerical impact is expected to be small or negligible in practice. Nonetheless, it would be desirable to eventually revisit this calculation in the BMHV scheme to ensure complete mathematical consistency.  

This work also represents the initial culmination of a year-long effort to develop functional methods for practical multi-loop EFT calculations~\cite{Fuentes-Martin:2023ljp,Born:2024mgz,Fuentes-Martin:2024agf}. While the SMEFT \befs could also be obtained using standard diagrammatic methods, the functional approach has proven especially powerful in managing the intricate bookkeeping of large operator structures that arise at intermediate stages of the calculation. We hope that this work demonstrates the maturity of the methods and encourages their use in other calculations.

\subsection*{Acknowledgments}

LB would like to thank Pol Morell for valuable crosschecks and constructive discussions. AET would like to thank Luca Naterop and Peter Stoffer for many helpful discussions. The work of LB and AET is funded by the Swiss National Science Foundation (SNSF) through the Ambizione grant ``Matching and Running: Improved Precision in the Hunt for New Physics,'' project number 209042. The work of JFM is supported by the grants EUR2024.153549, CNS2024-154834 and PID2022-139466NB-C21 (FEDER/UE) funded by the Spanish Research Agency (MICIU/AEI/10.13039/501100011033), and by the Junta de Andaluc\'ia grants P21\_00199 and FQM101.

\renewcommand{\thesection}{\Alph{section}}
\appendix 

\section{Relating the Mainz and Warsaw Bases} 
\label{app:basis_transformations}

In this appendix, we establish the relations between \befs in the Mainz basis, defined in Section~\ref{sec:SMEFTbasis}, and the Warsaw basis~\cite{Grzadkowski:2010es}. We provide a streamlined version of the basis change for simplicity; the interested reader can find additional details and a more general formulation in Appendix~\ref{app:eva_schemes}.
Throughout this appendix, we denote operators in our basis by $O_i$ and Warsaw operators by $Q_j$. The two sets are related by a transformation matrix $R$, defined as
    \begin{align}
    O_i &= R_{ij}(c)\, Q_j ,
    \end{align}
with indices $i,j$ running over the basis operators. Wilson coefficients in the Mainz and in the Warsaw basis are denoted by $C_i$ and $c_j$, respectively. They transform with the inverse matrix so that the Lagrangian remains invariant:
    \begin{align}\label{eq:WCRel}
    C_i &= c_j \, R^{\eminus 1}_{ji}.
    \end{align}

The transformation matrix $R$ is scale dependent, since it is a function of renormalizable couplings (cf. Section~\ref{sec:SMEFTbasis} and discussion below). The Warsaw basis \befs can thus be written as
    \begin{align}
    \frac{\dd c_j}{\dd t} = c_k \, R^{\eminus 1}_{ki}\, \frac{\dd R_{ij}}{\dd t} +  \beta_i(cR^{\eminus 1})\, R_{ij},
    \end{align}
where $t=\ln\mu$ and $\beta_i= \dd C_i / \!\dd t$ are the \befs in the Mainz basis. The argument $cR^{\eminus 1}$ indicates that the Wilson coefficients need to be converted into the ones in the Warsaw basis to express the r.h.s. as a function of the $c_i$ coefficients.\footnote{The two-loop \befs for the renormalizable couplings, needed to determine $\frac{\dd R}{\dd t}$, are provided in the ancillary material together with the \befs of the Wilson coefficients in the Mainz basis.} 

For convenience, we group the basis transformations into independent blocks:

\paragraph{Gauge-coupling rescaling} This class contains the operators that are rescaled by gauge couplings. It also includes the factor of $2$ in the $\SU(2)_L$ dipole operators, due to the use of $\SU(2)_L$ generators rather than Pauli matrices. The affected operators are
    \begin{equation}
    \begin{aligned}
    \boldsymbol{O}_\textrm{g-coup}&=
    \begin{pmatrix}
    O_{\ptilde G}&
    O_{\ptilde W}&
    O_{H\ptilde G}&
    O_{H\ptilde W}&
    O_{H\ptilde B}&
    O_{H\ptilde WB}&
    O_{\chi G}^{pr}&
    O_{\psi W}^{pr}&
    O_{\psi B}^{pr}
    \end{pmatrix},\\
    \boldsymbol{Q}_\textrm{g-coup}&=
    \begin{pmatrix}
    Q_{\ptilde G}&
    Q_{\ptilde W}&
    Q_{H\ptilde G}&
    Q_{H\ptilde W}&
    Q_{H\ptilde B}&
    Q_{H\ptilde WB}&
    Q_{\chi G}^{pr}&
    Q_{\psi W}^{pr}&
    Q_{\psi B}^{pr}
    \end{pmatrix},
    \end{aligned}
    \end{equation}
with $\chi= \{u,d\}$ and $\psi=\{e,u,d\}$, and the transformation matrix is\footnote{The minus signs arise from the different sign convention in the covariant derivatives; that is, expressed with our conventions some Warsaw operators would be defined with a minus sign. The CP-odd operators receive an additional sign because of the different sign convention for the Levi-Civita tensor used in~\cite{Grzadkowski:2010es}.} 
    \begin{align}
    R_\textrm{g-coup} = \mathrm{diag}\big(\tildepm g_s^3,\; \tildepm g_L^3,\; \tildemp g_s^2,\; \tildemp g_L^2,\; \tildemp g_Y^2,\; \tildemp \tfrac{1}{2}g_L g_Y,\; \eminus g_s,\; \eminus \tfrac{1}{2}g_L,\; \eminus g_Y\big).
    \end{align}
The same transformation applies to the conjugates of the dipole operators.

\paragraph{Gauge-index contractions} These transformations follow from the application of the $ \SU(N) $ Fierz identities~\eqref{eq:group_identities}. The relevant operators are:
    \begin{equation}
    \begin{aligned}
    \boldsymbol{O}_\textrm{g-index}&=
    \begin{pmatrix}
    O_{H\ell}^{{\scriptscriptstyle(\|)} \, pr} &
    O_{H\ell}^{{\scriptscriptstyle(\times)} \, pr} &
    O_{Hq}^{{\scriptscriptstyle(\|)} \, pr} &
    O_{Hq}^{{\scriptscriptstyle(\times)} \, pr} &
    O_m^{{\scriptscriptstyle(\|)}\, prst} &
    O_m^{{\scriptscriptstyle(\times)}\, prst} &
    O_n^{{\scriptscriptstyle(\|)}\, prst} &
    O_n^{{\scriptscriptstyle(\times)}\, prst}
    \end{pmatrix}, \\
    \boldsymbol{Q}_\textrm{g-index}&=
    \begin{pmatrix}
    Q_{H \ell}^{{(1)}\, pr} &
    Q_{H \ell}^{{(3)}\, pr} &
    Q_{H q}^{{(1)}\, pr} &
    Q_{H q}^{{(3)}\, pr} &
    Q_m^{{(1)}\, prst} &
    Q_m^{{(3)}\, prst} &
    Q_n^{{(1)}\, prst} &
    Q_n^{{(8)}\, prst} 
    \end{pmatrix},
    \end{aligned}
    \end{equation}
with $m=\{qq,\ell q\}$ and $n=\{ud,qu,qd,quqd\}$. The transformation matrix is
    \begin{align}
    R_\textrm{g-index}=
    \begin{pmatrix}
    1 & 0 & 0 & 0 & 0 & 0 & 0 & 0 \\
    \frac{1}{2} & \frac{1}{2} & 0 & 0 & 0 & 0 & 0 & 0 \\
    0 & 0 & 1 & 0 & 0 & 0 & 0 & 0 \\
    0 & 0 & \frac{1}{2} & \frac{1}{2} & 0 & 0 & 0 & 0 \\
    0 & 0 & 0 & 0 & 1 & 0 & 0 & 0 \\
    0 & 0 & 0 & 0 & \frac{1}{2} & \frac{1}{2} & 0 & 0 \\
    0 & 0 & 0 & 0 & 0 & 0 & 1 & 0 \\
    0 & 0 & 0 & 0 & 0 & 0 & \frac{1}{3} & 2 \\
    \end{pmatrix}.
    \end{align}
As in the previous block, since $R_\textrm{g-index}$ is real, the same transformations are obtained for the conjugates of the operators.

\paragraph{Higgs-box operator} This block replaces $Q_{H\Box}$ by $O_{HD}^{\scriptscriptstyle(\|)}$. The replacement requires field redefinitions and holds only on-shell. The operators involved in this replacement are
    \begin{align}
    \begin{aligned}
    \boldsymbol{O}_\textrm{box}&=
    \begin{pmatrix}
    \eminus\frac{1}{2}(H^\dagger H)^2 &
    O_H &
    O_{HD}^{\scriptscriptstyle(\|)} &
    O_{eH}^{pr} &
    O_{eH}^{*\, pr} &
    O_{uH}^{pr} &
    O_{uH}^{*\, pr} &
    O_{dH}^{pr} &
    O_{dH}^{*\, pr} 
    \end{pmatrix},\\
    \boldsymbol{Q}_\textrm{box}&=
    \begin{pmatrix}
    \eminus\frac{1}{2}(H^\dagger H)^2 &
    Q_H &
    Q_{H\Box} &
    Q_{eH}^{pr} &
    Q_{eH}^{*\,pr} &
    Q_{uH}^{pr} &
    Q_{uH}^{*\,pr} &
    Q_{dH}^{pr} &
    Q_{dH}^{*\,pr} 
    \end{pmatrix},
    \end{aligned}
    \end{align}
explicitly including the conjugates, and the transformation matrix is
    \begin{align}
    R_\textrm{box}=
    \begin{pmatrix}
    1 & 0 & 0 & 0 & 0 & 0 & 0 & 0 & 0 \\
    0 & 1 & 0 & 0 & 0 & 0 & 0 & 0 & 0 \\
    2\mu_H^2 & \lambda & \frac{1}{2} & \frac{1}{2}Y_e^{pr} & \frac{1}{2}(Y_e^\dagger)^{rp} & \frac{1}{2}Y_u^{pr} & \frac{1}{2}(Y_u^\dagger)^{rp} & \frac{1}{2}Y_d^{pr} & \frac{1}{2}(Y_d^\dagger)^{rp} \\
    0 & 0 & 0 & 1 & 0 & 0 & 0 & 0 & 0 \\
    0 & 0 & 0 & 0 & 1 & 0 & 0 & 0 & 0 \\
    0 & 0 & 0 & 0 & 0 & 1 & 0 & 0 & 0 \\
    0 & 0 & 0 & 0 & 0 & 0 & 1 & 0 & 0 \\
    0 & 0 & 0 & 0 & 0 & 0 & 0 & 1 & 0 \\
    0 & 0 & 0 & 0 & 0 & 0 & 0 & 0 & 1 \\
    \end{pmatrix}.
    \end{align}

\paragraph{Tensor-current operator} The transformation of the Warsaw basis tensor-current operator $Q^{(3)}_{\ell equ}$ into the scalar-current operator $O_{\ell uqe}$ is more subtle. These operators are related by a Fierz transformation, valid only in $d=4$, introducing an evanescent operator that can be removed through a shift of the physical couplings. At one loop, we have~\cite{Fuentes-Martin:2022vvu}:
    \begin{align}
    O_{\ell uqe}^{prst} &= -\frac{1}{2}\,Q_{\ell equ}^{{\scriptscriptstyle (1)}\, ptsr}-\frac{1}{8}\,Q_{\ell equ}^{{\scriptscriptstyle (3)}\, ptsr} +\frac{1}{16\pi^2}\bigg[\Big( \frac{85}{72} g_Y ^{ 2} +  \frac{15}{8} g_L ^{ 2} -  \frac{7}{3} g_s ^{ 2} \Big)Q_{\ell equ}^{{\scriptscriptstyle (1)}\, ptsr} \\ 
    &\quad + \Big( \frac{1}{12} g_s ^{ 2} - \frac{25}{288} g_Y ^{ 2} - \frac{3}{32} g_L ^{ 2} \Big)  Q_{\ell equ}^{{\scriptscriptstyle (3)}\, ptsr} -  \frac{1}{8} (Y_e^\dagger)^{up} (Y_u^\dagger)^{vs} Q_{eu}^{utvr} -  \frac{3}{16} (Y_e^\dagger)^{tu} (Y_u^\dagger)^{vs} Q_{\ell u}^{puvr}  \nonumber\\
    &\quad - \frac{3}{16} (Y_e^\dagger)^{up} (Y_u^\dagger)^{rv} Q_{qe}^{svut}  -  \frac{1}{32} (Y_e^\dagger)^{tu} (Y_u^\dagger)^{rv} Q_{\ell q}^{{\scriptscriptstyle (1)}\, pusv} +  \frac{1}{32} (Y_e^\dagger)^{tu} (Y_u^\dagger)^{rv} Q_{\ell q}^{{\scriptscriptstyle (3)}\, pusv}  \nonumber\\
    &\quad + \frac{1}{4} (Y_d^\dagger)^{us} (Y_u^\dagger)^{rv} Q_{\ell edq}^{ptuv} +  \frac{3}{8} g_L (Y_u^\dagger)^{rs} \Big( \xi_\mathrm{rp} -  1 \Big)  Q_{eW}^{pt} -  \frac{5}{8} g_Y (Y_u^\dagger)^{rs} \Big( \xi_\mathrm{rp} -  1 \Big) Q_{eB}^{pt}\bigg]. \nonumber
    \end{align}
From this expression it is straightforward to read off the transformation matrix $R_\textrm{tensor}$, whose inverse can be obtained perturbatively in the loop expansion. As already noted in~\cite{Fuentes-Martin:2022vvu}, this transformation is inherently ambiguous within the NDR scheme. We encode the dependence on the choice of reading point by the parameter $\xi_\mathrm{rp}$. The value $\xi_\mathrm{rp}=0$ corresponds to choosing the reading point starting from the Higgs interaction or the propagator coming after it, while $\xi_\mathrm{rp}=1$ applies for any other choice. In line with the discussion in Section~\ref{sec:PrescriptionsAndTricks}, we favor the prescription $\xi_\mathrm{rp}=1$.

\paragraph{Trivial replacements} All other operators coincide in both bases: $O_i = Q_i$. While we have retained most operator names in the new basis, we have renamed $Q_{H D}$ and $Q_{\ell equ}^{\scriptscriptstyle (1)}$ for notational consistency. The identical operators are now named $O_{HD}^{\scriptscriptstyle(\times)}$ and $O_{\ell equ}$, respectively. Clearly, the naming does not change the physics, and the transformation matrix for the whole block is simply the identity: $R_\textrm{trivial} = \mathds{1}$.

\section{Evanescent Schemes} \label{app:eva_schemes}

All renormalization schemes must reproduce the same physics. Furthermore, there are no $ \epsilon $-singularities associated with going from the quantum effective action to the full vacuum functional for the connected Green's functions. Thus, we have that  
    \begin{equation} \label{eq:phys_scheme_def}
    \mathcal{P}_{\mathcal{S}} \rstar_{\mathcal{S}} \Gamma_\mathcal{S}(\lambda_{\mathcal{S}}) \bigg|_{\epsilon =0}
    \end{equation} 
is independent on the scheme $\mathcal{S}$. This provides a way to relate various renormalization schemes. Any physical projection will agree down to factors of $ \epsilon $, so in fact one could replace $ \mathcal{P}_{\mathcal{S}} $ in~\eqref{eq:phys_scheme_def} with any other physical projection (from another scheme) without changing the four-dimensional limit.

\subsection{The finitely-compensated evanescent scheme}
In the finitely-compensated evanescent scheme, $ \Pf $, all contributions from the evanescent couplings are compensated by finite counterterms for the physical couplings in the physical subspace of the effective action. The finite counterterms are constructed to exactly cancel all physical contributions from the evanescent couplings. That is, the effective action satisfies  
    \begin{equation}\label{eq:Pf_def}
    \mathcal{P} \rstar_{\Pf} \Gamma_{\mathcal{P}} (g,\, \eta) = \mathcal{P} \rstar_{\Pf} \Gamma_{\mathcal{P}} (g,\, 0)+ \mathcal{O}(\epsilon) = 
    \mathcal{P} \rstar_{\Pbar} \Gamma_{\mathcal{P}} (g,\, 0)+ \mathcal{O}(\epsilon)\,.
    \end{equation}
The form of the effective actions is exclusively determined by the operator basis, so $ \Gamma_{\mathcal{P}}(g,\eta) = \Gamma_{\Pbar}(g,\eta) =\Gamma_{\Pf}(g,\eta) $. The second equality in \eqref{eq:Pf_def} is chosen to ensure that $ \Pf $ mimics a minimal subtraction scheme for the physical couplings at points with evanescent couplings.

Definition~\eqref{eq:Pf_def} implies that the divergent and finite one-loop counterterms for the physical couplings are 
    \begin{align}
    \delta_{\Pf}^{(1)}\!g_a(g, \eta) &= \delta_{\Pf}^{(1)}\!g_a(g, 0)  = \delta_{\Pbar}^{(1)}\!g_a(g, 0)\,, \\ 
    \Delta_{\Pf}^{(1)}\!g_a(g, \eta) \, Q_{\mathcal{P}}^{a} &= \mathcal{P} \rir \! \big[\Gamma^{(1)}_{\mathcal{P}} \!(g,\, 0) - \Gamma^{(1)}_{\mathcal{P}} \!(g,\, \eta) \big] \Big|_{\epsilon=0}\,,
    \end{align}
respectively. This relies on the observation that evanescent operators are rank $ \epsilon $, meaning that they cannot contribute to any one-loop divergence of the physical couplings. Additionally, finite contributions from the evanescent operators come solely from the divergence in the underlying loop-integrals, which ensures that these contributions are local and can be absorbed in a counterterm.

For the calculation of the $ \Pf $-scheme \befs, we will need to transition to and from the $ \Pbar $-scheme. We let $ (g_{\Pf(\Pbar)}, \eta_{\Pf(\Pbar)}) $ denote the renormalized couplings in the $ \Pf $-scheme ($ \Pbar$-scheme). At the special points with $ \eta_{\Pf} = 0 $ the finite counterterms $ \Delta_{\Pf} g_a(g_{\Pf}, 0) =0 $ and the $ \Pf $-scheme reduces to a minimal subtraction scheme. It is trivial that
    \begin{equation} \label{eq:Pf_to_Pbar_couplings}
    g_{\Pbar, a}(g_{\Pf}, \eta_{\Pf} = 0) = g_{\Pf,a}, \qquad
    \eta_{\Pbar, \alpha}(g_{\Pf}, \eta_{\Pf} = 0) = 0,
    \end{equation}
to all orders in the loop expansion. We will not worry about mapping any other coupling points (with $ \eta_{\Pf} \neq 0 $) to the $ \Pbar $-scheme, since physical predictions from the $ \Pf $ scheme are independent of $ \eta_{\Pf} $ at any rate~[cf.~\eqref{eq:Pf_def}]. 

On the other hand, we will need to shift from generic points in the $ \Pbar $-scheme to $ \Pf $, as this is a way to effectively decouple the evanescent couplings. There is no change in the renormalized evanescent couplings between the two schemes, so we parametrize  
    \begin{equation} \label{eq:Pbar_to_Pf}
    g_{\Pf,a}(g_{\Pbar}, \eta_{\Pbar}) \equiv g_{\Pbar,a} + \Delta_{\Pbar\to \Pf} g_a(g_{\Pbar}, \eta_{\Pbar}), \qquad \eta_{\Pf,\alpha}(g_{\Pbar}, \eta_{\Pbar}) = \eta_{\Pbar,\alpha}.
    \end{equation}
Obviously the shift of the physical couplings satisfy $ \Delta_{\Pbar\to \Pf} g_a (g, 0) = 0 $. The requirement~\eqref{eq:phys_scheme_def} that the two schemes produce the same physics informs the extraction of the shift. We have that
    \begin{equation}
    \mathcal{P} \rstar_{\Pbar} \Gamma_{\mathcal{P}} (g_{\Pbar},\, \eta_{\Pbar}) 
    = \mathcal{P} \rstar_{\Pf} \Gamma_{\mathcal{P}} (g_{\Pf},\, \eta_{\Pf})+ \ord{\epsilon}
    = \mathcal{P} \rstar_{\Pf} \Gamma_{\mathcal{P}} (g_{\Pf},\, 0)+ \ord{\epsilon},
    \end{equation}
using~\eqref{eq:Pf_def} in the second step. At one-loop order, one finds 
    \begin{equation}
    \Delta_{\Pbar\to \Pf}^{(1)}g_a(g, \eta) \, Q_{\mathcal{P}}^{a} = \mathcal{P} \rir \! \big[\Gamma^{(1)}_{\mathcal{P}} \!(g,\, \eta) - \Gamma^{(1)}_{\mathcal{P}} \!(g,\, 0) \big] \Big|_{\epsilon=0}= -\Delta_{\Pf}^{(1)}\!g_a(g, \eta) \, Q_{\mathcal{P}}^{a} .
    \end{equation}
In other words one extracts the finite counterterms from the renormalized couplings when changing to the $ \Pf$-scheme. This is essentially an accounting identity.

\subsection{Renormalization group equations}
We can now give a brief description on how to obtain the \befs in the different schemes. 

\paragraph{$ \Pbar $-scheme \befs} 
The $ \Pbar $-scheme is a minimal subtraction scheme, so the (finite) RG functions are fully determined by the simple poles of the coupling counterterms. The $ \ell $-loop \befs are given by
    \begin{equation}
    \beta^{(\ell)}_{\Pbar, a}(g, \eta) =  2 \ell\, \delta^{(\ell)}_{\Pbar,1}g_a(g, \eta)\,, \qquad 
    \beta^{(\ell)}_{\Pbar, \alpha}(g, \eta) =  2 \ell\, \delta^{(\ell)}_{\Pbar,1}\eta_\alpha(g, \eta)
    \end{equation}
for the physical and evanescent couplings, respectively. A simple relation then, at all loop orders.

\paragraph{$ \Pf $-scheme \befs} 
One way to obtain a formula for the $ \Pf $-scheme \befs is to transition from the $ \Pf $-scheme to the $ \Pbar $-scheme, do the running there, and then transitioning back to the $ \Pf $-scheme. By the equivalence between various schemes, this prescription has to reproduce the running directly in the $ \Pf $-scheme. In the $ \Pf $-scheme the running of the physical couplings is independent of the evanescent couplings; that is,\footnote{Technically, the equality need only be satisfied up to contributions that generate additional unphysical coupling rotations in flavor space. See~\cite{Thomsen:2025kka} for a detailed account of the role of such rotations in the RG.}
    \begin{equation}
    \beta_{\Pf, a}(g,\eta) = \beta_{\Pf, a}(g, 0)\,.
    \end{equation}
Otherwise physics could not be independent of evanescent couplings either. 
At vanishing evanescent couplings, the mapping of couplings from the $ \Pf $- to the $ \Pbar $-scheme is trivial as per~\eqref{eq:Pf_to_Pbar_couplings}. Equating the flow of the couplings in the $ \Pf $-scheme to that obtained through mapping to and running in the $\Pbar$-scheme (and then mapping back again) after an infinitesimal RG time step $ \dd t $ produces the equality 
    \begin{equation}
    g_a + \dd t\, \beta_{\Pf,a}(g, 0) = g_{\Pf,a} \big(g_b + \dd t\, \beta_{\Pbar,a}(g,0), \dd t\, \beta_{\Pbar,\alpha }(g,0) \big)\,.  
    \end{equation}
This equality is also visualized in Figure~\ref{fig:RG_schemes}. Linearizing in the change of RG time and applying~\eqref{eq:Pbar_to_Pf} produces the formula
    \begin{equation}
    \beta_{\Pf,a}(g, 0) = \beta_{\Pbar,a}(g, 0) + \beta_{\Pf,\alpha}(g, 0) \dfrac{\partial \Delta_{\Pbar \to \Pf } g_a}{\partial \eta_\alpha}(g,0)\,, 
    \end{equation}
relating the \befs in the two schemes. In the $ \Pbar$-scheme there is a flow into the evanescent space, and this has to be compensated with a shift to the physical couplings when mapping back to the $ \Pf $-scheme. 

At one-loop order then, the formula for the $ \Pf $-scheme \befs in terms of counterterms is
    \begin{equation}
    \beta_{\Pf, a}^{(1)} (g, 0)= 2 \delta^{(1)}_{\Pbar, 1} g_{a}(g, 0),
    \end{equation}
while at two-loop order, it is given by 
    \begin{equation}
    \beta_{\Pf, a}^{(2)} (g, 0)= 4 \delta^{(2)}_{\Pbar, 1} g_{a}(g, 0) + 2 \delta^{(1)}_{\Pbar,1} \eta_{\alpha}(g, 0) \dfrac{\partial \Delta^{(1)}_{\Pbar \to \Pf } g_{a} }{\partial \eta_{\Pbar,\alpha}}(g, 0) \,.
    \end{equation}

\subsection{Operator basis transformations}
So far, all calculations have assumed a particular choice of operator basis $ \mathscr{B}_{\mathcal{P}}$ implicit in the $\Pf$- and $\Pbar$-schemes. However, one might also choose to use another basis $ \mathscr{B}_{\mathcal{P}'} $ (with a different evanescent prescription $ \mathcal{P}'$) with associated schemes $\Pf'$ and $\Pbar'$. We will now see how to map the \befs between the $ \Pf^{(\prime)} $-schemes associated with the different basis choices. 

Physics in both schemes are independent of the renormalized evanescent couplings. The same goes for the \befs of the physical couplings. This ensures that we do not need to make any reference to the evanescent couplings in the mapping formulas between the two schemes, and the evanescent couplings can for all intents and purposes be taken to zero for convenience. When writing the argument $ f(g_{\Pf^{(\prime)}}) $ for some function, it is shorthand for $ f(g_{\Pf^{(\prime)}}, \eta_{\Pf^{(\prime)}}) $, where $ \eta_{\Pf^{(\prime)}} $ is essentially arbitrary.

Let $ \varphi $ be a map such that $ g_{\Pf} = \varphi(g_{\Pf'}) $, where the $ g_{\Pf} $ couplings (in the $ \Pf$-scheme) produces the same physics as $ g_{\Pf'} $ couplings (in the $ \Pf'$-scheme). In other words, the limit~\eqref{eq:phys_scheme_def} should agree in both schemes. We find at tree-level that
    \begin{equation}
    \varphi^{(0)}(g_{\Pf', a}) Q^a_{\mathcal P} = g_{\Pf', a} Q^a_{\mathcal P'} + \ord{\epsilon} = g_{\Pf', a} \mathcal{P} Q^a_{\mathcal P'} + \ord{\epsilon},
    \end{equation}
which lets us solve for $ \varphi^{(0)} $, taking it to be $ \epsilon $-independent. At one-loop order
    \begin{equation}
    \varphi^{(1)}(g_{\Pf', a}) Q^a_{\mathcal P} = \mathcal{P} \Big[ \rstar_{\Pf'}\Gamma^{(1)}_{\mathcal{P}'}(g_{\Pf', a}) - \rstar_{\Pf}\Gamma^{(1)}_{\mathcal{P}}\big(\varphi^{(0)}(g_{\Pf', a}) \big) \Big] + \ord{\epsilon},
    \end{equation}
using $ \mathcal{P} \mathcal{P}' = \mathcal{P} + \ord{\epsilon} $. Note that the physical part of $ \Pf^{(\prime)} $-scheme renormalized effective actions are independent of the evanescent couplings and so is $ \varphi^{(1)}(g_{\Pf', a}) $.
We let $ \varphi' $ denote the map from the $ \Pf $- to $ \Pf'$-schemes---the inverse map of $ \varphi $---such that $ \varphi(\varphi'(g_{\Pf})) = g_{\Pf}$. A loop expansion of this relation gives 
    \begin{equation}
    g_{\Pf,a} = \varphi^{(0)}_a \big( \varphi^{\prime(0)} \big), \qquad 
    \varphi^{(1)}_a \big( \varphi^{\prime(0)} \big) = - \varphi^{\prime(1)}_b \partial^b \varphi^{(0)}_a \big( \varphi^{\prime(0)} \big),
    \end{equation}
where $ \varphi^{\prime(\ell)} = \varphi^{\prime(\ell)}(g_{\Pf}) $.

Let us assume that we are given the physical \befs $ \beta_{\Pf',a} $ in the $ \Pf' $-scheme. This can be used to determine the $ \Pf $-scheme \befs $ \beta_{\Pf,a} $. For physics to be compatible between the schemes, running in the $\Pf$-scheme is equivalent to mapping to the $ \Pf'$-scheme, doing the running there, and then mapping back. Generally, we have 
    \begin{equation}
    \beta_{\Pf,a}(g_{\Pf}) = \dfrac{\partial \varphi_a(g')}{\partial g'_b} \beta_{\Pf',b}(g') \bigg|_{g' = \varphi'(g_{\Pf})}.
    \end{equation}
At one-loop order this gives 
    \begin{equation}
    \beta_{\Pf,a}^{(1)}(g_{\Pf}) = \dfrac{\partial \varphi^{(0)}_a(g')}{\partial g'_b} \beta^{(1)}_{\Pf',b}(g') \Bigg|_{g' = \varphi^{\prime (0)}(g_{\Pf})},
    \end{equation}
and at two-loop order, 
    \begin{multline}\label{eq:betaChangeOfBasis}
    \beta_{\Pf,a}^{(2)}(g_{\Pf}) = \dfrac{\partial \varphi^{(0)}_a(g')}{\partial g'_b} \beta^{(2)}_{\Pf',b}(g') + \dfrac{\partial \varphi^{(1)}_a(g')}{\partial g'_b} \beta^{(1)}_{\Pf',b}(g') \\
    + \varphi^{\prime (1)}_c(g_{\Pf}) \dfrac{\partial}{\partial g'_c} \! \left( \dfrac{\partial \varphi^{(0)}_a(g')}{\partial g'_b} \beta^{(1)}_{\Pf',b}(g') \right) \Bigg|_{g' = \varphi^{\prime (0)}(g_{\Pf})}.
    \end{multline}
Thus, the two-loop basis transformation of the \befs can be determined using the one-loop scheme transformation of the physical couplings.

\section{Evanescent Operators} \label{app:eva_ops}
While Section~\ref{sec:eva_precsription} describes the principles of our evanescent prescription, we list the action of our projection operator $\mathcal{P}$ on all redundant structures encountered in our calculation that are not in our physical operator basis. The corresponding evanescent operators are defined implicitly through~\eqref{eq:d_to_P_basis}. Our choice of $\mathcal{P}$ agrees with~\cite{Aebischer:2025hsx,Dekens:2019ept} (where the redundant structures overlap) such that no change of evanescent basis is needed when using their results to match to and run in the LEFT.   
We highlight the redundant structures that arise already at one-loop order because these evanescent operators are explicitly needed to perform the shift appearing in the two-loop \bef formula~\eqref{eq:two_loop_beta} and calculate the shift~\eqref{eq:RStarShift} needed to account for the discrepancy between $ \rstarbar_{\dbar} $ and $ \rstarbar_{\Pbar} $. At two-loop order, the evanescent operators are still part of our scheme choice, and any $\mathcal{O}(\epsilon)$ pieces in the reduction can contribute to the single poles of the physical counterterms and, thus, the \befs of the theory.

\subsection{Levi-Civita reduction}
The projection $\mathcal{P}$ removes Levi-Civita tensors by applying the reductions~\eqref{eq:lc_gamma_reduction} and~\eqref{eq:lc_lc_reduction}.
We list only the $(\Gamma \, P_\LL)$, $(\Gamma \, P_\LL \, \otimes \, \Gamma \,  P_\LL)$, and $(\Gamma \, P_\LL \, \otimes \, \Gamma \, P_\RR)$ operators for brevity. Flipping each projector in an expression leads to a single minus sign in the reduction because the Levi-Civita tensor is CP-odd.

\subsubsection{One-loop}
\begin{align}
    \mathcal{P} \Big[ \varepsilon_{\mu \nu \rho \sigma} \, \big( \gamma^{\sigma} \, P_\LL \big) \Big] &= i \big( \Gamma_{\mu \nu \rho} \, P_\LL \big), \\
    \mathcal{P} \Big[ \varepsilon_{\mu \nu \rho \sigma} \, \big( \Gamma^{\rho \sigma} \, P_\LL \big) \Big] &= 2i \big( \Gamma_{\mu \nu} \, P_\LL \big), \\
    \mathcal{P} \Big[ \varepsilon_{\mu \nu \rho \sigma} \, \big( \Gamma^{\rho \sigma \kappa} \, P_\LL \big) \Big] &= -2i \delta\du{\nu}{\kappa} \big( \gamma_\mu \, P_\LL \big) + 2i \delta\du{\mu}{\kappa} \big( \gamma_\nu \, P_\LL \big), \\
    \mathcal{P} \Big[ \varepsilon_{\mu \nu \rho \sigma} \, \big( \Gamma^{\nu \rho \sigma} \, P_\LL \big) \Big] &= -6 i \big( \gamma_\mu \, P_\LL \big).
\end{align}

\subsubsection{Two-loop}

\vspace{0.7em}
\sectionlikeParagraph{$\big(\varepsilon * \varepsilon \big)$-type:}
\begin{align}
    \mathcal{P} \Big[ \varepsilon_{\mu \nu \rho \sigma} \, \varepsilon^{\alpha \beta \gamma \sigma} \Big] &= (-3+d) 
    \begin{aligned}[t]
        \Big[
        &\big( \delta\du{\mu}{\gamma} \, \delta\du{\nu}{\beta} - \delta\du{\mu}{\beta} \, \delta\du{\nu}{\gamma} \big) \delta\du{\rho}{\alpha} +
        \big( \delta\du{\mu}{\beta} \, \delta\du{\rho}{\gamma} - \delta\du{\mu}{\gamma} \, \delta\du{\rho}{\beta} \big) \delta\du{\nu}{\alpha} + \\
        &\big( \delta\du{\nu}{\gamma} \, \delta\du{\rho}{\beta} - \delta\du{\nu}{\beta} \, \delta\du{\rho}{\gamma} \big) \delta\du{\mu}{\alpha}  
        \Big]\,,
    \end{aligned} \\
    \mathcal{P} \Big[ \varepsilon_{\mu \nu \rho \sigma} \, \varepsilon^{\alpha \beta \rho \sigma} \Big] &= (6-5d+d^2) \big( \delta\du{\mu}{\beta} \, \delta\du{\nu}{\alpha} - \delta\du{\mu}{\alpha} \, \delta\du{\nu}{\beta} \big)\,, \\
    \mathcal{P} \Big[ \varepsilon_{\mu \nu \rho \sigma} \, \varepsilon^{\alpha \nu \rho \sigma} \Big] &= (6-11d+6d^2-d^3) \, \delta\du{\mu}{\alpha}\,.
\end{align}

\sectionlikeParagraph{$\big( \varepsilon * \Gamma \big)$-type:}
\begin{align}
    \mathcal{P} \Big[ \varepsilon^{\mu \nu \rho \alpha} \, \big( \Gamma_{\mu \nu \rho \beta} \, P_\LL \big) \Big] &= -6 i \, P_\RR \, \delta\du{\beta}{\alpha}\,, \\
    \mathcal{P} \Big[ \varepsilon^{\mu \nu \rho \sigma} \, \big( \Gamma_{\mu \nu \rho \sigma} \, P_\LL \big) \Big] &= -24 i \, P_\LL \,,\\
    \mathcal{P} \Big[ \varepsilon^{\mu \nu \rho \alpha} \, \big( \Gamma_{\mu \nu \rho \beta \gamma} \, P_\LL \big) \Big] &= 0 \,,\\
    \mathcal{P} \Big[ \varepsilon^{\mu \nu \rho \sigma} \, \big( \Gamma_{\mu \nu \rho \sigma \alpha} \, P_\LL \big) \Big] &= 0 \,,\\
    \mathcal{P} \Big[ \varepsilon^{\mu \nu \rho \sigma} \, \big( \Gamma_{\mu \nu \rho \sigma \alpha \beta} \, P_\LL \big) \Big] &= 0\,, \\
    \mathcal{P} \Big[ \varepsilon^{\mu \nu \rho \sigma} \, \big( \Gamma_{\mu \nu \rho \sigma \alpha \beta \gamma} \, P_\LL \big) \Big] &= 0\,.
\end{align}

\sectionlikeParagraph{$\big( \varepsilon * \big[ \Gamma \otimes \Gamma \big] \big)$-type:}
\begin{align}
    \mathcal{P} \Big[ \varepsilon^{\mu \nu \rho \sigma} \, \big( \Gamma_{\mu \nu} \, P_\LL \, \otimes \, \Gamma_{\rho \sigma} \, P_\LL \big) \Big] &= 2i \, \big( \Gamma_{\mu \nu} \, P_\LL \, \otimes \, \Gamma^{\mu \nu} \, P_\LL \big)\,, \\
    \mathcal{P} \Big[ \varepsilon^{\mu \nu \rho \sigma} \, \big( \Gamma_{\mu \nu} \, P_\LL \, \otimes \, \Gamma_{\rho \sigma} \, P_\RR \big) \Big] &= 0\,, \\
    \mathcal{P} \Big[ \varepsilon^{\mu \nu \rho \sigma} \, \big( \Gamma_{\mu \nu \alpha} \, P_\LL \, \otimes \, \Gamma\ud{\alpha}{\rho \sigma} \, P_\LL \big) \Big] &= 0 \,,\\
    \mathcal{P} \Big[ \varepsilon^{\mu \nu \rho \sigma} \, \big( \Gamma_{\mu \nu \alpha} \, P_\LL \, \otimes \, \Gamma\ud{\alpha}{\rho \sigma} \, P_\RR \big) \Big] &= 0 \,,\\
    \mathcal{P} \Big[ \varepsilon^{\mu \nu \rho \sigma} \, \big( \Gamma_{\rho \alpha} \, P_\LL \, \otimes \, \Gamma\ud{\alpha}{\mu \nu \sigma} \, P_\LL \big) \Big] &= 0\,, \\
    \mathcal{P} \Big[ \varepsilon^{\mu \nu \rho \sigma} \, \big( \Gamma_{\rho \alpha} \, P_\LL \, \otimes \, \Gamma\ud{\alpha}{\mu \nu \sigma} \, P_\RR \big) \Big] &= 0 \,,\\
    \mathcal{P} \Big[ \varepsilon^{\mu \nu \rho \sigma} \, \big( \gamma^\alpha \, P_\LL \, \otimes \, \Gamma_{\mu \nu \rho \sigma \alpha} \, P_\LL \big) \Big] &= 0 \,,\\
    \mathcal{P} \Big[ \varepsilon^{\mu \nu \rho \sigma} \, \big( \gamma^\alpha \, P_\LL \, \otimes \, \Gamma_{\mu \nu \rho \sigma \alpha} \, P_\RR \big) \Big] &= 0\,.
\end{align}

\subsection{$\gamma$-reduction} \label{app:gamma_reduction}
The projection $\mathcal{P}$ reduces four-fermion Dirac structures to the four-dimensional chiral basis given in Table \ref{tab:dirac_basis}.
We list only the $(\Gamma \, P_\LL \, \otimes \, \Gamma \,  P_\LL)$ and $(\Gamma \, P_\LL \, \otimes \, \Gamma \, P_\RR)$ operators, since the reductions are invariant under simultaneous flip of all projectors in the equalities.

\subsubsection{One-loop}
\begin{alignat}{2}
    \mathcal{P} \Big[ \big( \Gamma_{\mu \nu} \, P_{\LL} \, &\otimes \, \Gamma^{\mu \nu} \, P_{\RR} \big) \Big] &&= (-4 + 5d - d^2) \big( P_{\LL} \, \otimes \,  P_{\RR} \big) ,\label{eq:LR_tensor_reduction}\\
    \mathcal{P} \Big[ \big( \Gamma_{\mu \nu \rho} \, P_{\LL} \, &\otimes \, \Gamma^{\mu \nu \rho} \, P_{\LL} \big) \Big] &&= (-6 + 7d - d^2) \big( \gamma_{\mu} \, P_{\LL} \, \otimes \, \gamma_{\mu} \, P_{\LL} \big), \\
    \mathcal{P} \Big[ \big( \Gamma_{\mu \nu \rho} \, P_{\LL} \, &\otimes \, \Gamma^{\mu \nu \rho} \, P_{\RR} \big) \Big] &&= (-2 + 3d - d^2) \big( \gamma_{\mu} \, P_{\LL} \, \otimes \, \gamma_{\mu} \, P_{\RR} \big).
\end{alignat}

\subsubsection{Two-loop}
\begin{alignat}{2}
    \mathcal{P} \Big[ \big( \Gamma_{\mu \nu \rho \sigma} \, P_{\LL} \, &\otimes \, \Gamma^{\mu \nu \rho \sigma} \, P_{\LL} \big) \Big] &&= 
    \begin{aligned}[t]
        &-d(-18+27d-10d^2+d^3) \big( P_{\LL} \, \otimes \, P_{\LL} \big) \\ &- 2(12-7d+d^2) \big( \Gamma_{\mu \nu} \, P_{\LL} \, \otimes \, \Gamma^{\mu \nu} \, P_{\LL} \big),
    \end{aligned} \\
    \mathcal{P} \Big[ \big( \Gamma_{\mu \nu \rho \sigma} \, P_{\LL} \, &\otimes \, \Gamma^{\mu \nu \rho \sigma} \, P_{\RR} \big) \Big] &&= (48-94d+59d^2-14d^3+d^4) \big( P_{\LL} \, \otimes \, P_{\RR} \big), \\
    \mathcal{P} \Big[ \big( \Gamma_{\mu \nu \rho \sigma \kappa} \, P_{\LL} \, &\otimes \, \Gamma^{\mu \nu \rho \sigma \kappa} \, P_{\LL} \big) \Big] &&= (120-202d+99d^2-18d^3+d^4) \big( \gamma_{\mu} \, P_{\LL} \, \otimes \, \gamma_{\mu} \, P_{\LL} \big), \\
    \mathcal{P} \Big[ \big( \Gamma_{\mu \nu \rho \sigma \kappa} \, P_{\LL} \, &\otimes \, \Gamma^{\mu \nu \rho \sigma \kappa} \, P_{\RR} \big) \Big] &&= (24-50d+35d^2-10d^3+d^4) \big( \gamma_{\mu} \, P_{\LL} \, \otimes \, \gamma_{\mu} \, P_{\RR} \big), \\
    \mathcal{P} \Big[ \big( C \, \Gamma_{\mu \nu} \, P_{\LL} \, &\otimes \, C \, \big(\Gamma^{\mu \nu}\big)^{\!\intercal} \, P_{\RR} \big) \Big] &&= 0
\end{alignat}
The last Dirac structure of this list contains the charge conjugation matrix and is needed to simplify counterterms generated by the double Weinberg operator insertions. It is seemingly at odds with~\eqref{eq:LR_tensor_reduction}; however, without the fermion flow to order the direction of each spinor line, an ambiguity arises: transposing one of the spin line before or after the reduction yields different signs, producing a contradiction for any non-zero physical projection. A similar issue was encountered in~\cite{Dekens:2019ept}.

\subsection{Schouten identities}
Schouten identities can only be applied to operators with summation over at least five different pairs of Lorentz indices, making them rarefied within the dimension-six SMEFT. Only two redundant operators that emerge in our calculations (both at two-loop order) require the application of Schouten identities: 
\begin{align}
    \mathcal{P} \Big[ B^{\mu \nu} \, W^{I\rho \sigma} \, W^I\du{\sigma}{\kappa} \, \varepsilon_{\mu \nu \rho \kappa} \Big] &= 0\,, \\
    \mathcal{P} \Big[ B^{\mu \nu} \, G^{A\rho \sigma}\, G^A\du{\sigma}{\kappa} \, \varepsilon_{\mu \nu \rho \kappa}\Big] &= 0\,.
\end{align}

\subsection{Fierz identities}
The Fierz identities relate four-fermion operators already with Dirac structures in the physical basis~\ref{tab:dirac_basis}. Contrary to the other reductions provided in this appendix, the Fierz identities must be applied judiciously to reduce operators to those in the operator basis. The reduction depends, among other things, on which fermion fields are involved. The set of identities is 
\begin{align}
    \mathcal{P} \Big[ \big(\Bar{\psi}_1 \gamma_\mu P_\RR \, \psi_4\big) \big(\Bar{\psi}_3 \gamma^\mu P_\RR \, \psi_2 \big) \Big] &= \big(\Bar{\psi}_1 \gamma_\mu P_\RR \, \psi_2\big) \big(\Bar{\psi}_3 \gamma^\mu P_\RR \, \psi_4\big), \\
    \mathcal{P} \Big[ \big(\Bar{\psi}_1 \gamma_\mu P_\LL \, \psi_4 \big) \big(\Bar{\psi}_3  \gamma^\mu P_\LL \, \psi_2 \big) \Big] &= \big(\Bar{\psi}_1 \gamma_\mu P_\LL \, \psi_2 \big) \big(\Bar{\psi}_3 \gamma^\mu P_\LL \, \psi_4\big), \\
    \mathcal{P} \Big[ \big(\Bar{\psi}_1  \gamma_\mu P_\RR \, \psi_4\big) \big(\Bar{\psi}_3 \gamma^\mu P_\LL \, \psi_2\big) \Big] &= -2 \big(\Bar{\psi}_1 P_\LL \, \psi_2\big) \big(\Bar{\psi}_3 P_\RR \, \psi_4 \big), \\
    \mathcal{P} \Big[ \big( \Bar{\psi}_1 \Gamma_{\mu \nu} P_\RR \, \psi_2 \big) \big( \Bar{\psi}_3 \Gamma^{\mu \nu} P_\RR \, \psi_4  \big) \Big] &= 4 \big( \Bar{\psi}_1  P_\RR \, \psi_2 \big) \big(\Bar{\psi}_3 P_\RR \, \psi_4\big) + 8 \big( \Bar{\psi}_1 P_\RR \, \psi_4 \big) \big( \Bar{\psi}_3 P_\RR \, \psi_2 \big), \\
    \mathcal{P} \Big[ \big( \Bar{\psi}_3 \gamma_\mu P_\LL \, \psi_1 \big) \big( \Bar{\psi}_4 \gamma^\mu P_\LL \, \psi_2 \big) \Big] &= 2 \big(\overline{\psi_1^\cc} P_\LL \, \psi_2 \big) \big( \Bar{\psi}_3 P_\RR \, \psi^\cc_4 \big), \\
    \mathcal{P} \Big[ \big(\Bar{\psi}_3 \gamma_\mu P_\RR \, \psi_1 \big) \big(\Bar{\psi}_4 \gamma^\mu P_\LL \, \psi_2\big) \Big] &= \big(\overline{\psi^\cc_1} \gamma^\mu P_\LL \, \psi_2 \big) \big( \Bar{\psi}_3 \gamma_\mu P_\RR \, \psi^\cc_4 \big).
\end{align}
One will combine these four-dimensional identities to obtain a complete reduction of any redundant operators to the physical basis. The only revision needed to adapt the resulting identities for $ d $-dimensions is to identify the difference between the redundant and physical operators with an evanescent operator.

\section{Divergent Field Rotations} \label{app:field_shifts}
Divergent flavor rotations are needed to ensure finite \befs; without them, the usual `t Hooft consistency conditions are violated (although the \befs still generate a correct finite flow of the observables). We introduce three new anti-Hermitian/imaginary objects for shorthand: a scalar $S$ and two flavor space tensors $A$ and $B$ defined as
\begin{align}
    S_{X,a} &= \Tr\!\left( C_{X} \, Y_{a}^{\dagger} - Y_a \, C_{X}^{\dagger}\right) \, , \\
    A^{pr}_{X,a} &= \left( C_{X} \, Y_{c}^{\dagger} - Y_a \, C_{X}^{\dagger}\right)^{pr} \, , \\
    B^{pr}_{X,a} &= \left( C_{X}^{\dagger} Y_{a} - Y_a^{\dagger} C_{X} \right)^{pr} \, ,
\end{align}
where $\{p,r\}$ denote flavor indices, while $a \in \{ u,d,e\}$ denotes the type of Yukawa coupling and $ X $ is the subscript in the dimension-six Wilson coefficient $ C_X $. Additional terms proportional to $i$ times a Wilson coefficient also appear. 

The field redefinitions take the form of $ \psi \to \psi + \omega \psi $, where $ \omega $ is an element of the Lie algebra associated with the flavor group $ \U(3)_\psi $.\footnote{For the purpose of these redefinitions $ \omega $ is conventionally taken to be an anti-Hermitian matrix~\cite{Herren:2021yur} contrary to the gauge groups, where one typically uses $ i $ times a Hermitian matrix.} 
We find suitable shifts of the SM fermion fields to be

\begin{align}
    q^p \to q^p - \frac{3}{8} \frac{\mu_H^2}{(16\pi^2)^2 \, \epsilon^2}
    \Big( A^{pr}_{uH,u} + A^{pr}_{dH,d} - g_Y^2 \, A^{pr}_{uB,u} + g_Y^2 \, A^{pr}_{dB,d} + \frac{3}{2} \, g_L^2 \, A^{pr}_{uW,u} + \frac{3}{2} \, g_L^2 \, A^{pr}_{dW,d} \Big) q^r \, ,
\end{align}

\begin{align}
    \begin{split}
        u^p \to u^p - \frac{1}{2} \frac{\mu_H^2}{(16\pi^2)^2 \, \epsilon^2} \Bigg\{
        &\frac{3}{2}\Big( B^{pr}_{uH,u} + g_Y^2 \, B^{pr}_{uB,u} - \frac{3}{2} \, g_L^2 \, B^{pr}_{uW,u} \Big) u^r + \\
        &\frac{3}{2}\Big( S_{eH,e} - \frac{3}{2} \, S_{uH,u} + \frac{3}{2} \, S_{dH,d} \Big) u^p + \\
        &i \Bigg( \frac{17}{3} \, g_Y^4 \, C_{H\widetilde{B}} + 9 \, g_L^4 \, C_{H\widetilde{W}} + 32 \, g_s^4 \, C_{H\widetilde{G}} \Bigg) u^p
        \Bigg\} \, ,
    \end{split}
\end{align}

\begin{align}
    \begin{split}
        d^p \to d^p - \frac{1}{2} \frac{\mu_H^2}{(16\pi^2)^2 \, \epsilon^2} \Bigg\{
        &\frac{3}{2}\Big( B^{pr}_{dH,d} - g_Y^2 \, B^{pr}_{dB,d} - \frac{3}{2} \, g_L^2 \, B^{pr}_{dW,d} \Big) d^r - \\
        &\frac{3}{2} \Big( S_{eH,e} - \frac{3}{2} \, S_{uH,u} + \frac{3}{2} \, S_{dH,d} \Big) d^p + \\
        &i \Bigg( \frac{5}{3} \, g_Y^4 \, C_{H\widetilde{B}} + 9 \, g_L^4 \, C_{H\widetilde{W}} + 32 \, g_s^4 \, C_{H\widetilde{G}} \Bigg) d^p
        \Bigg\} \, ,
    \end{split}
\end{align}

\begin{align}
    \begin{split}
        \ell^p \to \ell^p - \frac{3}{8} \frac{\mu_H^2}{(16\pi^2)^2 \, \epsilon^2} \Big( A^{pr}_{eH,e} + g_Y^2 \, A^{pr}_{eB,e} + \frac{3}{2} \, g_L^2 \, A^{pr}_{eW,e} \Big) \ell^r \, ,
    \end{split}
\end{align}

\begin{align}
    \begin{split}
        e^p \to e^p - \frac{1}{2} \frac{\mu_H^2}{(16\pi^2)^2 \, \epsilon^2} \Bigg\{
        &\frac{3}{2} \Big(B^{pr}_{eH,e} - g_Y^2 \, B^{pr}_{eB,e} - \frac{3}{2} \, g_L^2 \, B^{pr}_{eW,e} \Big) e^r - \\
        &\frac{3}{2}\Big( S_{eH,e} - 3 \, S_{uH,u} + 3 \, S_{dH,d} \Big) e^p + \\
        &i \Big( 15 \, g_Y^4 \, C_{H\widetilde{B}} + 9 \, g_L^4 \, C_{H\widetilde{W}} \Big) e^p
        \Bigg\} \, .
    \end{split}
\end{align}
These shifts are sufficient to remove the divergences~(\ref{eq:div_ye}--\ref{eq:div_yd}), showing that the double poles of the counterterms are consistent.

\sectionlike{References}
\vspace{-10pt}
\bibliography{References}

\end{document}